%%%%%%%%%%%%%%%%%%%%%%%%%%%%%%%%%%%%%%%%%%%%%%%%%%%%%%%%%%%%%%%%%%
%                                                                %
%   Perturbation Theory around Non--Nested Fermi Surfaces        %
%   I. Keeping the Fermi Surface Fixed                           %
%                                                                %
%   by Joel Feldman, Manfred Salmhofer, and Eugene Trubowitz     %
%                                                                %
%   August 1995, plain TeX, 20 postscript figures                %
%                                                                %
%%%%%%%%%%%%%%%%%%%%%%%%%%%%%%%%%%%%%%%%%%%%%%%%%%%%%%%%%%%%%%%%%% 
%%%%%%%
%%%%%%%     SELECT (a) YOUR POSTSCRIPT FILE SUFFIX AND (b) YOUR SYSTEM  NOW!
%%%%%%%
\newcount\system
%\global\system=1   % for textures 
%\global\system=2   % for msdos
\global\system=3   % for unix(dvips)
%\global\system=4   % for unix(dvips) scaled by a factor of 1.2
%\global\system=5   % for orphee dvips
%
%     Layout macros
%
\font\Cc=cmcsc10

\def\Thsty{\it}
\def\Lesty{\it}
\def\Cl#1{\centerline{#1}}
\magnification=\magstep1
\baselineskip=16pt
\def\Refskip{\vfill\eject}
\hfuzz=3pt
\overfullrule=0pt
\def\endpage{\vfill\eject}

\def\someroom{\removelastskip\par\vskip 30 true pt 
plus 10 truept minus 10 truept} %\noindent}
\def\hroom{\hglue 10pt }
\def\pni{\par\noindent}
\def\sni{\someroom\noindent}
\def\leftit#1{\par\noindent\hangindent=40pt\hangafter=1
           \hbox to 30pt{\hglue10pt#1\hss}\hglue 10pt\ignorespaces}
\headline={\ifodd\pageno\rightheadline \else\leftheadline\fi}
\def\rightheadline{\hfil}
\def\leftheadline{\hfil}
\footline={\ifodd\pageno\rightfootline \else\leftfootline\fi}
\def\rightfootline{\hss\tenrm\folio\hss}
\def\leftfootline{\hss\tenrm\folio\hss}
%
%
%    Standard Math--Macros
%
%
%   Numerals
%
\def\N{{\rm I\kern-.16em N}}
\font\twblk=cmss10
\font\tenblk=cmss8
\font\eiblk=cmss8
\def\Z{\mathchoice{{\hbox{\twblk Z\kern-.35emZ}}}
{{\hbox{\twblk Z\kern-.35emZ}}}
{{\hbox{\tenblk Z\kern-.30emZ}}}
{{\hbox{\eiblk Z\kern-.24emZ}}}}
\def\Q{{\rm \kern.25em\vrule height1.4ex
depth-.12ex width.06em\kern-.31em Q}}
\def\R{{\rm I\kern-.2emR}}
\def\C{{\rm \kern.25em\vrule height1.4ex
depth-.12ex width.06em\kern-.31em C}}
\def\K{{\rm I\kern-.2emK}}
\def\L{{\rm I\kern-.2emL}}
\def\bbbone{{\mathchoice {\rm 1\mskip-4mu l} {\rm 1\mskip-4mu l}    %\bbbone
{\rm 1\mskip-4.5mu l} {\rm 1\mskip-5mu l}}}
%
%  greek alphabet
%
\def\al{\alpha}
\def\be{\beta}
\def\ga{\gamma}
\def\de{\delta}
\def\ep{\epsilon}
\def\veps{\varepsilon}
\def\et{\eta}
\def\ze{\zeta}
\def\th{\theta}
\def\vth{\vartheta}

\def\ka{\kappa}
\def\la{\lambda}
\def\rh{\rho}
\def\si{\sigma}
\def\ta{\tau}

\def\ph{\phi}
\def\vphi{\varphi}
\def\ch{\chi}
\def\ps{\psi}
\def\om{\omega}
\def\vphi{\varphi}
\def\veps{\varepsilon}
%
% large
%
\def\Ga{\Gamma}
\def\De{\Delta}

\def\La{\Lambda}
\def\Si{\Sigma}

\def\Ph{\Phi}
\def\Ps{\Psi}
\def\Om{\Omega}
%
%   for fermionic reasons...
%
\def\alq{{\bar\alpha}}

\def\etq{{\bar\eta}}

\def\chq{{\bar\chi}}
\def\psq{{\bar\psi}}

%

%
% primes...
%
\def\alp{{\al ^\prime }}
\def\bep{{\be ^\prime }}

\def\tap{{\ta ^\prime }}

\def\Tp{{T^\prime}} 
\def\xp{{x ^\prime }}
\def\fp{{f^\prime}}
\def\fdp{{f^{\prime\prime}}}
\def\tp{{t^\prime}}
\def\Gp{{G^\prime}}
%
% calligraphic letters
%
\def\cA{{\cal A}}
\def\cB{{\cal B}}
\def\cC{{\cal C}}
\def\cD{{\cal D}}
\def\cE{{\cal E}}
\def\cF{{\cal F}}
\def\cG{{\cal G}}

\def\cJ{{\cal J}}
\def\cK{{\cal K}}
\def\cL{{\cal L}}
\def\cM{{\cal M}}
\def\cN{{\cal N}}
\def\cO{{\cal O}}
\def\cP{{\cal P}}
\def\cQ{{\cal Q}}
\def\cR{{\cal R}}

\def\cT{{\cal T}}
\def\cU{{\cal U}}
\def\cV{{\cal V}}

\def\cY{{\cal Y}}

%
% smaller numbers for subscripts
%
\def\zer{{\oldstyle 0}}
\def\one{{\oldstyle 1}}
\def\two{{\oldstyle 2}}
\def\tG{{\tilde G}}
\def\prP{{\bf P}}
\def\db{{\mkern2mu\mathchar'26\mkern-2mu\mkern-9mud}}
\def\tr{\hbox{ tr }}
\def\I2{{$I_2$}}
\def\del{\partial}

\def\ve#1{{\bf #1}}
\def\eof#1{e(\ve{#1})}

\def\tst#1{{{\theta_{#1}}^*}}

\def\nat#1{\{ 1,\ldots,#1 \} }
\def\natz#1{\{ 0,\ldots,#1 \} }
\def\abs#1{{\left\vert #1 \right\vert}}
\def\norm#1{{\left\Vert #1 \right\Vert}}

\def\Pol#1{\hbox{ pol}(#1)}
\def\dst{\displaystyle}

\def\tst{\textstyle}
\def\frac#1#2{\dst {#1\over#2}}     % fractions in displaystyle
\def\sfrac#1#2{{\tst{#1\over#2}}}   % fractions in textstyle    

\def\dbar{{\mkern2mu\mathchar'26\mkern-2mu\mkern-9mud}}

\def\Const{\hbox{ \rm const }}
\def\NOL{non--\-over\-lap\-ping}
\def\OL{{o\-ver\-lap\-ping}}
\def\SSI{{same scale insertion}}
%
%
%     Macros for numbering Chapters, Sections, Equations, Theorems etc.
%
\message{Numbers of Theorems etc. will be saved to file \jobname.txs}
\immediate\openout8=\jobname.txs
\newcount\sectno
\newcount\chapno
\newcount\equano
\newcount\theono
\newcount\refno
\sectno=0
\chapno=0
\equano=0
\theono=0
\refno=0
\def\eqhead{ }
\def\\{\backslash}
\font\bfe=cmbx12
\def\chapskip{\removelastskip\par\vglue 40pt}%\noindent}
\def\sectskip{\removelastskip\par\vskip 40pt}%\noindent}
\def\blwskip{\removelastskip\par\vskip 20pt}%\noindent}
\def\chap#1{\equano=0 \sectno=0 \theono=0 \global\advance\chapno by 1%
\def\eqhead{\number\chapno .}%
\chapskip\goodbreak\centerline{\bfe \eqhead \hglue 5pt #1}
\blwskip}%
\def\sect#1{\global\advance\sectno by 1
\sectskip\goodbreak%
\noindent{\bf \eqhead\number\sectno  \hglue 5pt #1}
\blwskip}%
\def\appendix#1#2{\par\chapskip\goodbreak\centerline{\bf Appendix #1. #2}%
\blwskip%
\equano=0\sectno=0\theono=0\def\eqhead{#1.}}
\edef\Raum{ }
\def\eqn{{\hbox{\global\advance\equano by 1}}%
\eqno (\eqhead\number\equano )}%
\def\EQN#1{\eqn\edef\Zwi{(\eqhead\number\equano )}%
\immediate\write8{EQN  \Zwi\Raum = \noexpand#1}
\global\let #1=\Zwi
}
\def\queq#1{$#1$}
\def\The#1{{\global\advance \theono by 1 \someroom \noindent% 
{\bf Theorem
\eqhead\number\theono }\hroom
\edef\Zwi{\eqhead\number\theono}
\immediate\write8{STM  \Zwi\Raum = \noexpand#1    Theorem }
\global\let#1=\Zwi
}}
\def\Pro#1{{\global\advance \theono by 1 \someroom \noindent% 
{\bf Proposition
\eqhead\number\theono }\hroom
\edef\Zwi{\eqhead\number\theono}
\immediate\write8{STM  \Zwi\Raum = \noexpand#1    Proposition }
\global\let#1=\Zwi}}
\def\Rem#1{{\global\advance \theono by 1 \someroom \noindent% 
{\bf Remark
\eqhead\number\theono }\hroom
\edef\Zwi{\eqhead\number\theono}
\immediate\write8{STM  \Zwi\Raum = \noexpand#1    Remark }
\global\let#1=\Zwi}}
\def\Cor#1{{\global\advance \theono by 1 \someroom \noindent%
{\bf Corollary
\eqhead\number\theono }\hroom
\edef\Zwi{\eqhead\number\theono}
\immediate\write8{STM  \Zwi\Raum = \noexpand#1    Corollary }
\global\let#1=\Zwi}}
\def\Def#1{{\global\advance \theono by 1 \someroom \noindent% 
{\bf Definition
\eqhead\number\theono }\hroom
\edef\Zwi{\eqhead\number\theono}
\immediate\write8{STM  \Zwi\Raum = \noexpand#1    Definition }
\global\let#1=\Zwi}}
\def\Lem#1{{\global\advance \theono by 1 \someroom \noindent% 
{\bf Lemma
\eqhead\number\theono }\hroom
\edef\Zwi{\eqhead\number\theono}
%\global\edef\Shwi{Lemma \Zwi \#1}
\immediate\write8{STM  \Zwi \Raum = \noexpand#1    Lemma }
\global\let#1=\Zwi
}}
\def\Proof{\someroom \noindent{\it Proof:}\hroom}
\def\endproof{\hfill
\hbox{\vrule width 7pt depth 0pt height 7pt} \someroom}

\def\refit#1{\par\noindent\hangindent=1.5cm\hangafter=1
           \hbox to 1cm{#1\hss}\hglue 0.5cm\ignorespaces}
\def\Ref#1#2{\refit{[#1]} #2}
\def\quref#1{[#1]}
\def\suffix{ps}
\def\ifundefined#1{\expandafter\ifx\csname#1\endcsname\relax}
\ifundefined{figdir}\def\figdir{}\fi
\def\setlabelsize{}
\newif\ifboxfigures
\boxfiguresfalse
\newdimen\boxrulewidth \newdimen\boxborderwidth
\boxrulewidth=1pt \boxborderwidth=2pt
%
% macro for drawing boxes
%
\def\boxit#1{\vtop{\hrule height\boxrulewidth%
\hbox{\vrule width\boxrulewidth\kern\boxborderwidth%
\vbox{\kern\boxborderwidth#1\kern\boxborderwidth}\kern\boxborderwidth%
\vrule width\boxrulewidth}%
\hrule height\boxrulewidth}}  
%
% Now for the definitions and main macro for figure inclusion.
%
\newcount\mgversion
\newdimen\pswidth  \newdimen\xleft
\newdimen\psheight \newdimen\ytop \newdimen\ybot
\newcount\justx \newcount\justy
\global\justx=0 \global\justy=0
\newdimen\vpos \newtoks\label 
\newread\labelfile \newdimen\xcoord \newdimen\ycoord
\newif\ifdoit 
\newbox\labox
\newcount\temp
\def\readdim#1{\global\read\labelfile to \temp
\global #1=\temp pt}
\def\figinsert#1{\par%  #1=filename
\openin\labelfile=\figdir#1.lbl                                              
\global\read\labelfile to\mgversion\message{#1}               
\readdim{\pswidth}                                     
\readdim{\psheight}                                    
\ifboxfigures\boxit{\fi\vbox to\psheight {\vfill
%%%
%%% NOTE: next lines may have to be changed for your DVIPS driver %%%
\ifnum\system=1
\fi%%textures
\ifnum\system=2
\fi%%msdos
\ifnum\system=3
\fi         %%unix:dvips
\ifnum\system=4
\fi         %%unix:dvips,scaled
\ifnum\system=1\special{postscriptfile  \figdir#1.\suffix  }\fi    %%textures
\ifnum\system=2\special{ps: plotfile \figdir#1.\suffix } \fi       %%msdos  
\ifnum\system=3\includegraphics{\figdir#1.\suffix} \fi             %%unix:dvips 
\ifnum\system=4\includegraphics{\figdir#1.\suffix}\fi
                                                             %%unix:dvips,scaled
\ifnum\system=5\includegraphics{\figdir#1.\suffix} \fi            %%orphee
%%%
\vskip-\psheight \setlabelsize%                                     
\hbox to\pswidth  {\hss}%                                            
\parindent=0pt\offinterlineskip                                       
\vpos=0 pt%                                                              
\loop\readdim{\xcoord}                                 
\ifdim \xcoord < -999pt \doitfalse\else\doittrue\fi                        
\ifdoit \readdim{\ycoord}                              
\global\read\labelfile to\justx                                       
\global\read\labelfile to\justy                                       
\global\read\labelfile to\label
\global\setbox\labox=\hbox{\label\hskip-0.3em}%    
\advance\vpos by-\ycoord                                              
\vskip-\vpos \vpos=\ycoord                                         
\hbox to\pswidth{\hskip\xcoord %                                 
\hbox to 0pt{\ifnum\justx>0\hss\fi%                                   
\vbox to0pt{%                                                         
\ifnum\justy<2\vss\fi%                                                
\copy\labox\kern0pt%  
\ifnum\justy>0\vss\fi}%                                               
\ifnum\justx<2\hss\fi}%                                               
\hss}%                                                                
\repeat%                                                              
\advance\vpos by-\psheight%                                           
\vskip-\vpos %                                                      
}\ifboxfigures}\fi\closein\labelfile}                                                  
%
% 
%    figcrop{<filename,w/o extension>} treats the first two labels as marking
%    the upper left and lower right corners of the figure. This is for
%    positioning purposes only. The figure may extend beyond the corners.
%    The corner markers are not printed.
%
%
\def\figcrop#1{\par%  #1=filename
\openin\labelfile=\figdir#1.lbl                                              
\global\read\labelfile to\mgversion\message{#1}               
\global\read\labelfile to\temp%read overall dimensions                                     
\readdim{\ybot}
\readdim{\xleft}%               read upper left point
\readdim{\ytop}
\global\read\labelfile to\justx%ignore
\global\read\labelfile to\justy%ignore
\global\read\labelfile to\label%ignore
\readdim{\pswidth}%            read lower right point
\global\advance\pswidth by -\xleft
\readdim{\psheight}
\global\advance\ybot by -\psheight
\global\advance\psheight by -\ytop
\global\read\labelfile to\justx%ignore
\global\read\labelfile to\justy%ignore
\global\read\labelfile to\label%ignore                                    
\ifboxfigures\boxit{\fi\vbox to\psheight{\vfill
%%%
%%% NOTE: next line may have to be changed for your DVIPS driver %%%
\ifnum\system=1
\fi %textures
\ifnum\system=2
\fi %msdos
\ifnum\system=3
\fi         %%unix:dvips
\ifnum\system=4
\fi         %%unix:dvips,scaled
\ifnum\system=1
\hbox to \pswidth{\kern-\xleft\special{postscriptfile \figdir#1.\suffix }\hfil}\fi
                                                              %textures
\ifnum\system=2
\hbox to \pswidth{\kern-\xleft\special{ps: plotfile \figdir#1.\suffix }\hfil}\fi
                                                              %mdos 
\ifnum\system=3
\hbox to \pswidth{\kern-\xleft\includegraphics{\figdir#1.\suffix}\hfil}\fi
                                                             %unix:dvips 
\ifnum\system=4
\hbox to \pswidth{\kern-\xleft\includegraphics{\figdir#1.\suffix}\hfil}\fi
                                                             %unix:dvips,scaled
\ifnum\system=5
\hbox to \pswidth{\kern-\xleft\includegraphics{\figdir#1.\suffix}\hfil}\fi %orphee
%%%
\vskip -\baselineskip
\vskip -\ybot 
\vskip-\psheight \setlabelsize%                                     
\hbox to\pswidth  {\hss}%                                            
\parindent=0pt\offinterlineskip                                       
\vpos=0 pt%                                                              
\loop\readdim{\xcoord}                                 
\ifdim \xcoord < -999pt \doitfalse\else\doittrue\fi                        
\ifdoit \advance \xcoord by -\xleft
\readdim{\ycoord}
\advance \ycoord by -\ytop                              
\global\read\labelfile to\justx                                       
\global\read\labelfile to\justy                                       
\global\read\labelfile to\label
\global\setbox\labox=\hbox{\label\hskip-0.3em}%    
\advance\vpos by-\ycoord                                              
\vskip-\vpos \vpos=\ycoord                                         
\hbox to\pswidth{\hskip\xcoord %                                 
\hbox to 0pt{\ifnum\justx>0\hss\fi%                                   
\vbox to0pt{%                                                         
\ifnum\justy<2\vss\fi%                                                
\copy\labox\kern0pt%  
\ifnum\justy>0\vss\fi}%                                               
\ifnum\justx<2\hss\fi}%                                               
\hss}%                                                                
\repeat%                                                              
\advance\vpos by-\psheight%                                           
\vskip-\vpos %                                                     
}\ifboxfigures}\fi\closein\labelfile}
%
%
%     \figplace{<filename w/o extension>}{<hor shift>}{<vert shift>}
%     moves to the right by <hor shift> and down by <vert shift>
%     and then applies \figcrop
% 
\def\figplace#1#2#3{
\openin\labelfile=\figdir#1.lbl
\ifeof\labelfile\immediate\write16{Can't find \figdir#1.lbl; I quit!}\end\fi 
\closein\labelfile
\vbox{\vskip #3 \hbox{\hskip#2 \figcrop{#1}}}}
              
\def\figdir{./}
%
% Text starts here
%
{\nopagenumbers
\rightline{August 1995}
\vglue 2 true cm

\Cl{\bf Perturbation Theory around Non--Nested Fermi Surfaces}
\Cl{\bf I. Keeping the Fermi Surface Fixed}
\vglue 2 true cm
\Cl{\Cc Joel Feldman$^a$, Manfred Salmhofer$^{ab}$, 
Eugene Trubowitz$^b$}
\vglue 1 true cm
\Cl{$^a$\sl Mathematics Department, University of British Columbia,
Vancouver, Canada V6T 1Z2}
\Cl{$^b$\sl Mathematik, ETH--Zentrum, CH--8092 Z\" urich, Switzerland}
\vglue 1 true cm
\vglue 1 true cm
\Cl{\sl This paper is dedicated to the memory of Ansgar Schnizer}
\vglue 1 true cm
\Cl{\bf Abstract}

{\noindent The perturbation expansion for a general class of 
many--fermion systems with a non--nested, non--spherical Fermi 
surface is renormalized to all orders. In the limit as the 
infrared cutoff is removed, the counterterms converge to a 
finite limit which is differentiable in the band structure.
The map from the renormalized to the bare band structure 
is shown to be locally injective. 
A new classification of graphs as \OL\ or \NOL\ is given, 
and improved power counting bounds are derived from it.
They imply that the only subgraphs that can generate $r$ factorials 
in the $r^{\rm th}$ order
of the renormalized perturbation series are indeed the ladder graphs
and thus give a precise sense to the statement that 
`ladders are the most divergent diagrams'. 
Our results apply directly to the Hubbard model at any filling except for
half--filling. The half--filled Hubbard model is treated in another place.}
\endpage}
\centerline{\bf Table of Contents}
\someroom\noindent
{\bf 1. Introduction and Overview}
\leftit{1.1} The Problem
\leftit{1.2} The Formal Perturbation Expansion  
\leftit{1.3} Assumptions
\leftit{1.4} Divergences and Hartree--Fock Theory
\leftit{1.5} Results
\leftit{1.6} Discussion
\someroom\noindent
{\bf 2. Renormalization and Convergence}
\leftit{2.1} Scale Decomposition and Power Counting
\leftit{2.2} Localization Operator
\leftit{2.3} Flow of Effective Actions
\leftit{2.4} Non--Overlapping Graphs
\leftit{2.5} Decomposition of the Tree of a Labelled Graph
\leftit{2.6} Improved Power Counting
\leftit{2.7} Convergence of the Renormalized Green Functions
\someroom\noindent
{\bf 3. The Derivative With Respect to the Band Structure}
\leftit{3.1} Integration by Parts
\leftit{3.2} Bounds for the Directional Derivative
\leftit{3.3} Convergence of the Derivative
\someroom\noindent
{\bf Appendix A: Volume Estimates}
\par\noindent
{\bf Appendix B: The One--Fermion Problem}
\vfill\eject
\def\AOne{\hbox{\bf A1}}
\def\ATwo{\hbox{\bf A2}}
\def\AThr{\hbox{\bf A3}}
\chap{Introduction and Overview}
\sect{The Problem}
\noindent
Consider the following problem in many--body physics.
Let $\La$ be a finite box in $d$--dimensional space, 
i.e. $\La \subset\R^d$ or $\La \subset \Ga $, where 
$\Ga $ is a lattice in $\R^d$, and let 
$c_{\si}(\ve{x} )$ and $c^+_\si (\ve{x} ) $ be fermionic annihilation and
creation operators obeying the canonical anticommutation
relations $\{ c_\si (\ve{x} ) , c^+_{\si^\prime} (\ve{x}^\prime ) \} =
\delta_{\si \si^\prime } \delta (\ve{x}-\ve{x}^\prime ) $
and let $\cF $ be the fermionic Fock space generated by this algebra \quref{BR}.
Let $H_\La=H_\zer +\la V $ be the operator on $\cF $ given by
$$
H_\zer = \sum_\si \int ds(\ve{x} ) c_\si^+ (\ve{x} ) 
( T + U ) c_\si (\ve{x} )
\eqn $$
where $T$ is an operator describing the one--particle kinetic energy,
$U$ is multiplication by a periodic potential, and $\int ds(\ve{x} )$ denotes 
$\int_\La d\ve{x} $ for a continuous system and $\sum_{\ve{x} \in \La }$
for a system on a lattice.  Let 
$n_\si (\ve{x} ) =  c^+_\si (\ve{x}) c_\si (\ve{x})$ be the number
operator at $\ve{x} $ for spin $\si$. The interaction
$$
V =
\sum\limits_{\si, \si^\prime}
\int ds(\ve{x} ) \int ds(\ve{\xp} )
n_\si(\ve{x} ) v_{\si\si^\prime} (\ve{x-\xp }) n_{\si^\prime} (\ve{\xp } )
\EQN\Vdef $$
is assumed to be short-ranged (see Assumption \AOne\ below). 
The Hamiltonian $H_\La$ describes many electrons in a crystal or on a
lattice, that interact with a stationary ionic background through $U$
and with each other through the pair potential $V$.
If the coupling strength of the electron--electron interaction 
$\la =0$, 
the electrons move independently according to the 
one-particle Schr\" odinger operator $ T + U(x) $. 
In the continuum system $T=-\Delta /2m$ is the Laplacean
and $U(x+\ga ) = U(x) $ for all $\ga \in \Ga $, where 
the lattice $\Ga $ is generated by $d$ linearly independent
vectors in $\R^d$ (e.g. $\Ga = \Z^d$);
in the case of a lattice system, 
$U=0$ and the kinetic energy $T$ is defined by 
the hopping matrix between the sites of the lattice. 
For $\la \neq 0$, the potential $V$ takes into account 
interactions such as screened electromagnetic interactions. 
A slight generalization of \queq{\Vdef} allows for inclusion of 
phonon--mediated interactions.

Let $\be  = 1 /kT$ be the inverse temperature and
define the grand canonical partition function $Z_\La $ as
$$
Z_\La = \tr e^{-\beta (H_\La - \mu N_\La)}
\eqn $$
where
$$
N_\La = \sum_\si \int\limits_\La ds(\ve{x})n_\si (\ve{x} ) 
\eqn $$
is the number operator, $\mu $ is the chemical potential
 and the trace is over Fock space.
For an observable $\cO$, i.e. a polynomial
in the fermion operators, the thermal expectation value 
is defined as
$$
\langle\cO \rangle_\La  = {1\over Z_\La }
\tr \left( e^{-\beta (H_\La-\mu N_\La)} \cO \right)
\EQN\Hamexp $$
The question we are interested in is whether the thermodynamic
limit $\cG=\lim\limits_{\La \to \infty}\cG_\La$ of the connected 
Green functions $\cG_\La = \langle c^+_{\si_\one} (\ve{x_\one}) \ldots 
c_{\si^\prime_m} (\ve{x}^\prime_m )\rangle_{\La,conn}$ , which are special
cases of $\cO $ above, exists and whether in infinite volume a weak-coupling
expansion
$$
\cG = \sum\limits_{r=0}^\infty
\la^r G_r
\eqn $$
can be used to determine the dependence of $\cG $ on $\la $.

For this question the most interesting, because most singular, case is
that of zero temperature, $T=0$.
For positive temperature or the finite volume
lattice case the expansion obtained by expanding the factor 
$e^{\la V}$ in $\la$  is convergent, but its radius
of convergence shrinks to zero in the thermodynamic 
and  zero-temperature limit: at $T=0$ and in
infinite volume, one can not even pose the question of
convergence of the expansion in $\la $ because 
the coefficients $G_r$ already diverge for $r\geq 3$. In the limit
$T \to 0$, \queq{\Hamexp} reduces to expectation values in the
ground state of the system, so physically the question is
about the nature of the many--particle ground state of the system and
the validity of perturbation theory to calculate
$n$-point-functions.
The radius of convergence of the unrenormalized expansion in finite volume
shrinks to zero as the volume goes to infinity. Thus, although the expansion
converges for the large but finite systems which these models
are to describe, this is true only if $\la $ is of order
1/volume, which is obviously unrealistic for any macroscopic
system. Consequently, the unrenormalized expansion will not give insight into
the properties of the ground state.

In this paper we consider formal perturbation theory.
That is, we study the thermodynamic limit
of the coefficient functions $G_r$. By an analysis similar to 
\quref{FT1}, the expansion is renormalized
so that these functions converge as the volume goes to infinity.
More precisely, we introduce a well--defined infinite volume model
obtained by cutting off the singularity at the Fermi surface
(i.e.\ introducing an infrared cutoff)  
and renormalized by including counterterms $K$ in the action, 
and then show that all coefficients $G_r$ have limits 
as the infrared cutoff is removed.
These counterterms are bilinear in the fermions and can therefore 
be viewed as a modification of $H_\zer$ (although they are treated 
as extra interaction vertices in the formal expansion). 
They also have finite limits as the infrared cutoff is removed.
The limiting counterterms reflect the modification of the band structure 
due to the interaction. 
The precise meaning of this will be discussed in much more detail below.
Although we do not go through the finite--volume bounds here, 
it will be clear from the way our bounds are derived that the same 
procedure can be applied to obtain an expansion in finite volume 
with coefficients that converge in the thermodynamic limit. 

Except for special cases, the renormalized expansion is, 
as an expansion in $\la $, not convergent but only locally 
Borel summable because the coefficients behave as $G_r \sim r!$. 
The occurence of these factorials indicates that the 
nonperturbative ground state may exhibit symmetry breaking. 
For example, if the interaction is attractive in the zero angular
momentum sector, this is the case \quref{FT2}.
One of the main results we shall prove here is that for a very wide 
class of models (and regardless of the sign of the interaction), 
the $r$ factorials in individual graphs come only from ladder diagrams. 

Renormalization has been done in \quref{FT1} for the continuum case
where $T=-\Delta/2m$ and $U=0$. We shall refer to this case as the 
spherical case since the band structure (defined below) has an 
$O(d)$ rotational symmetry.   
The procedure for removing the divergences
in the present case is similar to the spherical case
in that we have to renormalize two--legged insertions.
However, the present work is a nontrivial extension of \quref{FT1}
because in contrast to the spherical case the counterterms are not
constants. In brief, subtracting functions is much more complicated
than subtracting constants. In particular, the regularity properties 
of the counterterms are quite subtle. 

In the remainder of this introductory section, we give a
non--rigorous, physical discussion of why
divergences occur and how they may be removed by renormalization.
We hope that this will convince the reader, before going
through all the details, that the renormalization subtractions
are natural and the divergences of the naive expansion are 
artificial in these models.
We state our main results in Section 1.5
and then discuss their physical interpretation. 
Finally, we give an overview of the sections
containing the proofs. Every section begins with a brief 
explanation of what is done and how it fits into the general strategy.
\sect{The Formal Perturbation Expansion}
\noindent
The models have the formal functional integral representation
$$
P(\et , \etq ) = \int D\ps D\psq e^{\cA +(\etq , \ps )
+ (\et , \psq ) }
\eqn $$
where $\cA = - (\psq ,C^{-1} \ps ) - \la V$, $D\ps D\psq$ is the formal 
measure $\prod_{x,\al} d\ps_\al (x) d\psq_\al (x)$,
$$
(\psq ,C^{-1} \ps ) =
\int ds(x) ds(y) \sum\limits_{\al,\be}
\psq_\al (x) \left( C^{-1} \right) _{\al\be} (x,y)
\ps_\be (y),
\eqn $$
and
$$
V = \int ds(x) ds(\xp )
\sum\limits_{\al,\be,\alp,\bep}
\psq_\al (x) \ps _\be (x) \tilde v _{\al\be, \alp,\bep} (x,\xp )
\psq_\alp (\xp ) \ps_\bep (\xp )
,\eqn $$
where now $\int ds(x) F(x) $ stands for the integral
over the spatial variable $\ve{x} $ and imaginary time $\ta $,
$x = (\ta, \ve{x} )$, with an appropriate measure, e.g.
$$
ds(x) = d \ta d^d\ve{x}
\eqn $$
for a continuous system on $\lbrack 0,\beta \rbrack \times \R^d $ and 
$$
\int ds(x) F(x) = \int_0^\be d\ta \sum\limits_{\ve{x} \in \Ga}
F(\ta, \ve{x} )
\eqn $$
for a lattice system on $\lbrack 0,\beta \rbrack \times \Ga$, 
e.g.\ $\Ga=\Z^d$.
Here $\beta = 1/k_BT$ is the inverse temperature. 
The imaginary time is introduced to get a functional 
integral representation for the trace over Fock space in the 
standard way. The connected Green functions can formally be calculated
as derivatives of $\log P$ with respect to the sources $\eta $ and 
$\etq$.
 
In this paper,
we consider the limiting case $T=0$, so $\beta= \infty$
and the configuration spaces are $\R^{d+1}$ and $\R\times
\Ga$ (e.g.\ $\R\times \Z^d$), respectively.
The spin index $\al \in \{ \uparrow,\downarrow\}$,
the interaction is assumed to be translation
invariant, so that
$$
\tilde v _{\al\be,\alp\bep } (x,\xp ) =
v_{\al\be,\alp\bep} (\ta - \tap ,\ve{x} - \ve{\xp} )
,\eqn $$
and short--range, i.e. $v$ decreasing so fast that its Fourier transform
$\hat v$ is at least twice differentiable (see Assumption \AOne\ below). 
Note that we do not assume that it is instantaneous. For simplicity, 
we also assume that it is spin--diagonal, i.e.\
$v_{\al\be,\alp\bep} = \de_{\al\alp} \de_{\be\bep} v$. 
In contrast to the assumption about the decay of $v$, the latter 
assumption is merely for notational convenience and can easily 
be dropped.

One may imagine $v$ to arise from exchange of (quasi)particles
like photons or phonons and formalize this by a
Hubbard--Stratonovich transformation, introducing one or more
scalar fields with covariance $v$ so that the interaction
vertex is resolved as an exchange of fields and the interaction becomes
bilinear in the fermion fields. For the purposes of the
perturbation expansion we shall not need this. In particular
since we assume smoothness of $\hat v$, we shall not need a
cutoff on the interaction lines, and we shall often draw graphs
with four--legged vertices instead of ones with
interaction lines.

   For the lattice models, we take
$$
\left( C^{-1} \right)  (x, \xp ) =
\delta_{\al\be} \left( \delta_{\ve{x}\ve{\xp}}
\left( \partial_\tap - \mu \right) - T_{\ve{x}-\ve{\xp}}
\right) \delta (\ta - \tap )
,\eqn $$
where $\mu $ is the chemical potential and $T_{\ve{x}-\ve{\xp}}$
is the amplitude for hopping from site $\ve{x}$ to site $\ve{\xp}$, 
which we assume to be symmetric and short-ranged (see Assumptions
\ATwo\ and \AThr\ on $e$ below).

A model of particular interest that is easy to formulate but 
difficult to analyze is the Hubbard model, for which
$$
T_{\ve{x}} = \sum\limits_{\abs{\ve{y}}=1} t_{\ve{y}}
\delta_{\ve{x},\ve{y}}
\eqn $$
with $t_\ve{y}$ the so--called hopping parameters. 
In the simplest version of the model,
$t_\ve{y} = t$ is the same for all $\ve{y}$ of length one,
so the operator $T$ is just the discrete Laplacean on $\Z^d$, 
with the diagonal term omitted since it can be absorbed in the 
chemical potential $\mu$, 
and the interaction term is on-site and spin-diagonal,
$$
v_{\al\be,\alp\bep} (\ve{x} - \ve{\xp}) =
\delta_{\ve{x}\ve{\xp}}
\delta_{\al\be} \delta_{\alp\bep}
.\eqn $$
Various extensions of this model, e.g. with more 
complicated finite range hopping have been studied in
connection to high--temperature superconductivity.
For suitable values of the filling factor, they all fall into the class of 
band structures discussed here.
For a review of mathematically rigorous results about the 
Hubbard model, see \quref{L}.

   Formally equivalent to $P$, but in fact much more convenient
is the generating functional for connected amputated
Green functions
$$
\cG (\ps_e,\psq_e ) =
\log
{1\over Z} \int D\ps D \psq
e^{-(\psq , C^{-1} \ps )} e^{\la V (\ps+\ps_e , \psq + \psq_e ) }
\eqn $$
where the constant $Z$ takes out the field--independent term
so that $\cG (0,0) = 0$. $\cG$, as written above, is not a
well--defined object in infinite volume; it can  be made well-defined by
restricting to a finite volume $\La$, or by introducing a suitable 
cutoff. If the free
covariance $C$ is bounded and any power of it is integrable,
${1\over \abs{\La}} \cG_\La$ exists and is analytic in $\la $,
as was first observed by Caianiello. However, for any
realistic model, $C$ will not have this properties, unless 
cutoffs are imposed. The radius of convergence obtained using naive bounds
shrinks to zero when the cutoffs are removed, and establishing
analyticity uniformly in the cutoffs requires techniques
as in \quref{FMRT}. 

Our analysis is done in momentum space, where from now on 
momentum is short for Bloch's quasi--momentum, which 
can be used to label one--particle states 
because of the periodicity of the one--particle 
potential $U$. In infinite volume, momentum space 
is the first Brillouin zone $\cB$, i.e. the torus
$$
\cB = \R^d / \Ga^\#
\EQN\cBdef $$
where $\Ga^\#$ is the dual lattice to $\Ga $, 
e.g. $\Ga^\# = 2 \pi \Z^d$ for $\Ga = \Z^d$.
In finite volume, the momenta are in a finite subset of 
$\cB$, $\ve{p} = 2\pi \ve{n} /L$ with $\ve{n}\in \Z^d\cap \cB$ 
if the volume is a box of sidelength $L$.  
The eigenfunction expansions used to transform 
into momentum space are discussed briefly in Appendix B for the 
general case; for the purposes of this introduction, we just give
the formulas for the case of a lattice model on $\Z^d$, 
where we can simply do a Fourier expansion. 
The only changes in the general case are 
(of course) that the Brillouin zone will differ with the 
lattice and that the formulas for switching between 
position and quasi--momentum space involve  the eigenfunctions
of the one--particle Hamiltonian $H_\zer$ with the periodic 
potential.  
Under the Fourier transform 
$$\eqalign{
\ps (x) &= (2 \pi )^{-(d+1)}\int d^d{\ve{p}} dp_\zer
e^{-ip_\zer\ta + i \ve{px}} \hat\ps (p) \cr
\psq (x) &= (2 \pi )^{-(d+1)} \int d^d\ve{p} dp_\zer
e^{ip_\zer \ta -i\ve{px}} \hat\psq{p}
\cr}\EQN\Psourier $$
the quadratic part of the action becomes
$$
(\psq , C^{-1} \ps )  = (2 \pi )^{-(d+1)} \int d^d\ve{p} dp_\zer
\psq (p) (ip_\zer - e(\ve{p} ) ) \ps (p)
\eqn $$
where we have dropped the hats and introduced the band structure 
$$
\eof{p} = \veps ( \ve{p} ) - \mu
\EQN\eeqepsmu $$
where
$$
\veps (\ve{p }) = \int ds(\ve{x}) e^{-i\ve{px}}T_\ve{x}
,\eqn $$
and the interaction becomes, with $p_i = ((p_i)_\zer , \ve{p}_i)$, and
$$
\db ^{d+1} p = \db p_\zer \db^d \ve{p} =
{d p_\zer \over 2 \pi } {d^d\ve{p} \over (2 \pi ) ^d }
,\eqn $$
$$\eqalign{
V =  \int \db^{d+1} p_1 \ldots & \db ^{d+1} p_4
(2\pi )^{d+1}
\delta ((p_2+p_4 - p_1 -p_3)_\zer) \delta^\# (\ve{p_2-p_1+p_4-p_3}) \cr
& \hat v (\ve{p}_3 - \ve{p}_1 )
\psq (p_1)\ps (p_2) \psq  (p_3)\ps (p_4)
.\cr}\EQN\Interact $$
here $\delta ^\#$ is the delta function on $\cB$, more explicitly 
$$
\delta^\# (\ve{p} ) = {1 \over (2 \pi)^d}
\sum\limits_{\ve{x} \in \Z^d} e^{i \ve{px}}
= \sum\limits_{\ga \in \Ga^\#} \delta (\ve{p} + \ga )
\eqn $$
where the $\de $ on the right side denotes that on $\R^d$.
In general, the solution of the one-particle problem will produce
crossing bands. We exclude this case here, and 
we also introduce an ultraviolet cutoff that removes the high energy bands.
For the lattice systems, such a cutoff is
already built in as the lattice spacing; for continuous systems
it is not a real physical restriction 
since high energies do not occur in a crystal. 
If there are finitely many bands that do not cross, 
the band index is just a bookkeeping device dragged along, 
so, without loss, we restrict to the one--band case here. 

For $\lambda =0$, the fermions do not influence each other and the 
model is completely characterized by the covariance $C$,
$$
\check C(x) = \int \db^{d+1}p {e^{-ip_\zer\ta + i \ve{px}}
\over ip_\zer - \eof{p} }
\eqn $$
in the sense that all $2n$--point functions are determinants of
matrices with elements $ \check C(x_i - x_j )$.

The propagator in momentum space, $C(p)=e^{ip_\zer 0^+}/(ip_\zer - \eof{p} )$,
has a singularity at $p_\zer =0$ for all $\ve{p} \in S$,
where $S = \{ \ve{p} : \eof{p} = 0 \} $ is the Fermi
surface of the independent electron approximation. Although
the function $1/(ip_\zer - \eof{p} ) $ is in $L^{1 + \delta }_{loc}
(\R \times \cB ) $ for all $\delta \in [0, 1)$, graphs
in the perturbation expansion diverge because of the
singularity on $S$ and because in
the expansion, arbitrary powers of $C$ are integrated.
The numerator $e^{ip_\zer 0^+}$ is included in the standard way 
since we want to consider the expansion around the situation where 
all states inside the Fermi surface, i.e.\ those with $e(\ve{p}) <0$,
are already occupied. 

Expanding $\cG $ in a formal power series in $\la $, we can
write
$$\cG (\ps_e , \psq_e ) = \sum\limits_{r\geq 0} \la^r
\cG_r (\ps_e, \psq_e )
\eqn $$
with
$$\eqalign{
\cG_r (\ps, \psq) = \sum\limits_{m\geq 1}
\sum\limits_{\al_\one , \ldots , \al_m \atop
\alq_\one, \ldots , \alq_m } &
\int \prod\limits_{i=1}^{2m} \db^{d+1} p_i
(2 \pi ) ^{d+1} \delta^\# \left(
\sum\limits_{i=1}^m p_i - \sum\limits_{i=m+1}^{2m} p_i
\right) \cr
& \left(G _{2m,r}\right)
_{\al_\one, \ldots , \alq_m} (p_1, \ldots ,p_{2m-1} )
\prod\limits_{i=1}^m \ps_{\al_i} (p_i) \psq_{\alq_i} (p_{m+i})
,\cr}\EQN\Genconn $$
where the coefficient function $G _{2m,r}$ is totally
antisymmetric in the simultaneous exchange of momenta and spin indices (see Section 2.3).
Again, the $\de^\#$ is periodic with respect to $\Ga^\#$ 
in the spatial part of the momentum. The coefficient 
$G_{2m,r} $ can be expressed
in the usual way as a sum over values of connected
Feynman diagrams. The sum over $m$ runs over a finite
index set for each fixed $r$ because the number of vertices is $r$ and the graphs are connected with $2m$ external legs.

The Feynman graphs are similar to those in quantum electrodynamics:
there are two types of lines, namely fermion lines (drawn solid), 
carrying a direction, and interaction lines (drawn dashed). 
The vertices have two legs to which fermion lines can be connected 
(one incoming, one outgoing), and one leg for an interaction line. 
The action determines the assignment of propagators $C(p)$ to fermion lines,
$\hat v (p)$ to interaction lines,  
and momentum conservation delta functions to vertices.
Equivalently, one can replace two vertices that are joined by 
an interaction line by a single four--fermion vertex with exactly two incoming 
fermion legs and exactly two outgoing fermion legs. The graphs then have 
only four--legged fermion vertices and only fermion lines.
There is one notable difference between the cases $U=0$ and $U\neq 0$:
In the spherical case ($U=0$), where $\veps(\ve{p}) = \ve{p}^2/2m$,
$\ve{p} \in \R^d$. The corresponding ultraviolet problem 
(behaviour at large $|\ve{p}|$) 
was solved in \quref{FT1}. In presence of 
a crystal potential ($U\neq 0$), the integrals over the spatial 
part of the momentum are over the first Brillouin zone $\cB$,
which is a compact set. Thus there is no case of large $\ve{p}$ here. 
Momentum conservation at every vertex 
means conservation in $\cB$, as given by $\de^\#$ above. 
If one prefers to think of the momenta in $\R^d$, fixing 
momenta with $\de^\#$ means that at every vertex, there remains
a sum over $\ga \in \Ga^\#$. Although formally infinite, this
sum always contains only one nonzero term since there is a unique 
$\ga \in \Ga^\#$ that translates back a vector in $\R^d$ into
the fundamental domain of the translational group $\Ga^\# $. 
However, it is natural and simpler to consider
momentum space as the torus $\cB $ since $e$ is $\Ga^\#$--periodic.

For example, in the Hubbard model,
$$
\eof{p} = 2t \sum\limits_{i=1}^d \cos p_i - \mu
\EQN\Hubband $$
is the tight--binding band relation
and $\hat v (\ve{p} ) = 1$.

The much more general class of models and the range of chemical potential 
$\mu $ that we treat in this paper is given by the following assumptions.

\sect{Assumptions}
\noindent
We assume that the one--particle problem 
(discussed in Appendix B) is such that 
we have a Brillouin zone $\cB$ which is a $d$-dimensional torus
of type \queq{\cBdef}.  
We assume that $e=\veps-\mu$ (see \queq{\eeqepsmu})
is a continuous function on $\cB$ 
and that for some value $\mu_\zer$  of the chemical potential, 
the Fermi surface  
$$
S= \{ \ve{p} \in \cB : e(\ve{p} ) = 0\}
\eqn $$
has only a finite number of connected components.
Furthermore, there is $k \geq 2$ and a neighbourhood $\cN$ of $S$ such that:
\someroom
\leftit{\AOne} The interaction $\hat v \in C^k(\R\times \cB,\C)$. 
The sup norm
over $\R \times \cB$ of the first $k$ derivatives is finite.
\someroom
\leftit{\ATwo} The band structure $e \in C^k(\cN, \R )$,
and $\nabla e (\ve{p} ) \neq 0$ for all $\ve{p} \in S$.
\sni
The third assumption is a geometrical condition on the Fermi surface. 
It is very simple to understand and is fulfilled for generic surfaces. 
Let $n: S \to \R^d $, $\om \mapsto n(\om)=\sfrac{\nabla e}{|\nabla e|} (\om)$,
be the unit normal to the surface. By \ATwo, $S$ is a $C^{k}$ submanifold
of $\cB$, and $n$ is a $C^{k-1}$ unit vector field.
If $S$ consists of more than one connected component, choose a normal
field for any component. 
For $\om,\om' \in S$, define the angle between $n(\om )$ and $n(\om')$ by
$$
\th (\om,{\om'} ) = \arccos ( n(\om) \cdot n({\om'}) )
.\eqn $$
Let 
$$
\cD (\om) = \{ {\om'} \in S : \abs{n({\om}) \cdot n(\om')} =1 \}
= \{ {\om'} \in S : n({\om}) =\pm n(\om') \}
,\EQN\cDomdef $$ 
and
denote the $(d-1)$--dimensional measure of $A \subset S$ by $vol_{d-1} A$.
Also, for any $A\subset\R^d$ and $\beta> 0$ denote by
$
U_\beta (A) =\{ \ve{p}\in\R^d : {\rm distance}(\ve{p},A)<\beta \}
$
the open $\beta$-neighbourhood of $A$.
For fixed $\veps $ and $\mu_\zer $, we assume:
\someroom
\leftit{\AThr} There is an open interval $\cM $ around $\mu_\zer $ 
and there are strictly positive numbers
$Z_\zer, Z_\one, \rho, \beta_\zer $ and $\ka$ 
such that for all $\mu \in \cM $, the Fermi
surface $S=S(\mu)=\{\ve{p} \in \cB: e(\ve{p}) =0\}$ 
has the following properties: $S(\mu) \subset \cN$, and 
for all $\be \leq \be_\zer$ and all $\om \in S$,
\leftit{$(i)$} 
$vol_{d-1} \left( U_\be (\cD (\om)) \cap S \right) \leq Z_\zer \be^\ka $
\leftit{$(ii)$} if ${\om'} \not\in U_\be (\cD (\om) ) \cap S$,
then $\abs{\sin \th (\om,{\om'})}= \sqrt{1-(n(\om)\cdot n({\om'}))^2} 
\geq Z_\one \be^\rh$.
\sni
Throughout this paper, \AOne--\AThr\  will be assumed to hold, and
$\mu $ will be assumed to lie in the interval $\cM $ specified in \AThr.
We now explain what these assumptions mean.
\someroom
Assumption \AOne\ on $\hat v$ is a decay assumption in
position space, e.g.\ for an instantaneous interaction $V$ 
on a lattice system on $\Z^d$ and $k=2$, \AOne\ holds if
$$
\sum\limits_{\ve{x}\in \Z^d} \abs{\ve{x}}^2 \abs{V(\ve{x})}  < \infty
.\eqn $$
For continuous systems, \AOne\ is implied by a similar integral condition.

Assumption \ATwo\ excludes singular points. For example, a point 
$\ve{p} $ on $S$ where $\nabla e (\ve{p} ) = 0$ is called a 
van Hove singularity.

The condition that $e$ is continuously differentiable
is fulfilled for the case where $e$ comes from a Schr\" odinger
equation for the one--body problem with a regular periodic potential, if
there is no level--crossing.
Indeed, it is real analytic.
In lattice models with finite--range hopping, $e$ is analytic. 
However, infinite--range hopping is also allowed: $e\in C^k$ if the 
$k^{th}$ moment of the hopping amplitude exists, i.e.\
$\sum\limits_{\ve{x}} |\ve{x}|^k |T_{\ve{x}}|<\infty$.

Assumption \AThr\ is, more informally, that for every $\om \in S$ 
\leftit{$(i)$} the set of points ${\om'}$ where the normal $n({\om'})$ is parallel or
antiparallel to $n(\om )$, has positive codimension $\ka > 0$ in $S$
and 
\leftit{$(ii)$} if ${\om'}$ is not in the set $\cD (\om )$, where the normal is 
(anti)parallel to $n(\om )$, the angle between $n(\om )$ and $n({\om'})$ increases
with some power of the distance between ${\om'}$ and $\cD (\om ) $.
\someroom
Thus in order to violate these assumptions, the surface $S$ must have
flat regions or subsets where $\th (\om ,{\om'}) $ vanishes exponentially 
fast as $\abs{\om -{\om'}} \to 0$.
To illustrate \AThr, we draw an example of a Fermi surface that satisfies 
\AThr\ in $d=2$ (i.e.\ a Fermi curve) on $\cB = \R^2/2\pi\Z^2$
(the square bounds the fundamental region $[-\pi,\pi)^2$ for the torus $\cB$,
and the shaded areas indicate $e(\ve{p})<0$):
%
%   here comes a picture of a typical Fermi surface satisfying A3
%
\hfil\break
\null\hfill\figplace{typfs}{0in}{0in}\hfill\break
 
\noindent
\ATwo\ and \AThr\ imply the following bound, 
which we shall use in the proofs. 
\sni
{\bf Volume improvement estimate:} There is $\ep> 0 $ 
and there is a constant $C_{vol}$ such that for all $\mu \in \cM $    
and for all $\veps_\one>0$, $\veps_\two>0$, $\veps_3>0$
$$
I_2 ( \veps_1, \veps_2, \veps_3 ) \leq C_{vol}
\veps_\one \veps_\two {\veps_3}^\ep
\EQN\ImpVol $$
where
$$\eqalign{
I_2 ( \veps_1, \veps_2, \veps_3 )
= \sup\limits_{\ve{q}\in \cB}\; \max\limits_{v_\one, v_\two \in \{ 1, -1 \}}
\int\limits_{\cB \times\cB}
d^d \ve{p}_\one d^d \ve{p}_\two\;
&1\left(\abs{e(\ve{p}_\one )} < \veps_\one \right)
1\left(\abs{e(\ve{p}_\two )} < \veps_\two \right)\cr
&\times \;1\left(\abs{e(v_\one \ve{p}_\one \pm v_\two \ve{p}_\two + \ve{q} )}
< \veps_3 \right)
.\cr}\EQN\Itwodef $$
Here $1(E)$ denotes the indicator function of the event $E$, i.e.
$1(E) = 1$ if $E$ is true and $1(E)=0$ otherwise.
The additional factor ${\veps_3}^\ep$ will be called the
volume improvement factor. The function $I_\two $ allows us to 
give sharp bounds for arbitrary graphs based on a simple characterization 
of graphs (explained below).
\Pro{\NoNest} \AThr\ and \ATwo\ imply \ImpVol, with
$$
\ep \geq { \ka \over \ka + \rh }
.\EQN\EpsExp $$
\Proof See Appendix A. \endproof

\noindent
Assumption \AThr, and thus \queq{\ImpVol}, 
hold in particular if the set of filled states 
$\{ \ve{p} : e(\ve{p} ) \leq \mu \} $ is strictly convex and 
nowhere exponentially flat in the sense mentioned above. 
Thus the class of models with $\ep > 0$ contains all those where
the band structure is a strictly convex analytic function or a
strictly concave analytic function, because, by definition, the sets
$\{ \ve{p} : e(\ve{p} ) \leq \mu \} $ are then strictly convex
sets, and the Fermi surface is just the boundary of such a set.
By analyticity, exponential or complete flatness is excluded in
this case. This is obviously a very natural
condition since essentially all band structures of practical
importance in solid state models are strictly convex around the
band minimum, and so our results apply to the case where the
Fermi edge is just above the minimum of a band.

The proof we give in the Appendix also shows that
this non--nestedness is essentially a transversality
condition on the Fermi surface and its translates -- hence the need to have
some control over the set $\cD(\om)$, which is essentially the set
where the intersections would not be transversal. 
\someroom\noindent
{\bf Examples:} \leftit{1.} The spherical band structure
$e(\ve{p} ) = \ve{p}^2 /2m - \mu $ fulfills
all these hypotheses for any $\mu > 0$, with $\rh =1$ and
$\ka = d-1$. 
\leftit{2.} 
The Hubbard model with tight-binding 
band structure \queq{\Hubband} fulfills \AOne\ --\AThr\ for 
all $\mu \neq 0$, i.e. away from half--filling, with  $\ka = d-1$.
If the band is
either empty or full (cases which are of little physical interest),
the volume shrinks even faster in $d \geq 2$.
For the half--filled case $\mu =0$, both \ATwo\ and \AThr\ fail. \ATwo\ 
is not fulfilled because of the van Hove singularities at 
the boundary of $[-\pi , \pi] ^d$ and \AThr\ does not hold because the surface
has flat regions (in $d=2$ it is diamond-shaped).  This is an example 
where a non--generic (because flat) surface plays a role in 
a physical model. 

It is well--known that the half-filled band is a
very special case, and that this is due to the nesting we just discussed,
as well as to the presence of van Hove singularities. A physical way 
of understanding this is that the particle-hole symmetry restricts the shape 
of the Fermi surface.
More generally speaking, van Hove singularities
must always occur at some values of $\mu $ for topological
reasons: for generic $e$, the condition $\nabla e (\ve{p}) =0$
is satisfied at isolated points $\ve{p} \in \cB$. 
Thus there is a van Hove singularity for each value of $\mu$ for which 
the corresponding Fermi surface passes through one of these points. 
By way of contrast, nesting in the sense that \AThr\ fails
is a much more restrictive condition on $e(\ve{p})$. Stated
differently, a nesting condition requires fine--tuning of $e$.
The occurrence of flat parts of $S$ and van Hove singularities at
the same value of $\mu$ ($\mu =0$, half--filling) in the 
Hubbard model with the band structure \queq{\Hubband}
is accidential. They no longer occur
at the same value of $\mu$ if next--to--nearest neighbour hopping is allowed.
\leftit{3.} 
For $d=2$, $S = \{ (x,y) : x^{2n}+y^{2n} = 1\}$ is another example.
Here $\rh = 2n-1$ and, as in Examples 1 and 2, $\ka = d-1=1$.
As $n \to \infty$, $S$ approaches $\{ (x,y) : \abs{x} =1$
or $\abs{y} =1, \sqrt{x^2+y^2}\leq \sqrt{2} \}$, which is flat away
from its edges, and the lower bound for the 
volume improvement exponent $\ep $ goes to zero like $1/n$ by \queq{\EpsExp}.
\leftit{4.}
The two--torus imbedded in $\R^3$ is an example with $\rho =1 $ and
$\ka = 1$. The codimension $\ka $ is only $1$ in this example because
$\cD (\omega) $ may be a union of two circles for some $\omega$.
\leftit{5.} 
The surface $e^{-1/x^2} + e^{-1/y^2} = e^{-1} $ is an 
example where, due to the essential singularity at $(0,1)$, the 
condition \AThr\  does not hold. As discussed above, under some 
regularity conditions on the one--particle problem, such surfaces
are ruled out. We may well expect that they will not occur in any 
realistic model.  
\sect{Divergences and Hartree--Fock Theory}
\noindent
Under the assumptions stated above, the only source of divergences
in perturbation theory is exactly the same as in the spherical case,
where $e$ is given by $e(\ve{p} ) = \ve{p}^2 / 2m -\mu$, namely the
accumulation of powers of the propagator due to strings of
two--legged subdiagrams: the function $C(p) = (ip_\zer - e(\ve{p} ) )
^{-1} $ is singular on the set $\{ 0 \} \times S $. By the
assumption that $\nabla e$ does not vanish on $S $ and by compactness
of $S$, 
we can introduce coordinates $\rh = e(\ve{p} ) $ and
$\om$, where $\om $ parametrizes the submanifold
$S_\rh = \{ \ve{p} : e(\ve{p}) = \rh \} $ (this works in a neighbourhood
of $S=S_\zer$, i.e. for $\abs{\rh}\leq \rh_\zer $. Thus $\ve{p} =
\ph (\rh, \om ) $ in this neighbourhood, and for $\alpha < 2$,
$$\eqalign{
\int\limits_{\abs{ip_\zer - e(\ve{p}) } < \rh_\zer }
dp_\zer d^d\ve{p} {1 \over  \abs{ip_\zer - e(\ve{p}) }^\al }
&=
\int\limits_{\abs{ip_\zer - \rh } < \rh_\zer }
dp_\zer d\rh d^{d-1}\om {1 \over  \abs{ip_\zer - \rh }^\al}
\abs{\det \phi^\prime (\rh ,\om )} = \cr
&= \int\limits_0^{ \rh_\zer } {r dr \over r^\al } F(r)
\cr}\EQN\integr $$
where $F(r) = \int\limits_0^{2\pi} d\th \int d^{d-1}\om
\abs{\det \phi^\prime (r\cos\th ,\om )}$. The integral converges for
$\al < 2$, but since $F(r) \geq f_\zer > 0$ for all $r$, it diverges
for $\al \geq 2$. 
%
%   here comes Figure 1
%
\hfil\break
\null\hfil\figplace{fig1}{-1.5 truecm}{0in}\hfil\break
By the Feynman rules, graphs like the ones shown in Figure 1
diverge, because e.g. the value of the first one would be
$ \int dp_\zer \int d^d\ve{p}  C(p)^3 T(p)^2 \hat v (q-p)$
for external momentum $q$, which diverges because the third
power of the propagator appears, so $\al = 3$, and
$T(0,\ve{q} ) = \int d^{d+1}p \hat v((0,\ve{q})-p) C_h(p)$ 
will not vanish on the Fermi surface $S$ where
the propagator is singular. The proof that these two--legged
insertions are the only source of divergences of values of
individual graphs was given in \quref{FT1} for the spherical case,
and a similar result holds in the present case (see Section 2.1).
The only way a divergence could be absent is that the function
$T(p)$ also vanishes on the Fermi surface. However, this will not
happen by itself in general. Renormalization is done by subtracting
$(\ell T)(p)=T(0,\prP(\ve{p}))$ for any two--legged insertion $T(p)$, where
$\prP(\ve{p} )$ is the projection of the vector $\ve{p}$ onto the
Fermi surface $S $, for $\ve{p}$ in a fixed neighbourhood of $S$.
%
% here Figure 2
%

\midinsert
\hfil\break
\null\hfil\figplace{fig2}{0truecm}{0in}\hfil\break
\endinsert 

\noindent
The precise definition of $\prP$ is given in Section 2.2; it is defined 
by taking a vector field $u$ that is transversal to $S $
in the sense that $\abs{(u\cdot \nabla e) (\ve{p} )} \geq u_\zer > 0$
for all $\ve{p} \in S$, and taking $\prP(\ve{p} )$ to be the point
where the integral curve of $u$ through $\ve{p} $ intersects $S $
(see Figure 2).
The reader familiar with 
resummation methods based on the Schwinger-Dyson
equations in solid state theory, e.g. the Hartree--Fock method,
may ask why one never sees these divergences in the
integral equations corresponding to these approximations,
although they are said to resum part of those diagrams which appear
to be ill-defined in the formal perturbation expansion.
This point actually gives a hint at what renormalization in these
models does. Consider the Hartree--Fock approximation \quref{AGD},
as given by the integral equation for the two--point function
$$
\left( G_\two\right)_{\al\alp}(\ta_\one, \ve{x}_\one ,\ta_\two, \ve{x}_\two ) =
\left\langle \ps_\al (\ta_\one, \ve{x}_\one ) 
\psq_\alp (\ta_\two, \ve{x}_\two )\right\rangle
\eqn $$
which reads, denoting $x = (\ta , \ve{x} ) $
$$\eqalign{
G_\two ( x_\one , x_\two )
&= \check C(x_\one , x_\two ) +
\la \int ds(x) ds(\xp ) \check C( x_\one ,  x )  v(\xp -x )
G_\two ( x, \xp ) G_\two (\xp , x_\two )
\cr
& - \la
\int ds(x) ds(\xp ) \check C( x_\one ,  x )  G_\two (x , x_\two ) v(\xp -x )
\tr G_\two ( \xp , \xp )
\cr}\eqn $$
where the trace is over spin indices. Representing the free 
propagator $\check C$ by a thin solid line, interaction lines by 
dashed lines and the interacting propagator $G_\two $ by a thick line,
this equation can be depicted as
%
%   here comes Figure 3
%
\hfil\break
\null\hfil\figplace{hfeq}{-1.4 truecm}{0in}\hfil\break
from which it
is evident that by iterating the equation one produces a resummation
to all orders that includes graphs without polarization effects, in particular
some that are divergent in the formal perturbative expansion, as 
for instance the first one in Figure 1.
However, the whole point of the `resummation' is to avoid summation, 
instead making the ansatz
$$
\hat G_\two (p) = e^{ip_\zer 0^+}(i p_\zer - e(\ve{p} ) - \Si (p) ) ^{-1}
\eqn $$
and rewriting the integral equation in momentum space (in the
translation--invariant case) as
$$
\Si (q) = - \la \int d^{d+1} p {\hat v (q-p )e^{ip_\zer 0^+}
\over i p_\zer - e(\ve{p} ) - \Si (p)}  + 
\la \hat v(0) \tr \int d^{d+1}p {e^{ip_\zer 0^+} 
\over i p_\zer - e(\ve{p} ) - \Si (p)}
\EQN\HFse $$
For a reasonable function $\Si$, the singularity of $\hat G_\two $ is
again integrable by the argument of \queq{\integr}.
However, $\hat G_\two $ will be singular if $p_\zer =0$ and
$e(\ve{p} ) + \Si (0,\ve{p} ) = 0$, rather than if $p_\zer =0,\ e(\ve{p} ) = 0$,
so $S $ is not the Fermi surface of the interacting system.
If one attempted to seek the solution of \queq{\HFse} by an
expansion of $\Si$ in powers of $\la $ and exchanged summation and
integrals, one would run into divergent expressions for
well--defined integrals, such as
$$
\int d^{d+1} p \; \hat G_\two (p) \;``\; =\; "\; \sum\limits_{n=0}^\infty
\int d^{d+1} p {\Si (p) ^n \over  (i p_\zer - e(\ve{p} ))^{n+1} }
\EQN\bledsinn $$
These divergences are part of those of the unrenormalized perturbation
expansion.

The conclusion of this discussion is that in fact not
the subtractions, but the divergences are artificial, 
because they come from expanding a moving singularity in 
terms of a fixed one, and that the
counterterms that are added to the action to implement
renormalization have something to
do with $\Si (0, \ve{p} )$. This is the main idea; it remains to be
shown that this is really so for the exact theory, where there is
no such simple integral equation for the self--energy $\Si$ as
\queq{\HFse} for the Hartree-Fock approximation.
For instance, the Hartree--Fock resummation does not include any 
polarization effects
and thus differs from the exact result already in second order.
It is necessary to include those effects for renormalization, 
e.g. the second graph in Figure 1 also contributes a formally 
divergent term to $\cG_\two$ and thus needs to be renormalized.
The Hartree--Fock graphs will, however, turn out to be special 
in that they are the only two--legged graphs that are \NOL\ to all scales.
Also, the graphs contributing
to the (one--particle--irreducible) Hartree--Fock self-energy  
have the property that the external momentum can always be routed 
through an interaction line. Thus the degree of differentiability  
of the Hartree--Fock approximation to $\Si$, as defined by \queq{\HFse}, 
with respect to the external momentum is the 
same as that of the interaction $\hat v$. For the exact self--energy $\Si$, 
the answer is not so easy.
 
In the four--legged case, the \NOL\ graphs, 
i.e. those without improved power counting,
will turn out to be the ladder graphs that are known 
to produce symmetry breaking \quref{FT2, FMRT}. We will 
give an explicit bound that shows 
that only insertions of these four--legged diagrams can 
produce the factorials in the values of individual graphs.
The concept of improved power counting, together with this result,
makes precise the notion of ``leading divergences'' (see Section 2.7).

The subtractions are implemented by adding counterterms to the action.
These counterterms are of mass type, that is, they are bilinear in
the fermion fields. 
If both the band structure and the potential have spherical 
symmetry, any two--legged diagram contributes a value 
$T(p_\zer , \ve{p} )= T(p_\zer , \abs{\ve{p}} ) $ 
to the two--point function, i.e. spherical symmetry forces the 
function only to depend on $\abs{\ve{p}}$. Thus the 
subtracted terms are simply constants since 
$$
T(0, \prP (\ve{p} ) ) = T ( 0,\abs{\prP(\ve{p})} ) 
= T ( 0 , \sqrt{2 m \mu })
\eqn $$ 
for the spherical band structure $e ( \ve{p} ) = \ve{p}^2/2m - \mu $. 
Their sum produces a shift in the chemical potential \quref{FT1}, 
and the interpretation of renormalization in that case is that 
the interaction changes the radius of the Fermi sphere. 
In the case where the original band structure or the potential 
does not have spherical symmetry, $T(0, \prP(\ve{p} ) ) $ is
still a function of the spatial part of momentum. This is easy to 
understand since in that case the shape of the Fermi surface may change, 
but technically it is a complication because, in renormalization group
language, there is not only one relevant parameter but instead there are
infinitely many, needed to describe the shape of the surface. 
To cancel the divergences, the counterterms are chosen 
such that the interacting Fermi surface is held fixed.  
They determine the shift between
the noninteracting and the interacting Fermi surface and thus 
include part of the effects of the self--energy. 
%
%   Section on Results
%
\sect{Results}

\noindent
The long--distance behaviour of the free electron Green function 
$\check C(x-y)$ is a power law falloff in $\abs{x-y}$,
determined by the singularity of $C(p)$ in momentum
space. If one cuts off this singularity, i.e. forbids small 
values of the energy $e(\ve{p} )$, the Green function decays
exponentially, with a decay length $\sim 1/$energy. We do 
a multiscale analysis by decomposing into energy shells and 
successively integrating out fields in those energy shells.
This gives rise to a series of effective actions, which can 
also be viewed as the Green functions with an infrared cutoff
given by the energy scale. 
Let $M>1$ be a scale parameter (see Section 2.1), and 
$j \in \Z$, $j < 0$. The shell of scale $j$ around the 
Fermi surface is the set of $p$ for which 
$M^{j-2} \leq \abs{ip_\zer - e(\ve{p} )} \leq M^j$. 
We consider an infrared cutoff  on scale $M^I$ where $I > -\infty $, 
$I \in \Z$, $I < 0$ (see Section 2.1). We also call $I$ the 
infrared cutoff. 
Let $\cM $ be the interval given in \AThr\  and 
fix $\mu \in \cM $. 
Define the connected amputated renormalized $2m$ point Green functions $G_{2m}^I$ with infared cutoff $I$ as the formal power series 
$$
G^I_{2m} = \sum\limits_{\la =1}^\infty 
\la^r G_{2m,r}^I
\EQN\GImseries$$ 
where $G^I_{2m,r}$ is the renormalized $r$th order 
Green function (see (2.72)). 
Without going into the details, it is the modification of the 
connected Green function in \queq{\Genconn} where only the fields
with energy scale $\geq M^I$ are integrated over, 
and the interaction contains an additional term  $K^I$
that modifies the band structure. 
The term `renormalized' refers to this modification
since in the graphical analysis, $K^I$ appears as the 
counterterms. We also introduce an ultraviolet cutoff
that removes the higher bands. The $G^I_{2m}$ are 
analytic in $\la $ for $I > -\infty $, and we will show that 
the limit $I \to - \infty $, $G_{2m,r}$, of $G^I_{2m,r}$ exists, so that
in that limit \queq{\GImseries} becomes a well--defined {\it formal}
power series. For $s \geq 0$ and functions $F: \R \times \cB \times 
\{ \uparrow, \downarrow \}\to \C$ define the norms
$$
\abs{ F}_s = \sum\limits_{\al:\abs{\al} \leq s}
\sup\limits_{p \in \R \times \cB} \;
\max\limits_{\si,\si' \in \{\uparrow, \downarrow\}}
\abs{\del^\al F_{\si\si'} (p)}
\EQN\Sobotka $$
where $\al =(\al_\zer, \ldots , \al_d) \in \Z^{d+1}$ 
is a multiindex with $\al_i \geq 0$ for all $i$,
$\abs{\al} = \sum\limits_{i=0}^{d} \al_i$, and 
$\del^\al = \left({\del \over \del p_\zer}\right)^{\al_\zer} \ldots 
\left({\del \over \del p_d}\right)^{\al_d}$. Similarly, for functions
$u$ defined on $(\R \times \cB )^{n-1} \times \{ \uparrow, \downarrow \}^n$,
define 
$$
\abs{u}_s = 
\sup\{ 
\sum\limits_{\al_\one, \ldots \al_{n-1} \atop
\abs{\al_\one} + \ldots + \abs{\al_{n-1}} \leq s}
\abs{ D^\al u_A(p_1,\ldots,p_{n-1})}: 
p_i \in \R\times \cB , A \in \{ \uparrow , \downarrow\} ^n \}
\EQN\sNormdef $$
where $D^\al = \left({\del \over \del p_\one}\right)^{\al_\one} \ldots
\left({\del \over \del p_{n-1}}\right)^{\al_{n-1}} $,
and 
$$
\abs{u}^\prime =
\int\limits_{(\R \times \cB )^{n-1}}
d^{d+1} p_1 \ldots d^{d+1} p_{n-1}
\max\limits_{A \in \{ \uparrow , \downarrow \}^n }
\abs{u_A(p_1,\ldots ,p_{n-1})}
.\EQN\priNormdef $$
The self--energy $\Si^I=\sum\limits_{r\geq 1} \la^r \Si^I_r$ is given as a formal power series by
$$
\Si^I (\ve{p} ) = (1-G^I_\two C_I ) ^{-1} G^I_\two (\ve{p} ) 
\EQN\SelfI $$
where $C_I$ is the propagator with infrared cutoff $I$
(the inverse relation is $G^I_\two = \Si^I (1-C_I\Si^I )^{-1}$). 
\The{\Renorm} {\Thsty 
There is a formal power series
$$
K ^I(\ve{p} ) = \sum\limits_{r=1}^\infty K^I_r (\ve{p} ) \la^r
\eqn $$
such that for the interaction
$$
\cV = \la V + \int \db ^{d+1} p \psq (p) K^I (\ve{p} ) \ps (p)
\eqn $$
the following statements hold. For all $m \in \N $,
the infrared limit $I \to - \infty $ of the $G^I_{2m,r} $ exists.
More precisely, for every $r\geq 1$ there are 
$\Si_r \in C^1(\R\times \cB, \C )$ and $K_r \in C^1(\cB , \R ) $ and 
for all $m\geq 1$ there are $G_{2m,r}$ such that as $I \to -\infty$, 
\leftit{$(i)$} $G^I_{2,r} \to G_{2,r}$ in $\abs{\cdot }_\zer$,
\leftit{$(ii)$} 
$G^I_{2m,r} \to G_{2m,r} $ converges in
$\abs{\cdot } ^\prime $.
\leftit{$(iii)$} 
$\Si^I_r  \to \Si_r $ in $\abs{\cdot}_{1}$, 
and $(\ell \Si)(p) = \Si(0,\prP(\ve{p})) = 0$.
\leftit{$(iv)$} 
$K_r^I \to K_r $ in $\abs{\cdot}_1$.
\par\noindent
Moreover, the Green functions are locally Borel summable, that is, 
there are constants $\Ga_\two$, $\ka_\two$,$\si_\two$  and $\Ga_{2m}$
such that
$$\eqalign{
\abs{G_{2,r}}_\zer &\leq {\Ga_\two}^r\; r! \cr
\abs{K_r}_\one & \leq {\ka_\two}^r \; r! \cr
\abs{\Si_r}_\one & \leq {\si_\two}^r \; r! \cr
\abs{G_{2m,r}}^\prime &\leq {\Ga_{2m}}^r \; r!
.\cr}\EQN\Borel $$ 

}

\someroom\noindent
Thus, to all orders in $\la$, the Green functions of
the model with one-particle band structure $e + K$ can be calculated in
renormalized perturbation theory, and they are given by almost
everywhere finite functions of the independent external momenta
(the momentum conservation delta function is already taken out).
The self--energy is a continuously differentiable function of 
$p_\zer $ and $\ve{p}$, the counterterms $K_r$ are finite and 
continuously differentiable in $\ve{p}$. 
The counterterms $K_r^I$ are constructed recursively in $r$
(``order by order in the expansion in $\la $'') from (2.76), 
the diagrams that contribute are of self--energy type.
Since the amputated function $G_\two $ is first order in $\la$, 
i.e.\ the free propagator is subtracted from the two--point function before
amputating to get $G_\two$, the unamputated connected two point  
function indeed tends to $(ip_\zer - e(\ve{p} ) - \Si (p ))^{-1}$ 
in the limit $I\to -\infty$. Thus $\Si $ is the usual (Dyson) self--energy.  
Because $\Si(0,\ve{p})=(\ell \Si)(p) =0$ for all $\ve{p}\in S$,
 the interacting model with
one--particle band structure $e+K$ has the same Fermi surface at the 
given value $\mu $ of the chemical potential as the free model with 
band structure $e$. In other words, the effect of renormalization 
is indeed that the interacting Fermi surface is kept fixed. 
This is a much more delicate condition than in the spherical 
case where the function $K $ reduces to a constant, i.e. a 
shift $\de \mu$ in the chemical potential. 

The infrared limit of the renormalized expansion gives the same
convergent Greens functions if we choose a finite volume and
positive temperature, and the same conclusions hold, with functions
that have a limit as the volume tends to infinity and/or the
temperature goes to zero. The point of the renormalization in finite
volume and at finite temperature, where there are no divergences
in the loop integrals, is to rearrange the expansion in a way 
that uniformity in volume and temperature and 
convergence of the expansion coefficients for the Green 
functions in the thermodynamic and zero--temperature limit 
is achieved. Of course, by the above discussion, 
this rearrangement amounts precisely to keeping track of how 
the Fermi surface moves when the interaction is turned on.

If the bound for the coefficients \queq{\Borel} is saturated, 
the renormalized expansion has convergence radius zero, which 
in itself may not seem a very useful statement. However,  
if one is willing to go to a slightly more technical level and 
consider the representation of the Green functions as sums over 
values of Feynman graphs, the renormalization method also 
yields much more precise and detailed statements about 
when and why the series diverges. 
It is a well--known fact in renormalizable field theories 
that the only source of factorial growth of individual diagrams
is the marginal scale behaviour of insertions of four--legged 
subdiagrams. In this paper we show the stronger 
statement that if there are no {\it ladder} subdiagrams, the values of
all graphs are bounded without the $r$ factorial, i.e. we 
specify the set of those four--legged diagrams that can really 
produce factorials much more precisely. The meaning
of the term that in a given graph there are no ladder 
subdiagrams is defined in Section 2.5: they are the graphs that contain
no four--legged \NOL\ subdiagrams to any scale. This
statement is useful because the structure of these graphs is
given explicitly in Section 2.4. The four--legged \NOL\ diagrams
are ladder diagrams, also called bubble chains, where the fermion lines 
may be dressed with 
Hartree--Fock type corrections. However, any vertex corrections or 
polarization subdiagrams make the graphs overlapping and its 
scale sum convergent instead of marginally divergent. 
The detailed bound is stated and discussed in Section 2.7; 
it also depends on the tree decomposition of the graph. 
A short version is
\The{\Nice} {\Thsty 
Let $G$ be a graph contributing to $G_{2m,r}$, and denote
by $V(G)$ the norm of the scale sum of $Val(G^J)$, 
where $J$ is any labelling of $G$, and the norm depends on 
the number of external legs, as in Theorem \Renorm.  
If for any labelling $J$, $G^J$ does not contain any \NOL\ four--legged 
subdiagrams at any scale, then $V(G) \leq {V_m}^r$, where $V_m$ is a constant 
independent of $r$. 
}  

\someroom\noindent
In other words, Theorem \Nice\ means that a single graph in the $n^{th}$
order of perturbation theory can have
value $\sim n!$ only if it contains ladder subdiagrams. All other 
four--legged insertions do not produce any factorials in the value
of single graphs.
This suggests that only insertions of ladder diagrams
can change the behaviour of the correlations, i.e. the properties 
of the ground state, qualitatively, and that all other 
corrections are analytic in the coupling. A nonperturbative proof
of this requires control over the sum of all graphs, 
hence an implementation of the Pauli principle, which
has been done in $d=2$ spatial dimensions in \quref{FMRT}.
 
In the case of a strictly convex Fermi 
surface, only the behaviour at transfer momentum zero can lead 
to factorials in the values of individual graphs 
because at all other values of the momentum, 
the surface intersects transversally with its translate or at least
there is a curvature effect that implies the absence of a singularity. 
In the spherical case the existence of a singularity at zero 
transfer momentum has been shown to be responsible for the 
occurrence of off-diagonal long range order in the ground state
\quref{FT2,FMRT}. If the Fermi surface is transversal to its 
negative, the ladder graphs are nonsingular and analyticity in 
$\la $ holds in infinite volume \quref{FKLT}.  

The classification of graphs into \OL\ and \NOL\ ones that will 
be introduced in Section 2.4 may seem technical at first;  
it is, however, natural since the graphs that are \NOL\ to all scales
are 
the dressed ladder graphs in the four--legged case and the Hartree--Fock 
graphs in the two--legged case. The four-legged \NOL\ graphs are the only ones
that do not show improved power counting behaviour, and in this
sense their resummation is a resummation of the 
`leading divergences'. 

Note, however, that Theorem 1.3 is a statement about the 
behaviour of values of single graphs, and does not require 
any resummation. Therefore it holds irrespective of the sign
of the coupling (on which the existence
of solutions to the gap equations from resummation depends).
Also, it holds for the general class of non--flat Fermi surfaces 
given by our Assumptions \AOne--\AThr, and not just for strictly 
convex Fermi surfaces.

It is technically necessary to do an expansion with a fixed
interacting Fermi surface, to prevent the problems described
above when one expands a moving singularity in terms of a fixed one.
In order to construct a model with a given one--particle band structure
and to see how the Fermi surface moves under the interaction,
and also to clarify the relation between the counterterm function
$K$ and the self--energy, we have to study the map $e \mapsto
E=e+K $ further and show that it is invertible. To invert this
map, one would like to take a derivative of $K$  
with respect to $e$. It is not obvious that such a derivative exists 
since $K$ is a functional of $e$ obtained by integrating factors
of $1/(ip_\zer - e(\ve{p}) )$, and taking a derivative  
produces a square of the denominator, and thus potentially a singularity, 
since the square of the propagator is not locally integrable. 
However, the volume improvement bounds allow us to take this derivative.
The latter is also necessary to get information about the dependence
on the chemical potential $\mu$, since the expansion has so far been
done at a fixed value of $\mu $,
which then fixes the Fermi surface. Different values of $\mu $
give rise to different Fermi surfaces, and in the case without
spherical symmetry, different also means of different shape.
The renormalization would be useless if it worked only 
pointwise in $\mu $. In other words, it is important to establish some
continuity properties in $\mu$. We show that the counterterms and
thus the self--energy and the value of any graph, are continuously
differentiable in $\mu$. More generally, we prove that
this $C^1$ property holds for derivatives 
with respect to $e$, i.e. we allow for much more general variations
of the band structure than just a shift by a constant.
Let 
$$
D_h K_r^I (e,h) = {\del \over \del \al } K_r^I(e+\al h)\vert_{\al=0}
\EQN\Dhdef $$
be the directional derivative of $K_r^I$ with in direction $h$.

\The{\Direct} {\Thsty 
If \AOne\ -- \AThr\  hold, then 
$\lim\limits_{I \to -\infty} D_h K_r^I (e,h) $ exists for all $r\geq 1$ and 
$$
\abs{\lim\limits_{I \to -\infty} D_h K_r^I (e,h)}_\zer 
\leq \Const(r)\; \abs{h}_\zer
\eqn $$
}
  
\Cor{\Mucont} {\Thsty 
If \AOne\ -- \AThr\  hold, then 
the counterterms $K$ are continuously differentiable
functions of the chemical potential $\mu$.}

\someroom\noindent
To convert this statement about directional derivatives into one 
about derivatives as bounded linear operators \quref{D} and to consider 
varying $e$, not just $\mu $, we have to be more specific about the set 
of allowed 
$e$'s. Let $\emptyset \neq \cN \subset \cB $ be open. 
For $k \geq 0$, denote the Banach spaces
$(C^k (\cB, \R ) , \abs{\; \cdot \; }_k)$ by $\cC^k$ and  
$(C^k (\overline{\cN}, \R ) , \abs{\; \cdot \; }_k)$ by $\cC^k_\cN$.
For $1 \leq \si \leq d-1$, $g_\two>g_\zer > 0$, and $g_3 >0$ let 
$$\eqalign{
\cA_\two (\si, \cN, g_\zer,g_\two,g_3)=\{& e\in\cC^2_\cN:\abs{e}_\two<g_\two,
S(e) = \{ \ve{p} \in \cB : e(\ve{p} ) = 0\} \subset \cN, \cr
& \abs{\nabla e (\ve{p} )} > g_\zer \hbox{ for all } \ve{p} \in \cN, 
\hbox{ and } n: S(e) \to S^d,\cr & 
\om \mapsto n(\om )={\nabla e \over \abs{\nabla e}}(\om )
\hbox{ satisfies: for all } \om \in S, \cr 
&\hbox{rank } dn(\om ) \geq \si, \hbox{ and all nonzero eigenvalues }m \cr
&\hbox{of $dn$ satisfy } \abs{m} > g_3 \}.
\cr}\EQN\Atwodef $$
Here $ S^d=\{a \in \R^d: \abs{a}=1\}$, and $dn$ is the derivative of $n$ 
with respect to\ $\om \in S$. In other words, $dn$ is  the 
derivative of $n$ tangential to the surface $S$
(note that $dn(\om)$ is a quadratic
matrix since the dimension of $S(e)$ is $d-1$, and $n(\om) \in S^d$).
Let $\cL$ be the space of bounded linear operators from
$\cC^2_\cN $ to $\cC^0_\cN $.  
\The{\Kabl} {\Thsty
Let $1 \leq \si \leq d-1$ and $g_\two > g_\zer > 0$. Then $\cA=\cA_\two (\si,\cN,g_\zer,g_\two)$ 
is open in $\cC^2_\cN$. For all $ e \in \cA$, \ATwo\ and \AThr\  hold, 
with $\ka = \si $ and $\rho =1$. For all $e \in \cA $ and all $r \geq 1$,
$D_h K_r^I (e,h) = \left( K_r^I\right)'(e) h$ with 
$\left( K_r^I\right)'(e) \in \cL$, and there is $K_r' \in \cL$ such that
$$
\norm{\left( K_r^I\right)'(e) - K_r'}_\cL \to 0 
\hbox{ as } I \to -\infty
.\eqn $$
The function $K_r: \cA \to \cC^0_\cN$ is differentiable in $e$, and its 
derivative is given by $K_r' \in \cL$. The map $e \mapsto K_r' (e) $ 
is continuous on $\cA$, and there is $C_r>0$ such that for all $h \in \cC_\cN^2$
$$
\abs{K_r' (e) h}_\zer \leq C_r \abs{h}_\zer
\EQN\ofifi $$
$C_r$ is independent of $e\in \cA$. 
}
\sni
The bound \queq{\ofifi} is the most subtle result of this paper. 
Note that no derivative acts on $h$ on the right side of \queq{\ofifi}. 
Because of that, $K_r'$ extends
uniquely to a bounded linear operator on $\cC_\cN^0$, and we can prove
\The{\Inject} {\Thsty 
Let $R\geq 1$, $\la \in \R$, and for $e \in \cA $ let 
$E_\la^{(R)} (e) = e+ \sum\limits_{s=1}^R \la^s K_s (e)$.
If 
$$
\sum\limits_{r=1}^R C_r \abs{\la}^r < 1
,\eqn $$
then $E_\la^{(R)}$ is injective on every convex subset of $\cA$.
That is, if $e_\one, e_\two \in \cA $ with 
$E_\la^{(R)}(e_\one ) = E_\la^{(R)}(e_\two )$, and if 
$(1-s) e_\one + s e_\two \in \cA$ for all $s \in [0,1]$,
then $e_\one = e_\two $. 

}
\sni
Since $\cA $ is open, the maximal ball around any $e \in \cA $ 
is such a convex subset. Thus $E_\la^{(R)}$ is locally injective. 
The significance of Theorem \Inject\ for the problem of self--consistent
renormalization is discussed in the next section. 

The set $\cA$ of Theorem \Kabl\ is more restricted than the set 
of all $e$ satisfying \AThr. It comprises the case of a strictly 
convex Fermi surface, or that of a torus or a cylinder, but, e.g.,
not Example 3 of Section 1.3. More generally, the specification 
of a set of $e$ for which $dn(\om)$ may vanish for some $\om $ 
on $S$, but for which the exponent $\rho \geq 1$ is still uniform 
in $e$, requires the existence of more derivatives ($k>2$) of $e$.
This can be formulated, but for conciseness, we restricted to the
simplest case here. The reader may construct his own generalizations;
the essential requirement is that the constants $u_\zer$ (defined in 
Chapter 2) and the volume improvement exponent $\ep$ must be 
uniform on the set. The reason we gave \AThr\  as a separate 
assumption is that it is more general and can be checked 
without trouble in examples; for instance, it is easy to see
that for the Example 3, $e(\ve{p} ) = p_\one^{2n}+p_\two^{2n} -\mu$, 
there is an open $\mu $--interval with the desired properties, 
and this already suffices to prove Theorems \Renorm\ -- \Mucont.

Finally, we define the Hartree--Fock approximation as the sum over all 
graphs that are \NOL\ on all scales; equivalently, these are 
the graphs produced by iterating the Hartree--Fock integral 
equation \queq{\HFse}. 
This resummation also defines a map 
$e \mapsto e+H $, where $H$ are the counterterms in the 
Hartree-Fock approximation. 
\The{\HFbrav} {\Thsty 
The map $e \mapsto e+H$ is invertible in every fixed 
order in perturbation theory. } 
\sni
This Theorem is easy to prove; we shall discuss its motivation
in the next section. 
\sect{Discussion}
\noindent
The interpretation of renormalization is thus: the 
unrenormalized Green functions diverge because it is
wrong to assume that both the band structure and the Fermi surface 
stay fixed when the interaction $\la V$ is 
turned on. In reality, they respond to the interaction --
if the surface is fixed, the band structure changes, and vice versa  
(this is similar to the situation in KAM theory
where the frequencies and actions of quasiperiodic orbits cannot both 
stay fixed under a perturbation). 
To do the expansion, we prefer not to let the Fermi surface move,
since the moving of the singularity produces the divergences 
discussed above. Instead we allow for a change in the band structure $e$. 
The function $K(\ve{p})$ contains the terms that are necessary to prevent
the surface from moving under the perturbation. 
This function depends on the vector field $u$ which we 
used to define the projection onto the Fermi surface. 
The dependence on $u$ amounts to a reparametrization 
of the Fermi surface and has no physical consequences. 

It has long been known in solid state theory that 
self--energy effects have to be taken into account to 
avoid divergences in perturbation theory. In many 
accounts this is described as self--consistent renormalization, 
with the idea that if the free two--point function is
expressed in terms of the exact two--point function
everywhere, two--legged insertions disappear, since they 
arose from self--energy terms. This procedure is usually 
called `self--consistent' renormalization and described in 
words in the literature. Often, one then goes on to describe
particular approximations, such as the Hartree Fock approximation. 
Since none of these approximations is exact, none of them 
removes all the divergences, and one point of our analysis 
is that we give a clear procedure how to do this to all 
orders in perturbation theory: the divergences are removed 
by fixing the Fermi surface. Self--consistent renormalization 
is then achieved by inversion, i.e. solving the equation 
$$
E=e+K(e,\la )
\EQN\EiseplK $$
for $e$ in terms of a given $E$. 
It is really a separate step. 
Once this is done, the combination of renormalization and
inversion allows one to determine how the Fermi surface moves
when the band structure is fixed. 

Obviously, the solution of \queq{\EiseplK} 
requires some knowledge of the regularity properties of the 
map $K$, which is a map between function spaces. 
We have established enough regularity to show 
that  $id + K $ is locally injective. In other words, we have proven that
for any interacting band structure, there is locally at most one 
bare band structure that has the same Fermi surface, 
i.e. uniqueness of the solution.  
The existence (surjectivity) proof requires more detailed 
bounds and more stringent assumptions and will appear in a sequel paper.

Note that this regularity problem does not just arise 
because we use counterterms
to do the expansion correctly. Any attempt to `consider only 
skeleton graphs first and then replace the free propagator by the interacting 
one on all lines' also requires the inversion of the map $e \mapsto e + \Si $, 
and the regularity problem is thus similar to ours (only harder, since 
$\Si $ also depends on $p_\zer $). 
In brief, in any way of looking at the system, there is the question how 
regular the self-energy, and thus the interacting Fermi surface, is.
In the heuristic discussion after \queq{\HFse} that motivated why the 
divergences are artificial, this question was postponed by the assumption
that $\Si$ is ``reasonable'', so that \queq{\integr} can be used to show 
well-definedness of the left side of \queq{\bledsinn}. However, 
$\Si$ is not a function one is free to choose or make assumptions about.
It is determined by the interaction, and therefore its regularity has 
to be proven. We have proven that $\Si$ and $K$ are $C^1$ (Theorem \Renorm).
Inverting \queq{\EiseplK} requires at least $K \in C^2$. The proof of this
is will appear in another paper.

Regularity of the Hartree--Fock approximation to $\Si$ (Theorem \HFbrav), 
is easily shown: it is obvious that the external momentum
can be routed through an interaction line in every Hartree--Fock graph, 
so the Hartree--Fock self--energy has the same regularity properties
as the interaction potential.  

Since improved power counting plays a central role in the 
technical analysis done here and since the facts on which it is based
are not specific to our multiscale analysis and therefore 
have wider applications, we describe briefly how it 
comes about. A way to understand power counting is to weight 
the growth of the propagator in the vicinity of its singularity $S$
against the smallness of the volume of shells around the Fermi surface,
where it becomes large. We use a scale decomposition where momentum 
space is cut into shells around the Fermi surface, as sketched in 
Figure 3 (a). It is easy to see that the $\ve{p}$--volume of a shell 
in which (say) $2^{j-1} \leq \abs{e(\ve{p})} \leq 2^{j}$ ($j<0$)
is bounded by a constant times $2^{j}$ (see also Section 2.1). 
It is also easy to deduce the integrability 
properties of $C$ that we discussed above by 
weighting this volume against the growth of $\abs{C}$ in a summation 
over shells:
$$\eqalign{
\int\limits_{|ip_\zer-e(\ve{p})| \le \sfrac{1}{2}} dp_\zer d\ve{p} \;
\frac{1}{|ip_\zer-e(\ve{p})|^\al} & = \sum\limits_{j<0} 
\int\limits_\R dp_\zer \int\limits_\cB d\ve{p} \; 
\frac{1(2^{j-1}< |ip_\zer -e(\ve{p})| \le 2^j)}{|ip_\zer-e(\ve{p})|^\al} 
\cr
&\leq \sum\limits_{j< 0} 2^{(1-j)\al}
\int\limits_{-2^j}^{2^j} dp_\zer  
 \int\limits_\cB d\ve{p} \; 
1(2^{j-1}< |e(\ve{p})| \le 2^j) 
\cr
&\leq \Const 2^{\al} \sum\limits_{j< 0} 2^{-j\al} \; 2^j \; 2^j \cr
& = \Const 2^{\al}\sum\limits_{j< 0} 2^{j(2-\al)} 
,\cr}\eqn$$
which converges if $\al<2$.
Up to this point, this is just a rewriting of \queq{\integr}. However, 
the geometry of these shells has important consequences for nontrivial 
graphs, which we discuss now. 
%
%   here comes Figure 3
%
\midinsert
\hfil\break
\null\figplace{fig3}{0in}{0in}\hfil

\centerline{{\it Figure 3}}
\endinsert

Beyond lowest order, 
the momentum assignments in graphs with at least two lines consist 
of linear combinations of the loop momenta with the external momenta, 
e.g.\ $\ve{p}$ and $\ve{p+q}$, where $\ve{p}$ is a loop momentum. 
On scale $j$, both $\ve{p}$ and $\ve{p+q}$ 
must be in a shell of thickness $2^{j}$ around $S$. 
The volume of the full shell is of order $2^{j}$. 
On the other hand, for most values of $\ve{q} \in\cB$, the intersection 
of $S$ and $\pm S + \ve{q}$ will be transversal. Thus the support
of the integrand will have a volume which is much smaller, 
roughly by a factor $2^{j}$,  
since the volume is not that of an entire shell any more, but that
of a transversal intersection of two shells around $S$ (see Figure 3 (b)). 
However, for those values of $\ve{q}$ where the intersection of the 
shell with its translate by $\ve{q}$ is not transversal, 
e.g.\ for $\ve{q} =0$, or the translation shown in Figure 3 (c), 
there is no gain at all, 
i.e.\ there is no uniformity in $\ve{q}$. 
The improved power counting bound is based on the observation that 
if there is no nesting in the sense that \AThr\ holds, then the 
set of $\ve{q}$ for which 
the intersection is not transversal has small volume itself.
So, if $\ve{q} = \ve{k} + \ve{Q}$ where $\ve{k} $ is another loop 
momentum, and if there was no gain in the integration 
over $\ve{p}$, $\ve{k}$ must be in a set of small volume.   
This restriction produces an additional `volume improvement factor' 
$2^{\ep j}$ in the second loop integration over $\ve{k}$ 
(this also applies to the surface drawn in Figure 3).
Thus, in the double integral
over $\ve{p}$ and $\ve{k}$ that appears in \queq{\Itwodef}, there will 
{\it always} be an improvement factor $2^{\ep j}$, which is
uniform in $\ve{Q}$. Therefore, $\ve{Q}$ may be an arbitrary 
combination of loop momenta and external momenta, and it is not 
necessary to keep track of all complications of the momentum flow in
general graphs to extract the improvement factor. 
It is only necessary to find out which graphs have this volume gain,
i.e.\ contain a factor $I_\two$ as a subintegral. 
Obviously, they must have at least two loops, 
but the above condition that  $\ve{q} = \ve{k} + \ve{Q}$ 
with another loop momentum $\ve{k}$ means also 
that there must be a fermion line in which two loop momenta flow
(the two loop momenta $\ve{p}$ and $\ve{k}$ flow
in the line with momentum $\ve{p}+\ve{q}= 
\ve{p}+\ve{k}+\ve{Q}$). This is now a purely graph theoretical 
question. The class of graphs for which such a line exists 
is precisely that of 
\OL\ graphs defined in Section 2.4. The \NOL\ graphs are classified
explicitly in the two-- and four--legged case (it is not 
hard to generalize the characterizations given in Section 2.4 
to graphs with more that four external legs, but we do not need that here).
The volume improvement bound \queq{\ImpVol} is proven under the 
hypotheses \ATwo\ and \AThr\ in Appendix A by the argument 
outlined above. 

Note that the above transversality and no--nesting 
arguments require $d \geq 2$. In $d=1$, improved power counting is 
absent. This is one reason why one--dimensional many--fermion models behave 
differently from the higher--dimensional ones.

The proofs in Section 2.4 are elementary and independent of the scale 
decomposition. Indeed, the only property of the 
model that is used in Section 2.4 is that all vertices have an 
even incidence number, which is true in our class of models since
the interaction is a four--fermion interaction (see also Figure 4 
in Section 2). 
For vertices with an odd incidence number, a similar classification 
can be done.  

The implementation of these graphical statements for the 
volume gains in the problem with the full scale structure 
is done in Section 2.5 and 2.6 and used 
thereafter to prove the stated Theorems. Some of these proofs are not short, 
but they are in principle an application of the simple ideas stated above.

\someroom\noindent
{\bf Acknowledgements:} It is our pleasure to 
thank Horst Kn\" orrer for 
many discussions. M.S. would also like to thank 
Michael Aizenman and Thomas Spencer for a discussion, and 
Michael Aizenman for hospitality at Princeton University, where 
part of this work was done. 
The research of M.S. was supported by NSERC, an Otto--Hahn fellowship of the 
Max--Planck Gesellschaft zur F\" orderung der Wissenschaften, and by the 
funding agency FOM. The research of J.F. was supported in part by the
Natural Sciences and Engineering Research Council of Canada.

\vfill\eject

\chap{Renormalization and Convergence}

\noindent
In this chapter, we set up the renormalization flow and
define the localization operator that is used to subtract the
value of two--legged diagrams on the Fermi surface. 
We then develop
one of the main technical tools, the graph structure Lemmas that
are used to extract volume improvement factors systematically
for any labelled graph. 
We use this to show an improved power counting bound, 
and then show  that the renormalized Green functions 
converge in every order in perturbation theory, and 
that the only four--legged graphs
which do not obey improved power counting are 
the ladder graphs.

We start with some elementary remarks that follow from the 
assumptions. By \ATwo, $S$ is a compact $(d-1)$ -- dimensional 
$C^k$--submanifold of $\cB$. Let
$$
U_\veps (S) =\{ \ve{p} : \exists \ve{q} \in S {\rm\ with\ } |\ve{p-q}|<\veps \}
.\eqn $$
Then there is $\de $ such that
$G_\zer = \sup \{ |\nabla \eof{p}|:\ve{p} \in U_{2\de} (S) \}$ 
is finite, and such that
$g_\zer = \inf\{ |\nabla \eof{p}|:\ve{p} \in U_{2\de} (S) \} > 0$.
Let $u$ be a unit vector field on a neighbourhood $U_\de (S)$ of
$S$. We call $u$ transversal to $S$ if there is $u_\zer >0$ such that
for all $\ve{p} \in S$, $\nabla \eof{p} \cdot u(\ve{p} ) \geq u_\zer > 0$. 
Denote the integral curve of $u$ passing through 
$\ve{p} \in S$ by $\ga_{\ve{p}} $, that is,
$\ga_{\ve{p}} : (-\veps , \veps ) \to \cB$, 
$t \mapsto \ga_{\ve{p}}(t)$,
$\ga_{\ve{p}} (0) = \ve{p}$, 
and for all $t \in (-\veps , \veps )$,
${\del \over \del t} \ga_{\ve{p}} (t) = u (  \ga_{\ve{p}} (t))$.

\Lem{\uProps}  {\Lesty  
Assume \ATwo.
\leftit{$(i)$} There is a $C^\infty $ vector field $u$ transversal to $S$, 
and there is $\veps > 0$ such that
$\Psi : S \times (-\veps , \veps ) \to
\Psi\left( S \times (-\veps , \veps )\right)
\subset  \cB$, defined by
$\Psi ( \ve{p}, t ) = \ga_{\ve{p}}(t ) $,
is a $C^k$--diffeomorphism.
\leftit{$(ii)$} 
There are $\de > 0$ and $u_\zer \in (0,1) $ such that $\overline{U_{2\de} (S)}
\subset \Psi\left( S \times (-\veps , \veps )\right)$, and such that
for all $\ve{q} \in U_{2\de} (S)$ :
$0 < {g_\zer \over 2} \leq u_\zer \leq 
\nabla e (\ve{q}) \cdot u(\ve{q}) \leq G_\zer$.
\leftit{$(iii)$}
Define the functions $\ta : \overline{U_{2\de} (S)} \to \R $ and
$\prP : \overline{U_{2\de} (S)} \to S $ as follows.
For $\ve{q} \in  \overline{U_{2\de} (S)}$
$$
\left( \prP (\ve{q} ), \ta(\ve{q}) \right) = \Psi^{-1} (\ve{q} )
.\eqn $$
In other words, $\ga_{\prP (\ve{q})}(\ta (\ve{q}) ) = \ve{q}$.
Then
$$
\ve{q} = \prP (\ve{q}) + \int\limits_0^{\tau (\ve{q} )}
u(  \ga_{\prP (\ve{q})} (t)) dt
,\EQN\Interpol $$
so $\abs{\ve{q} - \prP (\ve{q})} \leq  \abs{\tau (\ve{q} )}$ and
$$
\abs{\ve{q} - \prP (\ve{q})} \leq  {1\over u_\zer} \abs{e(\ve{q} )}
\eqn $$
Furthermore, 
$u_\zer \leq { e(\ve{q}) \over \ta (\ve{q})} \leq G_\zer $.
\leftit{$(iv)$} 
Let $\ve{p} \in U_\de (S)$, 
$\rh = e(\ve{p} ) $ and $\om = \prP (\ve{p} )$.
The map $\ch : \ve{p} \mapsto (\rh, \om )$ 
is a $C^k$--diffeomorphism
from $U_\de (S)$ to a subset of $\R \times S$. Denoting
its  inverse map by $\ve{p} (\rh , \om )$, 
there are constants $A_\zer $ and $A_\one$ such that 
the Jacobian $J(\rh , \om )= \det \ve{p}' (\rh, \om )$
obeys
$$
\sup\limits_{\ve{p} \in U_\de (S)}
\abs{J(\rh, \om)} \leq 
{1 \over u_\zer } A_\zer
\eqn $$
and its derivative $\del J$ obeys
$$
\sup\limits_{\ve{p} \in U_\de (S)}
\abs{\del J(\rh, \om)} \leq 
{1 \over {u_\zer}^2 } A_\one
\eqn $$
$A_\zer $ depends on $\de $, $u_\zer$, and $\abs{u}_\one$,
$A_\one $ also depends on the second derivative of $u$.

}

\Proof $(i,ii)$ We show that $u \in C^\infty $ transversal to $S$ exists 
even if $e$ is only $C^1$. For $\ve{p} \in U_{\de_\zer} (S)$ let 
$n(\ve{p} ) = {\nabla e(\ve{p}) \over \abs{\nabla e(\ve{p} )}}$, 
then $n$ is continuous in $\ve{p}$. So for all $\ve{p} \in S$ there is
$r(\ve{p} ) > 0$ such that $n(\ve{p} )\cdot n(\ve{p}') > {1 \over 2}$ 
for all $\ve{p}' \in U_{2r(\ve{p})} (\ve{p})$. Since $S$ is compact, 
the covering $\left( U_{r(\ve{p})}  (\ve{p})\right)_{\ve{p} \in S }$
contains a finite subcovering by $V_i = U_{r(\ve{p_i} )} (\ve{p_i})$,
$i \in \{ 1, \ldots , n\}$, and there is $\de > 0$ such that 
$U_{3\de }(S) \subset \bigcup\limits_{i=1}^n V_i$. $U_{3\de} (S) \subset \cB$
is open, hence a $C^\infty$ submanifold of $\cB$. Choose a $C^\infty$ 
partition of unity $(\ch_i)_{i}$ that is subordinate to the cover $V_i \cap
U_{3\de }(S)$ and that obeys supp $\ch_i \subset U_{r(\ve{p_i} )} (\ve{p_i})$
for all $i$. Define
$$
u(\ve{p} ) = \sum\limits_{i=1}^n \ch_i (\ve{p} ) n(\ve{p}_i)
\eqn $$
then $u \in C^\infty (U_{3\de} (S) , \R^d )$ and by construction of $\ch_i$
$$
u(\ve{p} ) \cdot \nabla e(\ve{p} ) = 
\sum\limits_{i: \ve{p} \in U_{r (\ve{p}_i)} (\ve{p}_i)}
\ch_i (\ve{p}) \abs{\nabla e (\ve{p} )}\; n(\ve{p}_i ) \cdot n(\ve{p} )
\geq {g_\zer \over 2} \sum\limits_i \ch_i (\ve{p} ) 
=  {g_\zer \over 2}
.\eqn $$
$(i)$ is now obvious since $S$ is a $C^k$--submanifold of $\cB$, and 
$(ii)$ is clear by construction of $u$. \pni
$(iii)$ Eq. \queq{\Interpol } holds by definition of the map $\Psi$
and that of the integral curve $\ga $. It obviously implies
$$
\abs{\ve{q} - \prP (\ve{q})}  \leq  
\abs{
\int\limits_0^{\tau (\ve{q} )}
u(  \ga_{\prP (\ve{q})} (t)) dt 
}\leq
\abs{\tau (\ve{q} )}
\eqn $$
since $u$ is a unit vector field. If $e(\ve{q})\geq 0$, then
$\tau (\ve{q} )\geq 0$ and, since $e(\prP (\ve{q} )) =0$,
$$
e(\ve{q} ) =
\int\limits_0^{\tau (\ve{q} )}{d\hfill\over dt}
e(\ga_{\prP (\ve{q})} (t)) dt
= \int\limits_0^{\tau (\ve{q} )}
(u\cdot \nabla e)(\ga_{\prP (\ve{q})} (t)) dt
\geq u_\zer \ta(\ve{q} )  \geq u_\zer \abs{\ve{q} - \prP (\ve{q})}
.\eqn $$
The case $e(\ve{q})\leq 0$ is similar.

\noindent
$(iv)$ The map is a diffeomorphism 
because it is the composition of $\Ps$ 
with the inverse of 
$(\om, \ta )\mapsto (\om , \rh ) = (\om , e (\Ps ( \om ,\ta )))$ 
and because 
$$
{\del \rh \over \del \ta } (\om , \ta ) = 
\left( \nabla e \cdot u \right) (\Ps ( \om , \ta ))
\EQN\drhdta $$
so that ${\del \rh \over \del \ta } \geq u_\zer $
in $\Ps^{-1}(U_\de (S))$. this also implies the bounds for the 
Jacobian and its derivative.  
\endproof

\Rem{\grade} The choice $u = {\nabla e \over \abs{\nabla e} }$
has most of the above properties, with $u_\zer  = g_\zer$, 
but it is only $C^{k-1}$ if $e$ is $C^k$, and then
the maps $\Psi $ and $\ch$ are only $C^{k-1}$. 
In particular, with this choice of $u$, finiteness of $A_\one $ 
requires $k \geq 3$ in \ATwo.

\sect{Scale Decomposition and Power Counting}
\noindent
Let $\ep $ as in \AThr, $u_\zer $ and $\de $ as in Lemma \uProps\ and
$M>\max\{ 4^{1/\ep}, {1 \over u_\zer \de}\}$. 
Then $\abs{e (\ve{p} )} \leq M^{-1} $ implies 
$\ve{p} \in U_\de (S)$. 
Let $a \in C^\infty (\R_\zer^+, [0,1])$ be such that
$$
a(x) = \cases{0 & for $x \leq M^{-4}$ \cr
              1 & for $x \geq M^{-2}$\cr}
\EQN\cutadef $$
and $a^\prime (x)> 0 $ for all $x \in (M^{-4}, M^{-2})$. Set
$$
f(x) = a(x) -a({\textstyle{x\over M^2}}) =
\cases{ 0       & if $x \leq M^{-4} $         \cr
        a(x)    & if $ M^{-4} \leq x \leq  M^{-2} $  \cr
        1-a({x\over M^2}) & if $ M^{-2} \leq x \leq 1$ \cr
        0       & if $x \geq 1$,   \cr
}\EQN\cutfdef $$
so that, for all $x > 0$,  $f(x) \geq 0 $ and 
$$
1-a(x) = \sum\limits_{j=-\infty}^{-1} f(M^{-2j}x)
.\EQN\prtunity $$ 
Calling $f_j(x) = f(M^{-2j}x)$,
$$
\hbox{supp } f_j  = \lbrack M^{2j-4}, M^{2j} \rbrack
,\eqn $$
and for all $x \geq 0$,
$$
f_j(x) f_{j^\prime}(x) = 0  \hbox{ if } \abs{j - j^\prime } \geq 2
.\EQN\Nullzwei $$
Defining
$$
C_j(x,y) = {f(M^{-2j}\abs{x+iy}^2) \over ix-y} =
{f_j(\abs{x+iy}^2) \over ix-y}
,\EQN\Cjxydef $$
we decompose
$$
{e^{ip_\zer 0^+} \over ip_\zer - e(\ve{p} ) } = 
{e^{ip_\zer 0^+}a(p_\zer^2 + e(\ve{p})^2)
\over ip_\zer - e(\ve{p} ) } +
e^{ip_\zer 0^+} \sum\limits_{j<0} C_j (p_\zer, \eof{p} )
.\eqn $$
For the purposes of the present paper, we discard the ultraviolet
end of the model by removing the first term in this sum, 
in other words, taking $(1-a)/(ip_\zer - e(\ve{p}))$ as propagator.
The infrared singularity, which is the physically relevant 
feature of the problem, is unchanged.  
 
\Lem{\Simpl}  {\Lesty 
For all $j < 0$
\leftit{$(i)$}
 $\abs{C_j}_\zer \leq M^{-j+2}$. More precisely, for all $p=(p_\zer,\ve{p})$,
$$
\abs{C_j(p_\zer,\eof{p} ) } \leq M^{-j+2}\,
1\left(\abs{ip_\zer+\eof{p}} \in \lbrack M^{j-2}, M^j \rbrack \right)
.\eqn $$
\leftit{$(ii)$} 
Let $A_\zer $ as in Lemma \uProps\ and $A=\max\{ 1,2A_\zer \int_S d\om \} $.
$$
\int\limits_{\cB} d^d\ve{p}\ 
1(\abs{\eof{p}}\leq \veps ) \leq
{A \over u_\zer} \veps
\eqn $$
and 
$$
\abs{C_j}' = \int\limits_{\R \times \cB} dp_\zer d\ve{p}
\abs{C_j(p_\zer,\eof{p})} \leq {2 A M^2\over u_\zer} M^j
\EQN\Cjpbd $$ 
In particular, taking 
$$
K_\zer = {2AM^2\over u_\zer}
,\EQN\Kzerodef $$
$\abs{C_j}_\zer \leq K_\zer M^{-j} $ and
$\abs{C_j}' \leq K_\zer M^j$ 
\leftit{$(iii)$} For any multiindex $\al $ 
with $s=\abs{\al} \leq k$, there is a constant 
$W_s $ depending on $\abs{e}_s$ and $M$ such that
$$
\abs{D^\al C_j (p_\zer , e ( \ve{p} ) )} \leq 
W_s M^{-(s+1)j} 1(\abs{ip_\zer - e(\ve{p} )} \in [M^{j-2},M^j])
.\eqn $$

} 

\someroom 
\noindent
This Lemma implies that
for any $0> I > - \infty $, any power of the propagator
$\sum\limits_{j \geq I} C_j $ is integrable and so values
of connected graphs evaluated according to the above Feynman rules,
but with this cutoff propagator instead of $1/(ip_\zer - \eof{p})$,
are finite, and $C^\infty $ in $p_\zer$ and $e$.

The bounds given in the Lemma are similar to those in the
spherical case of \quref{FT1,FT2}, and so the power counting is
the same as in Lemma III.1 of \quref{FT2}. The dimension
$\delta_l$ (see \quref{FT2}, eq.(III.5)), is  $\delta_l=1$.

We now state the analogue of the abstract power counting lemma of
\quref{FT2}. For the moment, we refer the reader to \quref{FT2} for 
details about
labelled graphs and the associated trees; 
they will be explained in more detail in Section 2.3.
Let $G$ be a connected graph with an even number
$E$ of external lines, and two-- and four--legged vertices.  
Let $L(G) $ be the set of internal lines of $G$ and
$J:L(G) \to \{ z \in \Z : 0 > z \geq I \}$, $l \mapsto j_l$
be a labelling of $G$, which assigns a scale to each line of $G$. 
Construct the tree $t=t(G^J)$ associated to the labelled
graph $G^J$ as follows \quref{FT2}. The forks $f$ of the tree are the
connected components $G^J_f$ of all the subgraphs 
$\{\ell\in L(G^J) : j_\ell\ge h\}$ with $h\le -1$. The subgraphs are 
partially ordered by inclusion to form $t(G^J)$. The scale of a fork is defined
by
$$
j_f=\min\{j_\ell : \ell\in L(G^J_f)\}
$$

Define, in analogy to eqs (III.9) and (III.10) of \quref{FT2}
$$\eqalign{
D_f &= |L(G_f^J)| - 2 \left(|V(G_f^J )| -1 \right) \cr
\Delta_f &= - {1 \over 2}
\abs{\{ l: l \hbox{ internal line of } G^J, 
l \hbox{ external line of } G^J_f
\} } \cr
\Delta_v &= - {1\over 2} \abs{ \{ l: l \hbox{ internal line of } G^J, 
v\in l\}
}\cr}\EQN\Ddeldef $$
where $V(G)$ is the set of vertices of $G$.
The value of the graph $G^J$ is defined as
$$\eqalign{
\delta ^\# (p_{out}-p_{in} ) Val(G^J)  &=
\sum_{spins \;\al}
\int \prod\limits_{l \in L(G)} \db ^{d+1} p_l\ C_{j_l} (p_l)
\de_{\al_l\al_l^\prime} \cr
&\times \prod\limits_{v \in V_4 (G)}
\de^\# (p_{out,v}-p_{in,v} ) 
\left(\cU_v\right)_{\al_\one^{(v)} \ldots 
\al_{4}^{(v)} }(q_v )\cr
&\times \prod\limits_{v \in V_\two (G)}
\de^\# (p_{out,v}-p_{in,v} )\th_v (q_v )
\cr}\eqn $$
where $L(G)$ is the set of internal lines of $G$, $V_4(G)$ is the
set of four--legged vertices of $G$ and $V_\two (G)$ is the set of
two--legged vertices of $G$, 
$\th(q_v)$ is the function associated to
a two--legged vertex $v$, $\cU $ the vertex function associated to 
a four--legged vertex $v$, and the momenta $q_v$ are given in terms
of external and loop momenta by the momentum conservation at every 
vertex. The spin indices on a line $l$ and a vertex $v$ are the same
if $l$ goes into $v$ or out of $v$, 
and the symbol ``sum over spins'' 
indicates that they are summed over.
\Lem{\Powco}  {\Lesty 
Let $K_\zer $ be as in Lemma \Simpl. Then
$$
\abs{Val (G^J)}_\zer \leq  (4 K_\zer)^{\abs{L(G)}}
\prod\limits_{v\in V_\two (G)} \abs{\th_v}_\zer
\prod\limits_{v\in V_4 (G)} \abs{\cU_v}_\zer
\; M^{D_\phi j_\phi}\;
\prod\limits_{f>\phi } M^{(j_f-j_{\pi (f)})D_f}
\EQN\herkoemml $$
and
$$
\abs{Val (G^J)}' \leq (4 K_\zer)^{\abs{L(G)}}
\prod\limits_{v\in V_\two (G)} 
\left(\abs{\th_v}_\zer M^{-j_{\pi (v)}} \right)
\prod\limits_{v\in V_4 (G)} \abs{\cU_v}_\zer
\; S_{int} S_{f,ext} S_{v,ext}
\eqn $$
where $\abs{\; \cdot \; }'$ is defined in \queq{\priNormdef},
$$
S_{int} = \prod\limits_{f>\phi, internal } M^{D_f
(j_f - j_{\pi (f)})}
\eqn $$
where the product is only over those forks of $t(G^J)$ such
that $G_f^J$ does not contain any external vertices, and
$$
S_{f,ext} = \prod\limits_{f>\phi, external }
M^{\Delta_f (j_f - j_{\pi (f)})}
\eqn $$
where the product is over those forks $f$ of $t(G^J)$ such that
$G_f^J$ contains an external vertex of $G$, and
$$
S_{v,ext} = \prod\limits_{v, external}
M^{\Delta_v (0-j_{\pi (v)})}
\eqn $$
where the product is over those vertices of $G$ to which
an external leg is joined.
$\pi (v) $ is the highest fork such that $G_f $ contains $v$ and $\pi(f)$
is the predecessor fork of $f$, that is, the fork of $t(G^J)$ immediately below 
$f$.

}

\Proof See \quref{FT2}. An improvement of \queq{\herkoemml} will be shown in 
Section 2.6. \endproof

\Rem{\dummy} By definition, 
$$
D_f =
\left(2 |V_4 (G_f)| + |V_\two (G_f)| - {E_f\over 2}\right) 
- 2 (|V_4 (G_f)| + |V_\two (G_f)| -1) =
{1\over 2} (4 - E_f ) - |V_\two (G_f)|
.\eqn $$
If the graph $G$ has no two--legged 
vertices, and if no {\it internal} subgraph $G_f^J$ 
(i.e. $G_f^J$ contains no external vertices)
has $E_f=2$, then
$\Delta_v \leq -1/2$ and $\Delta_f \leq - 1/2$, and
$$
D_f = {1\over 2} (4 - E_f ) \leq 0
\eqn $$
for all internal forks, 
so the scale sum $\sum\limits_J\abs{Val (G^J)}'$
where $J$ runs over all labellings of $G$ compatible with a
fixed tree $t$ \quref{FT1,FT2} will be finite.
This is the rigorous counterpart of
the remark in the Introduction that only insertions of
two--legged diagrams give rise to divergences.
Renormalization will be done by subtracting
the value of the two--legged subgraphs on the Fermi surface.
For this we need to introduce a projection onto the Fermi
surface.

\sect{Localization Operator}
\noindent
The localization operator implements the projection
onto the Fermi surface for functions defined on $\R \times \cB$, and
it is used to define the subtractions needed for renormalization.
This projection can be defined in various ways, and so the
localization operator is not uniquely determined. 
In the spherically symmetric case, 
there is exactly one choice that is rotationally invariant.
Moreover, it does not matter which projection is chosen 
because rotational invariance implies 
that the value of any two--legged diagram $T(p_\zer,\ve{p})$ 
depends only on $p_\zer $ and $\abs{\ve{p}}$. 
In the case without spherical symmetry, there is no such independence 
and hence no canonical choice of the projection, 
although the geometrically most natural one seems to be  
that which projects along integral curves of $\nabla e$.
We project $\ve{p} $ onto $S$ differently, by moving it along the integral curve 
of the fixed vector field $u$ transversal to $S$ (see Lemma \uProps).
This yields bounds in better norms than using $\nabla e$ 
(see Remark \grade) because 
$\nabla e \in C^{k-1}$, but $u$ may be chosen in $C^\infty$. 

\Def{\LocDef}Let $\de $ be as in Lemma \uProps\ and let 
$\ch \in C^\infty  (\cB ,\lbrack 0,1\rbrack )$
obey $\ch (x) = 1 $ for $x \in U_{\de}(S)$ and
$\ch (x) = 0 $ for $x \not\in U_{2\de }(S)$.
Let $\prP $ be as in Lemma \uProps\ $(iii)$.
For functions $T: \R \times \cB \to X$, X any linear space, define
$$
\left(\ell T \right) (q_\zer,\ve{q} ) =
\cases{ T(0,\prP (\ve{q} ))\ch (\ve{q}) &
                      if $\ve{q} \in U_{2\de} (S)$ \cr
        0 & otherwise
.\cr}\eqn$$

\someroom
\noindent
If $T \in C^p ( \R\times \cB, X)$, then
$\ell T \in C^q (\cB ,X)$, where $q=\min \{ k,p\}$. 

\Lem{\Lieder} {\Lesty 
Let $\cD_u= u \cdot \nabla$ be the Lie derivative
with respect to $u$, 
and $T$ be differentiable on $\R \times U_\de (S)$
with a bounded derivative. 
In terms of the coordinates $(\rh , \om )$ introduced in 
Lemma \uProps, 
$$
(\ell T ) (q_\zer  , \ve{q} (\rh , \om )) = 
T(0, \ve{q} (0, \om ))
,\EQN\ellrho $$
$$
{\del \over \del \rh } T(q_\zer , \ve{q} (\rh , \om )) = 
\left({\cD_u T \over \cD_u e} \right)
(q_\zer , \ve{q} (\rh , \om ))
,\EQN\rhcDu $$
In particular ${\cD_u} \prP =0$  and 
$${\cD_u} \ell \mid_{U_{\de}(S)}= 0
.\eqn $$
For all $q=(q_\zer, \ve{q} ) \in \R \times U_\de (S)$, 
$$
\abs{(1-\ell ) T(q)} \leq 
{\sqrt{2} \over u_\zer} \abs{iq_\zer - e(\ve{q} )} \;
\max\{ \abs{\del_\zer T}_\zer , \abs{\nabla T}_\zer \}
.\EQN\RTaylor $$

}

\Proof By the chain rule
$$\eqalign{
{\del \over \del \rh } T(q_\zer , \ve{q} (\rh , \om )) 
&=\nabla T(q_\zer,\ve{q}(\rh,\om ))\cdot{\del \ve{q}\over \del \rh }(\rh,\om)\cr
&=\nabla T(q_\zer,\ve{q}(\rh,\om ))\cdot{\del\over \del \rh }\ga_{\tau(\rh)}(\om)\cr
&=\nabla T(q_\zer,\ve{q}(\rh,\om ))\cdot u(\ve{q}(\rh,\om )){\del\tau\over \del \rh }(\rh)\cr
}$$
So \queq{\rhcDu} follows from \queq{\drhdta}.
$\cD_u \prP =0$ then follows immediately from 
 \queq{\rhcDu} with $T(q_\zer,\ve{q})=P(\ve{q})$. 
In other words, because the projection $P(\ve{q}(\rh,\om))=\om$ is constant
along the integral curves of $u$, and the Lie derivative $\cD_u$ is
a directional derivative tangent to these integral curves, we have
$\cD_u \prP =0$.
If $\ve{q} \in  U_{\de}(S) $, $\ch ( \ve{q}) =1$, so
$$
({\cD_u} \ell T ) (q_\zer, \ve{q} ) = 
(u \cdot \nabla ) T (0, \prP (\ve{q} )) =
({\cD_u} \prP \cdot \nabla T) (0,P(\ve{q}) = 0 
\eqn $$
and 
$$
\abs{(1-\ell )T(q)} = \abs{T(q_\zer , \ve{q} ) - T(0, \prP (\ve{q} ))}
\leq \abs{q_\zer} \, \abs{\del_\zer T}_\zer + 
\abs{\ve{q} -\prP (\ve{q})} \abs{\nabla T}_\zer
\eqn $$
so \queq{\RTaylor} holds by Lemma \uProps\ $(iii)$.
\endproof

\noindent
To put the localization operation into contact with the flow of
effective actions, we define its action on a linear subspace
of the Grassmann algebra given by ``connected'' polynomials 
of even degree.
To define this subspace, we introduce some notation. 
The fermions in our model carry an index $\xi=(\al , p_\zer ,\ve{p})$,
where $\alpha \in \{ \uparrow ,\downarrow \} $, 
$p_\zer  \in \R $, and $\ve{p} \in \cB $
for infinite volume and zero temperature. For temperature $T>0$,
$p_\zer \in ( 2\Z +1 ) \pi T$. For a periodic box of side $L$,  
$ \ve{p} \in \cB \cap {2 \pi \over L} \Z ^d$.
For $\xi = (\al , p_\zer , \ve{p} )$ we denote
$\ps  (\xi ) : = \ps  _ \alpha (p_\zer , \ve{p})$ and
similarly for $\psq  $. We also write
$X = \{ \uparrow , \downarrow \} \times \R \times \cB $
and $\int _{X} ds (\xi ) F (\xi ) : = \sum\limits_{\alpha \in
\{ \uparrow , \downarrow \} } \int _{ \R } d p_\zer  \int _{ \cB } d^d
{ \ve{p} }  F( \alpha ,p_{o} , \ve{p} )$ (and their obvious variations 
for $T>0$ or finite volume). 
\Def{\Pain} We say that $Q \in \cQ^{k}_c$ iff
$Q=(Q_{2m,r})_{m \ge 0 , r \ge 1 }$, where for all $r \ge 1$ 
and $m \ge 0$,
\leftit{$(i)$} 
$Q_{2m,r} : X^{2m-1} \times \{ \uparrow , \downarrow \} \rightarrow \C$,
$( \xi _{1} , \ldots , \xi _{2m-1} , \alpha _ {2m} ) \mapsto Q _ {2m,r}
( \xi _ {1} , \ldots , \xi _ {2m-1} , \alpha _ {2m} )$ is $C^k$ and all
derivatives up to order k are bounded uniformly on
$X^{2m-1} \times \{ \uparrow ,\downarrow \} $,
\leftit{$(ii)$} for all $r \ge 1$ there is $\bar{m} (r) \ge 0$ such that
for all $m \ge \bar {m} (r)$ , $Q_{2m,r} = 0$.
\leftit{$(iii)$} $Q_{2m,r}$ is antisymmetric under permutations of momenta and spins, i.e.
$\cA Q_{2m,r} = Q_{2m,r}$ , where $\cA$ is the following operation. Define
$p_{2m} = ( (p_ {2m}) _{\zer} , \ve{p} _ {2m} ) \in \R \times \cB $ by 
$$
p_{2m} = \sum_{i=1}^{m-1}
 (p_{i} - p_{m+i} ) + p_{m} 
\eqn $$ and define
$\xi _{2m} = (\alpha _{2m} , p_{2m} )$. Then
$$\eqalign{
\cA Q (\xi _{1}, \ldots , &\xi _{2m-1}, \alpha _{m} ) = {1 \over (m!)^2}
\sum_{\pi ,\sigma \in Perm(m)} sign ( \pi \sigma ) \, \cr
&Q ( \xi _ {\sigma (1)} , \ldots  , \xi _ {\sigma (m) } ,
\xi_{m+\pi (1)}, \ldots  , \xi _{m + \pi (m-1)} , \alpha _ {m+\pi (m)} )
.\cr}\EQN\Antisy$$
\leftit{$(iv)$} The polynomial in the Grassmann algebra associated to
$Q \in \cQ ^{k}_c$ is the formal power series in $\la$
$$\eqalign{
Q(\ps, \psq )  = \sum _{r=1}^{\infty} \la^{r} & \sum _{m=0}^{\bar m (r) - 1}
\int\limits_{X^{2m}} \prod\limits_{i=1}^{2m} ds ( \xi _{i} )
\de ^{\#} \left( \sum _{i=1}^{m} (p_{i} - p_{i+m} ) \right) \cr
& Q_{2m,r} (\xi _{1} , \ldots , \xi _{2m-1} , \al _{m} )
\left( \prod\limits_{i=1}^{m} \psq (\xi _{m+i}) \ps (\xi _{i}) \right)
.\cr}\EQN\Sugg $$
Every fixed order in $\la $ is a polynomial in
the Grassmann variables. For convenience of notation, 
we sometimes write
$Q_{\al _{1} \ldots \al _{m}, \al_{m+1}, \ldots , \al_{2m} }^{(r)}
(p_{1}, \ldots , p_{2m-1})$ for 
$Q_{2m,r} (\xi _{1} , \ldots ,\xi_{2m-1} ,
\al _{m} )$. In this notation, the quadratic $(m=1)$ term in 
$Q(\ps ,\psq )$ is given by the formal power series
$$
\sum _{r=1}^{\infty} \la ^r \sum _{\al_{1} , \al_{2}} \int d^{d+1}p\
\psq _{\al _{1}} (p) Q_{\al _{1} \al _{2}}^{(r)} (p) 
\ps _{\al _{2}} (p).
$$

\Def{\LocDef} The localization operator
$\ell : \cQ_{c}^{k} \rightarrow \cQ_{c}^{k}$ is defined as follows.
For $Q \in \cQ _{c}^{k}$ and all $r \ge 1$
$$\eqalign{
(\ell Q)_{2m,r} &= 0 \hbox{ if }m \ge 2 \cr
(\ell Q)_{2,r} ((\al _{1} , p_{1}), \al _{2} ) &= Q_{2,r} ((\al _{1},
(0, \prP (\ve{ p} _{1})), \al_{2}) \cr
(\ell Q)_{\zer ,r} &= Q _{\zer ,r}
.\cr}\EQN\Locdef $$
In other words, for $Q(\ps ,\psq )$ given by \queq{\Sugg},
$$
(\ell Q) (\ps ,\psq ) = \sum _{r=1}^{\infty} \la ^r \left(
Q_{\zer ,r} +
\int d^{d+1} p\ \psq _{\al _{1}} (p) Q_{\al _{1}\al {2}}^{(r)}
(0, \prP (\ve{p})) \ps _{\al _{2}} (p)\right)
$$

\goodbreak

\sect{Flow of Effective Actions}
\noindent
We review briefly the definition of effective actions and
their flow, as given, e.g. in [FT2]. We introduce a cutoff 
that regulates the fermion propagator by restricting its support
away from the Fermi surface, so that the formally divergent 
integrals discussed above are convergent as long as the 
cutoff is present. This can also be done in finite volume and the 
infrared cutoff can be removed before taking the volume to infinity. 
The flow is used to study the dependence of the Green functions
on the cutoff as the latter varies. The propagator is decomposed
linearly into a sum of slice propagators that are supported
in thin shells around the Fermi surface. Because the decomposition
is linear, the flow has a semigroup structure
that allows one to view the Green functions as 
effective interactions where the fields with momenta 
that are away from the Fermi surface by an amount given by 
the cutoff are integrated out. 
Let $I\in \Z $, $I < 0 $,
be the infrared cutoff and decompose the cutoff propagator
$$
C= \sum\limits_{-1\geq j\geq I} C_j
\eqn $$
Define $\cG^\cV_I$ by
$$
e^{\cG^\cV_I (\chi, \chq)} =
{1 \over Z_I} \int d\mu_C (\ps,\psq)
e^{\cV (\ps+\chi , \psq+\chq )}
\eqn $$
where $d\mu _C$ denotes the Gaussian ``measure'', i.e.
the linear functional on the Grassmann
algebra generated by the $\ps $ and $\psq $ defined to vanish for
odd monomials and determined by its values for even monomials,
which are
$$
\int d\mu_C (\ps , \psq )
\prod\limits_{i=1}^n \ps_{\al_i} (x_i ) \psq_{\be_i} (y_i) =
\det \left( C_{\al_i \be_j}(x_i,y_j)\right)_{1\leq i,j \leq n}
.\eqn $$
In our case $C_{\al\be}(x,y)= \de_{\al\be} \check C (x-y)$, so, 
using the Fourier expansion \queq{\Psourier}, we get in momentum 
space (in the sense of distributions)
$$
\int d\mu_C (\ps , \psq ) 
\prod\limits_{i=1}^n \ps_{\al_i} (p_i) \psq _{\alq_i} (\bar p_i)
= \det \left( \de_{\al_i\alq_j} \de (p_i - \bar p_j)
C(p_i) \right)_{ij}
\EQN\Gauss $$
where 
$$
C(p) = \sum\limits_{-1 \geq j \geq I } C_j ( p_\zer , e(\ve{p}))
.\eqn $$
$\cG^\cV_I$ is the generating functional for connected, amputated
Green functions with infrared cutoff $I$ and vertices given
by $\cV $, because, formally, a shift in the integration variables,
$$
\cG^\cV_I (\chi, \chq)= -(\chq , C^{-1} \chi ) +
\log {1 \over Z_I} \int d\mu_C (\ps,\psq)
e^{-(\psq , C^{-1} \chi) - (\chq , C^{-1} \ps ) + \cV (\ps , \psq )}
,\eqn $$
indicates that $C^{-1}\ch$ and $\chq C^{-1}$ appear
as source terms. The effect of the $C^{-1}$ is that  
propagators associated to external lines are removed. This is, 
by definition, the procedure to get Green functions that are 
amputated by the free propagator.
 
The unrenormalized expansion has $\cV $ being the
bare interaction. For the renormalized expansion we will
allow $\cV$ to depend on $I$ because the
counterterms will be $I$-dependent. The factor
$$
Z_I = \int d\mu_C (\ps,\psq)
e^{\cV (\ps , \psq )}
\eqn $$
ensures that $\cG^\cV_I (0,0) =0$. We now define precisely the fluctuation integrals used for the flow. 
\Def{\Fluctint} 
\leftit{$(i)$} Let $\cU \in \cQ_c^k$ and the covariance $C$ 
be a bounded integrable $C^k$ function on $\R \times \cB $. Define
$$
\cR (C,\cU ) (\chi, \chq ) = \log {1 \over \ze (C , \cU )}
\int d\mu_C(\ps, \psq ) e^{\cU ( \ps+\chi, \psq+\chq )}
\eqn $$
where
$\ze (C, \cU ) = \int d\mu_C(\ps, \psq ) e^{\cU ( \ps , \psq )}$
so that $\cR (C,\cU )(0,0)=0$. Also, define
$$
\cE (C , \cU ) (\chi,\chq ) = \cR (C,\cU ) (\chi, \chq )   - 
\left( \cU (\chi, \chq ) - \cU (0,0) \right)
\eqn $$
\leftit{$(ii)$} 
Let $G$ be a connected graph with $n$ vertices $v_\one, \ldots v_n$
and $2 m$ external legs such that every vertex $v_i$ has $m_i$ 
ingoing and $m_i$ outgoing legs 
(incidence number $2m_i$), and let 
$$\eqalign{ 
\cU_{v_i}: X^{2m_i -1} \times \{\uparrow ,\downarrow \}&\to \C 
\cr
(\xi_\one, \ldots , \xi_{2m_i -1}, \al_{2m_i} ) & \mapsto
\cU_{v_i} (\xi_\one, \ldots , \xi_{2m_i -1}, \al_{2m_i} )
\cr}\eqn $$
satisfy $(i)$ and $(iii)$ of Definition \Pain. 
Let $J: L(G) \to  \{ z \in \Z : z < 0 \} $ be a labelling of $G$. 
The value of $G^J$ is defined as the function 
$Val(G^J)(C, \cU_{v_\one}, \ldots , \cU_{v_n}):
X^{2m -1} \times \{ \uparrow , \downarrow \}  \to \C$, 
determined by  
$$\eqalign{
\de^\# \left( \sum\limits_{i=1}^m (q_i - q_{m+i})\right) 
& Val(G^J)(C, \cU_{v_\one}, \ldots , \cU_{v_n})
(\et_\one , \ldots , \et_{2m-1}, \be_{2m} )  = \cr
& = \sum\limits_{spins}\int \prod\limits_{l \in L(G)} 
\left( C_{j_l}((p_l)_\zer ,e(\ve{p}_l))_{\al_l, \bar\al_l} 
 d^{d+1} p_l \right) \cr
& \prod\limits_{i=1}^n 
\de^\# \left( \sum\limits_{k=1}^{m_i} (p^{(i)}_k - p^{(i)}_{m+k})\right) \cr 
& \cU_{v_i} \left( 
(p^{(i)}_\one , \al^{(i)}_\one ) , \ldots , 
(p^{(i)}_{2m_i-1} , \al^{(i)}_{2 m_i -1}), \al_{2m_i}^{(i)} \right)
,\cr}\EQN\ValU $$
where $\et_i = ( q_i ,\be_i )$, $\sum_{spins}$ means that 
all $\al_k^{(i)}$ are summed over $\{ \uparrow, \downarrow \}$. 
If the line $l$ joins the outgoing leg $k$ of vertex $v_i$ to the 
incoming leg $k^\prime $ of $v_{i^\prime}$, then 
$\al_l=\al_k^{(i)}$, $\alq_l=\al_{k^\prime}^{(i^\prime )}$, 
and the momenta $p_k^{(i)} = p_{k^\prime}^{(i^\prime)} = p_l$.
\leftit{$(iii)$} 
The set of all connected graphs with $2m$ external legs and
with $n$ vertices $v_\one , \ldots , v_n$ where $v_i$ has 
$m_i$ incoming and $m_i$ outgoing legs (incidence number
$2 m_i$)  is denoted by $Gr (n,m; m_\one, \ldots , m_n )$.
When $n=1$, the graphs are required to have at least one internal 
particle line.
\Lem{\Well} {\Lesty 
\leftit{$(i)$} 
$Val(G^J)(C, \cU_{v_\one}, \ldots , \cU_{v_n})$
is well--defined and a $C^k$ function of the external momenta.
\leftit{$(ii)$} 
$\cE $ and $\cR$ are well--defined formal power series
in $\la$. They map $\cQ_c^k $ to $\cQ_c^k$.
\leftit{$(iii)$}  Expanding in powers of $\cU$, 
$$
\cE (C_j, \cU ) = \sum _{n=1}^{\infty } {1 \over {n !}} 
\cE ^{(n)} (C_j, (\cU , \cU , \ldots , \cU ))
\eqn $$ 
($\cU$ appears $n$ times), 
and expanding the $\cE^{(n)}$ in $\la $ and the fields as well,
$\cE$ has the expansion (see also Definition \Pain, $(iv)$)
$$\eqalign{
\cE (C_j, \cU ) (\ch , \chq ) &=
\sum\limits_{r=1}^\infty \la^r 
\sum\limits_{m=1}^{\bar m(r)}
\int\prod\limits_{i=1}^{2m} ds(\eta_i)\cr
& \de^\# \left( \sum\limits_{i=1}^m (q_i - q_{m+i} ) \right)
\prod\limits_{k=1}^m \left( \chq (\et_k) \ch (\et_{k+m} ) \right)
\cr
& E_{j,2m,r} (\cU ) (\et_\one , \ldots , \et_{2m-1}, \be_{2m} )
\cr}\EQN\Ejmr $$
where  $\et_k = (q_k, \be_k )$, and
the kernels $E_{j,2m,r}$ are the following sum of 
values of Feynman diagrams:
$$ 
E_{j,2m,r}(\cU ) = \sum_{n=1}^\infty {1\over n!} 
E^{(n)}_{j,2m,r} (\cU , \ldots , \cU )
\EQN\Ejmrn $$
with
$$\eqalign{
E^{(n)}_{j,2m,r} (\cU_{v_\one} ,\ldots , & \cU_{v_n} )
(\et_\one , \ldots ,  \be_{2m} ) =
\sum_{r_\one, \ldots , r_n \geq 1 \atop r_\one + \ldots +r_n =r}
\sum\limits_{m_\one , \ldots , m_n \geq 1} 
\sum\limits_{G \in Gr (n,m; m_\one, \ldots m_n )}
\cr 
sign(G) \; \cA \; & 
Val G^{(j)}(C, \cU_{v_\one,r_\one}, \ldots , \cU_{v_n,r_n})
(\et_\one , \ldots , \be_{2m} )
\cr}\EQN\Loong $$ 
Here $\cA $ denotes 
the antisymmetrization operator defined in \queq{\Antisy}, 
$sign(G)\in \{ 1, -1\}$ is a sign factor 
determined by the structure of $G$, 
$(j)$ denotes the labelling $j_l=j$ for all $l$,
and $ \cU_{v_k,r_k}$ denotes the coefficient of order $r_k$ 
in the formal expansion of $\cU_{v_k}$ in powers of $\la $.
The sum over the number of effective interaction vertices $n$ 
in \queq{\Ejmrn} is a finite sum with $n\le r$. The sums in  \queq{\Loong}
are finite because the interaction $\cU \in \cQ_c^k$ 
and in particular satisfies $(ii)$ of Definition \Pain. }

\Rem{\dummy}
Although lengthy, \queq{\Loong} is easy to interpret: At every 
scale, the Green function is expanded in a formal power series in 
$\la$. In every order in $\la $, the functional is expanded in 
powers of the external (unintegrated) fields $\ch $ and $\chq$. 
The term with $m$ factors of $\ch$ and $m$ factors of $\chq$ 
contributes to the $2m$--point function, and  is given by the
sum over all connected Feynman diagrams with $2m$ external legs, 
built from the effective vertices $\cU$. 
\Proof The momentum conservation delta functions at every vertex 
can provide a set of loop momenta using some choice of a 
spanning tree for $G$ in the standard way. Since $G$ is connected, 
only the global momentum conservation delta function remains. 
It is then obvious by the properties of the integrand to see 
that the function that multiplies this delta function is $C^k$ 
and so $(i)$ follows. Since $(ii)$ follows from $(iii)$ and $(i)$, 
it suffices to show $(iii)$. Although this
is quite standard, we briefly describe how the expansion in 
Feynman graphs comes about since it is also very simple.
By definition, $\cE^{(n)} $ takes out the terms proportional 
to $\cU_{v_\one} \ldots \cU_{v_n}$, so it is obviously linear
in every $\cU_{v_k}$. Inserting the expansion for every 
$\cU \in \cQ_c^k$, we obtain the sum over the $r_i$ and $m_i$. 
Note that by definition of $\cQ_c^k$, all sums over $r_i$ 
start at $1$ and therefore $\cE $ and $\cR$ 
contain no zeroth order terms in $\la $.
Furthermore, $m_i \leq \bar m_i (r_i) $ so all sums contain 
only finitely many terms.
The rules for Gaussian integration \queq{\Gauss} then join 
outgoing ($\psq $) legs of $\cU_v$ to ingoing ($\ps $) legs 
of $\cU_{v^\prime}$. The result can be translated into a sum over 
Feynman graphs by joining $v$ and $v^\prime $ by a line and using 
the definition of the value of a graph given above. 
Since the logarithm is taken, only connected graphs 
contribute (see, e.g. \quref{GJ}). Consequently
$$
\bar m(r)\le \max_{n,r_\one, \ldots , r_n \geq 1 \atop r_\one + \ldots +r_n =r}
\sum_{i=1}^n[\bar m_i(r_i)-1]
$$
\endproof
\Rem{\dummy} In fact, because of the fermionic nature of the fields,
defining suitable norms on $\cQ _{c}^{k}$,
analyticity holds in a disk $\{ \vert \la \vert < \la _{\zer} \}$,
where $\la _{\zer }$ depends on the cutoff $I$.
Since we consider the formal expansion only, we do not
need to make use of that here.
\sni
The flow is now obtained by successively integrating out 
the momenta of shells around the Fermi surface. 
Since $C$ is a sum of covariances, 
the Gaussian measure factorizes into a product  $\prod_{j=I}^{-1}d\mu _{C_j}$, 
and $\cG^\cV_I$ can be written as the endpoint
of the sequence 
$$
\cG^\cV_j = \cR \left( \sum_{i=j}^{-1} C_j , \cV \right)
\EQN\endPoint $$ 
The sequence starts with $\cG^\cV_\zer = \cV $ and may be 
obtained by iteration of
$$
\cG^\cV_j  = \cR ( C_j, \cG^\cV_{j+1})  =
\cG^\cV_{j+1}  + \cE (C_j, \cG^\cV_{j+1}) 
\EQN\Recu $$
The recursion can be summed to get, assuming $\cV (0,0)=0$, 
$$
\cG^\cV_{j} (\chi, \chq ) = \cV (\chi, \chq )
+ \sum\limits_{-1 \geq i \geq j}
\cE (C_j, \cG^\cV_{j+1}) (\chi, \chq )
.\EQN\Inteflow $$
Lemma \Well\ implies
\Lem{\goodFlow} {\Lesty 
Let $e \in C^{k} (\cB , \R )$, $k \ge 1$, and assume
\ATwo. If the initial interaction $\cV \in \cQ_c^k$,
then for any scale $j$, $\cG_j^{\cV} \in \cQ _{c}^{k}$.}
\someroom\noindent
Taking the initial interaction to be the bare one,
$\cV = \lambda V$, yields the sequence of unrenormalized
effective actions which diverges as $I \to -\infty $ 
for the reasons discussed in the Introduction. 

The renormalized Green functions are constructed by modifying 
the interaction such that the Fermi surface of
the interacting system, that is, the singular surface of the
interacting fermion propagator, stays fixed. This requires a
specific choice of $\cV$ which we denote as $\cG_\zer^I$, the $I$
indicating the dependence on the infrared cutoff. Using the
similar notation $\cG_j^{\cG_\zer^I} = \cG_j^I$  for the $\cG_j$
obtained from this interaction by \queq{\endPoint}, 
we require, as a condition on $\cG_\zer^I$,
$$
\ell \cG_I^I = \ell \cG_\zer^I + \sum\limits_{I\leq i \leq -1}
\ell \cE ( C_i , \cG^I_{i+1} ) = 0
,\eqn $$
so
$$
\ell \cG_\zer^I = - \sum\limits_{I \leq i \leq -1 }
\ell \cE (C_i, \cG_{i+1}^I )
.\EQN\Gzeq $$
Since all $\cG_j^I$ are functionals of $\cG_\zer^I$, this is not
a definition but an equation to be solved by $\cG_\zer^I$. There 
are further conditions on $\cG_\zer^I$: We want the form of 
the interaction to be similar to the original one. 
Only terms bilinear in the fermion fields shall be 
generated:
$$
(1-\ell ) \cG_\zer^I = (1-\ell ) \lambda V = \lambda V
.\eqn $$
\queq{\Gzeq} can be solved order by order
in $\lambda $, that is, as a formal power series in $\la $,
$$
\cG_\zer^I = \sum\limits_{r=1}^\infty \cG_{\zer,r}^I \lambda ^r
,\eqn $$
as follows. All $\cG_j^I$ are formal power series in $\la$, 
with no ${\rm zero}^{\rm th}$ order term since they are connected Green functions
(and since the free part is subtracted from the two--point function). 
One proceeds inductively in $r$, the order in $\la $, in 
\queq{\Gzeq}. To get the left side in order $r$ 
only counterterms in $\cG^I_{i+1}$ up
to order $r-1$ are needed on the right side of the equation.
No graph contributing to the right hand side of \queq{\Gzeq} can
consist of a single two--legged vertex with no internal lines.
The left side of
\queq{\Gzeq} can simply be used to give a recursive definition 
for the counterterms.
\Def{\RenExp} The generating functional for the 
renormalized Green functions is obtained 
by the flow \queq{\Inteflow} with  initial interaction
$\cG_\zer^I = \la V + \cK ^I$, where $V$ is the interaction 
given by \queq{\Interact} and the counterterms $\cK^I$ are defined 
as a formal power series in $\la $ by 
$$
\cK^I (\ch ,\chq ) = -\ell \sum\limits_{I \leq i \leq -1}
\cE (C_i , \cG_{i+1}^I ) (\ch ,\chq ) 
,\eqn $$
and we shall call their formal power series expansion in terms of $\la $ 
the renormalized perturbation expansion.
The expansion coefficients of $\cG^I$ given by Definition \Pain\ $(iv)$ 
 are the renormalized, amputated, connected Green 
functions. More explicitly, the $r^{\rm th}$ order $2m$--point function
on scale $j \geq I$, $G_{j,2m,r}^I$, is obtained by replacing
the $\cE(C_j,\cU) $ by $\cG_j^I=\cG_j^{\cG_\zer^I}$ and the $E_{j,2m,r}$ by $G^I_{j,2m,r}$
in \queq{\Ejmr} and the functions
$G^I_{2m,r}$ from Section 1.5 are defined as 
$$
G^I_{2m,r}=G^I_{I,2m,r}
\eqn $$ 
\sni 
The counterterms are of the form
$$
\cK^I (\ch , \chq ) = \sum_\al\int\limits_{\R \times \cB}
\db^{d+1} p\ \chq_\al (p ) K^I (\ve{p} ) \ch_\al (p) 
\eqn $$
where $K^I$ is a formal power series in $\la $, 
$$
K^I (\ve{p} ) = \sum\limits_{r=1}^\infty \la^r K^I_r (\ve{p} )
.\eqn $$ 
The $G^I_{j,2m,r}$ are all of order $r\ge 1$ 
in the coupling $\la$. 
In particular, the two--point function has
the ${\rm zero}^{\rm th}$ order propagator subtracted. Hence the 
formula \queq{\SelfI} for the self--energy.
The recursion formula \queq{\Recu} 
can be written for the kernels $G^I$  as 
$$
G_{j,2m,r}^I - G^I_{j+1,2m,r} 
= E_{j,m,r} (\cG^I_{j+1})
\EQN\Kernflow $$
To show convergence of the renormalized Green functions
in the limit as the cutoff $I$ is removed, $I \to -\infty$, 
it is convenient to arrange \queq{\Inteflow} in the form 
$$
\cG_j^I = \la V  +
\sum\limits_{i\geq j} (1-\ell ) \cE ( C_i , \cG_{i+1}^I ) +
\sum\limits_{i< j} (-\ell ) \cE (C_i, \cG_{i+1}^I ) 
.\EQN\rcflow $$
Iteration of this equation for $\cG _{j}^{I}$ generates a tree
structure, corresponding to layers of $\cG _{j}$. Expanding this out
to scale zero, one recovers the scaled graph $G^{J}$ from Lemma \Powco.
For the unrenormalized expansion, the scales of lines in $G_{f'}$ are
strictly higher than those in $G_{f}$ if $f' > f$ on the tree.
In case of the renormalized expansion, this holds for r-forks,
generated by the second term in \queq{\rcflow}. The third term in 
\queq{\rcflow} gives rise to the c-forks of the tree. The
scales of a c-fork $f$ are summed
from $I$ to $j_{\pi (f)}$ since $i \leq j$ in the third term of 
\queq{\rcflow}.

The semigroup structure of the renormalization flow is obvious from
the way it is defined by fluctuation integrals. It is a consequence of
 the linear decomposition of the covariance $C$ into a sum of $C_{j}$'s.
It allows one to interpret the formula for $\cG _{j}$ in various ways.
By definition, $\cG_j$ is the amputated connected Green function 
with infrared cutoff $j$. Alternatively, one can also view 
$\cG _{j+1}$ as an effective action, i.e.\ the $G^I_{j+1,2m} $ are 
vertex functions of effective interaction vertices with $2m$ external 
legs. These vertices are connected by $C_{j}$--lines to form 
the effective action on scale $j$, $\cG_j$.
The process of expanding different parts of the tree, or equivalently, 
expanding the effective vertices in terms of higher--scale objects, 
can be done to various degrees.
One can choose to iterate selected parts of the tree, 
i.e. resolve selected vertices up to a certain higher scale 
or ``trim the tree'' at a fork $f$ by regarding the subgraph
$G_{f}^{J}$ as a vertex with $E(G_{f}^{J})$ external lines and vertex
factor $\de ^{\#} (p_{f, out} - p_{f, in}) Val (G_{f}^{J})$.
We shall make use of three variations on this theme, 
which we now briefly describe. 

\Rem{\Reso} $(i)$ 
Resolve every vertex up to scale zero, as described above; 
this gives sums over values of the standard labelled
graphs $G^{J}$ of Lemma \Powco.
More precisely, this leads to the following formula  
for the amputated connected Green functions:
$$
G^I_{2m,r} = \sum_{j\geq I}
\sum_{t } \; \prod_{f \in t} \; {1 \over n_f!}
\sum_{G} \;
\sum_{J \in \cJ (t,j)} \;
Val (G^J)
\EQN\GjIform $$
where (as follows from \queq{\Ejmr}, \queq{\Ejmrn},
see also Section VI of \quref{FT1})
the second sum is over all planar trees $t$ with 
$r$ leaves. The root is denoted $\ph$, 
and for each fork $f$, $n_f\geq 1$ is the number of upward branches. 
($n_f= 1$ is possible because we do not use normal ordering).
The factorial is that from \queq{\Ejmr}.
The sum over graphs $G$ runs over all $G$ 
compatible with $t$, that is, connected graphs with
$2m$ external legs, $r$ ordered vertices, constructed 
according to the Feynman rules of the model. 
The leaves of $t$ correspond to the four--legged 
interaction vertices of $G$.
For any fork $f \in t$, there is a connected subgraph 
$G_f$ of $G$, such that the quotient graph $\tG(\{ f\} )$
(obtained by replacing all $G_g$ with $\pi (g) =f$ by 
effective vertices) has $n_f$ vertices.   
The set $\cJ (t,j)$ of scale families $J$ consists of all 
$\left( j_f \right) _{f \in t}$ ordered 
according to the partial ordering given by the tree $t$, 
$$\eqalign{
\cJ (t,j) =  \{ (j_f)_{f \in t} : j_\ph = j,
&\hbox{ if } f\in t \hbox{ is not a c-fork, }
  j_f \in \{ j_{\pi (f)}+1 , \ldots , 0\} \cr
&\hbox{ if } f \hbox{ is a c-fork, } 
 j_f \in \{ I, \ldots , j_{\pi (f)} \}
\} 
\cr}\EQN\cJdef $$ 
This definition is understood recursively, i.e.
the root scale $j_\ph $ is fixed to $j$; if $f$ is a $c$--fork 
with $\pi (f) = \ph$, then $j_f$ runs from $I$ to $j_\ph$. 
If $\pi(f) = \ph $, but $f$ is not a $c$--fork, $j_f$ runs 
from $j_\ph +1 $ to $0$. This assignment of scales is now continued
upwards on the tree, determining the range of $j_f$ in terms of 
$j_{\pi (f)}$ and the r/c label on the fork.  
All leaves $b$ of the tree $t$ have scale zero, and the vertices 
of $G$ associated to them are the interaction vertices of the 
original action.
The labelling of the graph $G$, $\ell \mapsto j_\ell $, is:
all lines $\ell $ in $\tG (\{ f \} )$ get scale $j_\ell =j_f$. 
Finally, $Val (G^J)$ is defined according to the 
Feynman rules for labelled graphs, 
with a propagator of scale $j_l$ associated to each line $l$.
In our case, there is no hard/soft labelling for the lines
because we do not use normal ordering.  
\pni $(ii)$ 
Resolve everything except for one-particle irreducible two- and four-
legged insertions. More algorithmically, let $G$ be a graph contributing
to $\cG _{j}$. For every vertex $v$, $\cU _{v}$ is again the sum of 
values of graphs on scale $\ge j+1$. 
If $v$ has $\le$ 4 legs and is $1PI$ leave it. Otherwise
repeat the same procedure for the graph whose value is $\cU _{v}$.
Continue to resolve until all graphs that are not
resolved are 1PI (for details, see Section 2.7).
The result is a labelled graph $\Gp $ that has no nontrivial
one--particle--irreducible 
two-- or four--legged subdiagrams but instead two-- and 
four--legged vertices with scale--dependent vertex functions.
This will be used to trace back the factorials in values of
individual graphs (the reason for their occurrence are the 
\NOL\ four-legged subgraphs) and to order the inductive proofs, 
since the scale--dependent vertex functions are themselves 
values of subgraphs of lower order. 
The vertices are scale--dependent because the trimming 
procedure splits the summation over $\cJ$. Trimming a tree $t$ 
at a fork $\ps$ decomposes $t$ into two subtrees $t_\one$ and $t_\two$ with
$\ps $ is a leaf of $t_\one$
and $t_\two $ is rooted at $\ps $. Then 
$$\eqalign{
\cJ (t, j) = \{  (j_f)_{f \in t} : 
J_\one &= (j_f)_{f \in t_1} \in \cJ_\one = \cJ (t_\one , j ) \cr
\hbox{ and } & J_\two = (j_f)_{f \in t_2} \in 
\cJ_\two = \cJ (t_\two , j_\ps ) \}
\cr}\EQN\cJsplit $$
The vertex function $\cP_w V_w$ of the vertex $w$ in $\Gp $ 
that corresponds to $G_\ps^J$ is on scale $j_\ps$, 
and it is obtained by summing over the scales in $\cJ_\two$, 
keeping those in $\cJ_\one $ fixed, 
$$
V_w = \sum_{J_2 \in \cJ_2} Val (G_\ps^{J_2})
.\EQN\Vwdef $$
The projection $\cP_w \in \{ \ell, 1-\ell, 1\}$ is given by the 
r/c labelling of the forks of $t$.
\pni
$(iii)$ 
Resolve according to families of \NOL\ subtrees rooted at forks
belonging to two--legged diagrams; this will
play a major technical role in the estimates of the derivative
with respect to the band structure $e$. For details, see Section 2.5. 
\sni
For further reference, we give the formula for $K_r^I$ explicitly,
$$
K_r^I (\ve{p})= - \sum\limits_{G}
\sum\limits_{j=I}^{-1} \sum\limits_{t \sim G}
\prod\limits_{f \in t} {1 \over n_f !}
\sum\limits_{J \in \cJ (t,j)} Val (G^J) (0, \prP (\ve{p}))
.\EQN\KrIform $$
Note that $K_r^I$ actually only depends on $\prP (\ve{p} ) \in S$. The sum over 
$G$ is over all two--legged and one--particle irreducible (1PI) graphs
$G$ with $r$ interaction vertices. The graphs have to be 1PI since
 $e(\prP (\ve{p} ))=0$ implies $C_j ( 0 , 0) = 0$, 
and since the value of a 1P--reducible graph would contain such a factor.

\sect{Non-Overlapping Graphs}
In this section, we give an explicit characterization of two-- and
four--legged graphs that do not contain any \OL\ loops. These
graphs turn out to be  dressed bubbles in the four--legged case 
and graphs of the type encountered in the Hartree--Fock resummation
in the two--legged case. 

To make contact with the graph structure in our problem, and for convenience
of the reader, we show explicitly how certain low--order diagrams look
when the interaction lines are collapsed to four--fermion vertices,
in Figure 4. Graph numbers 1, 5 and 6 each contain two loops which do not
overlap. The last three graphs each contain two loops that do overlap.
\midinsert
\null\hfil\figplace{fig4}{0in}{0in}\hfil
\centerline{\it Figure 4}
\endinsert

We wish  to single out those graphs which have \OL\ loops. Their value
contains a volume integral that can be bounded by the function $I_2$,
defined in \queq{\Itwodef}, which gives an additional convergence factor
in scale sums.
This will serve to show that derivatives converge and that
a large class of 4-forks is actually not marginal, that is, that the
power counting behaviour $D_f = 0$ is not saturated.
For the graphs without \OL\ loops, there is no such improvement. 
But these graphs have a rather special structure (see Figure 4). 
In particular, the momentum of the external line will not enter in 
any of the loop lines if the graph is two--legged and \NOL.
The two Lemmas in this Section characterize graphs $G$ that have no \OL\ loops
explicitly for $E(G)=2$ or $4$. They are stated for the more general class
of graphs that arise naturally when expanding the fluctuation integral
for the effective action at some scale (see \queq{\ValU} and below it).

In the following, let $G$ be a connected graph constructed from particle lines
and generalized vertices $v$ that all
have an even incidence number $E_v \ge 2$. Such graphs occur naturally in
the flow of effective actions.  
We call $G$ one--particle irreducible (1PI)
if any internal particle line of
$G$ can be cut without disconnecting the graph.
We also use $L(G)$ for the set of all internal lines of $G$,
$E(G)$ for the set of external lines of $G$, 
 $V(G)$ for the set of all vertices, and, $V_k(G)$ for
the set of all vertices with incidence number $k$. For $v\in V(G)$,
$G-v$ denotes the graph in which $v$ and all lines going into $v$ are
deleted. For $l\in L(G)$, $G-l$ denotes the graph in which
only the line $l$ is removed (but not its endpoints).
Denote the set of directed lines of $G$ by $\cL (G)$, 
$$
\cL (G) = \{ (\ell,v,w) \in L(G)\times V(G)\times V(G) : 
 \ell {\rm\ connects\ }v {\rm\ and\ } w \}
.$$
\Def{\PathDef}
\leftit{$(i)$} 
Let $n_\one, n_\two \in \N_\zer$, $n_\one \leq n_\two$.
A path $P$ in $G$ is a map
$P: \{ n_\one, \ldots , n_\two \} \to \cL(G)$, 
$n \mapsto (v_n, w_n)$, such that 
for all $n \in   \{ n_\one, \ldots , n_\two -1\}$, 
$w_n = v_{n+1}$, and such that 
each vertex of $G$ is visited at most 
once by $P$.
\leftit{$(ii)$}
A loop in $G$ is a map $P: \natz{s} \to \cL (G)$
such that $P\mid_{\natz{s-1}}$ is a path, 
and (in the notation of $(i)$), $w_s=v_\zer$, and
the line from $v_{s} $ to $w_s$ is a line of $G$
(the case $s=0$ is a line from a vertex to itself, 
also called `self--contraction').
\leftit{$(iii)$}
The trace $\th(P)$ of the path or loop $P$ is defined as the
subgraph consisting of lines and vertices visited by $P$.
\leftit{$(iv)$} We say that two loops $P_\one $ and $P_\two $ are
independent if
their traces are distinct, $\th(P_\one ) \neq \th (P_\two )$.

\vglue 0.5 true cm
\someroom\noindent
For example, under Definition \PathDef, 
\hbox to 0pt{\hss\includegraphics{nopath.ps}\hss}
\hskip 40pt plus 5pt 
is not a path because it is self--intersecting.
However, 
\hbox to 0pt{\hglue 3pt
\vbox to 0pt{\vglue 10 pt\includegraphics{loop.ps}\vss}
\hss}
\hskip 30pt plus 5pt is a loop consisting of one line 
(a `self--contraction'). 
 
\someroom
\Rem{\Paths} \leftit{$(i)$} We sometimes write the path as a finite sequence
$(P(n_\one ) , \ldots , P(n_\two ))$.
\leftit{$(ii)$} If $P$ is a path, so is its inversion $P^{-1}$,
defined as going over the same lines as $P$, 
but in opposite direction. If $P$ is a loop, so is
its shift by $m$, $P_m$, defined as $P_m(l) = P(l-m $ mod $s)$.
\leftit{$(iii)$} Usually a loop is defined to be an element of the
first homology group $H_\one (G, \Z )$. For the purposes of the following
analysis of \OL\ and \NOL\ graphs, it does not really matter which of the
definitions one takes. 
\leftit{$(iv)$} Let $T$ be a spanning tree for a graph $G$. Let $\ell \in L(G)
\setminus L(T)$. Then the subgraph of $G$ gotten by taking the union of $\ell$
with the linear subtree of $T$ that joins the endpoints of $\ell$ is the trace
of a loop under Definition \PathDef.
\Def {\ItwoDef} $G$ is called \OL\ if there
is a line of $G$ which is part of two independent loops.
\Rem{\BasGra}
\leftit{$(i)$} If $G$ is \NOL\ and $S$ a connected subgraph, then
$S$ is \NOL.
\leftit{$(ii)$} If $G$ is \NOL, $\tG $ a quotient of $G$
obtained by replacing  a connected subdiagram with a vertex,
then  $\tG $ is \NOL.
\leftit{$(iii)$} If $G$ is connected and  $S$ a subgraph that is \OL,
then  $G$ is \OL.
\leftit{$(iv)$} If $G$ is a \NOL\ graph, and $\tG $ is obtained from $G$
by forming a self--contraction of two external legs of a vertex $v$ of $G$, 
then $\tG $ is \NOL.
\leftit{$(v)$} If $G$ contains a subgraph consisting of two vertices
$v_\one $ and $v_\two$, joined
by $n\geq 3$ lines $l_\one , \ldots , l_n$, then $G$ is \OL.
\Proof $(i)$, $(iii)$ and $(iv)$ are obvious.
$(ii)$ Let $\tG $ a quotient of $G$ obtained by
replacing a connected subdiagram $H$ by a vertex. Let $\tG$ be \OL.
Then there are two independent loops $L_\one $ and $L_\two $ in $\tG$ 
that have a
line $\ell \in \tG $ in common. As a path in $G$, $L_\one$ either crosses
at most one external vertex of $H$, in which case $L_\one $ is still a loop in
$G$, or it stops at two distinct external vertices of $H$. Since $H$
is connected, there is a path connecting these vertices, and the composition
of $L_\one $ with this path is a new loop in $G$ that still contains $\ell$.
Similarly, $L_\two $ either is already a loop in $G$ or can be completed to
one, and so $G$ is also overlapping. This shows $(ii)$.
$(v)$ Let the vertices be $v_\one $ and $v_\two$.
Since $n \geq 3$, the loop $L_\one $ going from $v_\one $ to $v_\two $
over $l_\one $ and back over $l_\two $ and the path $L_\two $ going
from $v_\one $ to $v_\two $ over $l_\one $ and back over $l_3$ are
independent. Both contain $l_\one$. So the subgraph is \OL, and the same
follows for $G$ itself by $(iii)$. \endproof

\midinsert
\null\hfil\figplace{fig5}{0 truecm}{0in}\hfil
\centerline{\it Figure 5}
\endinsert

%
% two-legged case
%
\Def{\GSTDef} Let $G$ be a connected graph with two external legs and
$N$ vertices all having even incidence number.
\leftit{$(i)$} If $G_\one , \ldots G_n$ are two--legged graphs, the string
$G_\one \ldots G_n$ is the graph shown in Figure 5 (a). The $G_i$ may
be two--legged vertices (i.e.\ vertices with incidence number two).
\leftit{$(ii)$} $G$ is called a self-contracted two-legged (ST) diagram if $G$
consists only of one two--legged vertex with two external legs or
if $G$ has exactly one vertex $v_\one $ to which both external legs
of $G$ connect, all other vertices have two legs and the remaining
legs of $v_\one$ are joined pairwise by strings of two--legged vertices
to form loops. See Figure 5 (b).
\leftit{$(iii)$} A generalized ST diagram (GST) with $N$ vertices is defined
recursively: if $N=1$, $G$ is an ST diagram.
If $N\ge 2$ and GST are defined for all $N^\prime\le N-1$,
a GST with $N$ vertices is a graph such that $G$ has exactly one
{\it external} vertex $v_\one $ to which the two external legs of $G$ join,
and all other legs of $v_\one $ are joined by strings of GST with
at most $N-1$ vertices, to form loops (we call that 
`generalized self--contractions'). For an example, see Figure
5 (c) and (d); in (c) the GST insertions are marked by crosses.

%
%  Lemma for two--legged nonoverlapping graphs
%

\Lem{\TwoL} {\Lesty 
Let $G$ be a connected graph with two external legs and
all vertices of $G$ having an even incidence number.
If $G$ is \NOL, it is a string of GST graphs.
If $G$ is \NOL\ and 1PI, then $G$ is a GST graph.}
\Proof The second statement is obvious, given the first. 
To prove the first, do induction in the number of vertices $N$.
For $N=1$, an $E=2$ graph with one vertex must obviously be an ST diagram.
It is instructive to look at $N=2$ first. There are two cases:
(1) only $v_1$ has incident external legs, and 
(2) $v_1$ and $v_2$ both join to an external leg.
%
%   here comes Figure 6
%
\midinsert
\centerline{\figplace{fig6}{-1truecm}{0in}}
\centerline{\it Figure 6}
\endinsert

\smallskip
\noindent
(1) Denote the incidence number of $v_1$ by $n_1$ and that of $v_2$ by $n_2$.
Since two legs of $v_1$ are external and every self-contraction binds
two legs, there must be an even number $n$ of lines between $v_1$
and $v_2$ (see Figure 6). If $n\ge 4$, $G$ is \OL\ by Remark \BasGra\ $(v)$
(there are $n-1\ge 3$ independent loops containing any of the lines 
between $v_1$ and $v_2$). So $n=2$, which means that the graph is a GST.
\pni
(2) $n_1$ and $n_2$ are even, and $v_1$ and $v_2$ each bind one external leg
of $G$.
Since self-contractions bind an even number of lines, $v_1$ and $v_2$
must be joined by an odd number $n$ of lines. If $n\ge 3$, 
$G$ would be \OL\ by Remark \BasGra\ $(v)$. 
So $n=1$, and $G$ is a string of two ST graphs.

Let $N \geq 2$ and assume the Lemma to be true for \NOL\ graphs with
$N^\prime$ vertices, $N^\prime\le N-1$, and let $G$ be a \NOL\ graph
with $N$ vertices and $E=2$. Call the vertices where the external legs
join external vertices.
%
%   here comes Figure 7
%
\hfil\break
\null\hfil\figplace{fig7}{0 true in}{0in}\hfil

\noindent
\pni
(1) If there is only one external vertex, $v_\one$, $G$ takes the form
shown in Figure 7 (a).
Decomposing the subgraph $B=G-v_\one $ into its connected components
$C_1,\ldots,C_\ell$ we see that $G$ must be as drawn in Figure 7 (b).
Denote the number of lines joining $v_1$ and $C_k$ by $n_k$.
Let $k\in\{1,\ldots,\ell\}$.
Since all vertices in $C_k$ have an even incidence number and 
legs are joined pairwise to form
internal lines of $C_k$, the number $n_k$ of external legs of $C_k$ 
is even. As in the case
$N=2$, if $n_k\ge 4$, the subgraph consisting of $v_\one $ and $C_k$
shown in Figure 7 (c) is \OL\ by Remark \BasGra\ $(v)$ and $(ii)$.
By Remark \BasGra\ $(iii)$, so is $G$.
Therefore $n_k=2$ for all $k\in\{1,\ldots,\ell\}$
and, by Remark \BasGra\ $(i)$, being a subdiagram of the \NOL\
graph $G$, $C_k$ is \NOL\ and two--legged with even-legged vertices
and at most $N-1$ vertices. By the inductive hypothesis, $C_k$ is a
string of GST, so $G$ is a GST by definition.
\pni
(2) If there are two external vertices $v_\one $ and $v_\two$, let
$G_\one = G - v_\two $ be the graph obtained from $G$ by deleting
$v_\two $ and all the lines going into it. Let $C_\one , \ldots ,
C_r$ be the connected components of $G_\one$, where $C_\one $ contains
$v_\one$. Then $G$ takes the form drawn in Figure 8 (a).
Consider the quotient graph $G_\two $ where all $C_k$ are replaced
by vertices $c_k$ (see Figure 8 (b)). Denote the
number of lines from $v_\two $ to $c_k$ by $n_k$. Then for all
$k\geq 2$, $n_k$ must be even, since all the vertices of the graph
$C_k$ have even incidence number. Since 
both $c_\one$ and $v_\two$ join to one
external leg, the number $n_\one $ of lines between them must be odd.
If $n_\one \geq 3$ or for any $k \geq 2$, $n_k \geq 4$ ,
$G_\two $ is \OL\ by Remark \BasGra\ $(v)$. So $n_\one = 1$ and $n_k =2$
for all $k \geq 2$, and $G$ takes the form shown in Figure 8 (c).
Thus for all $k \in \nat{r}$, $C_k$ is a two--legged \NOL\ graph
with at most $N-1$ vertices. By the inductive hypothesis,
all the $C_k$ are GST graphs or strings of GST graphs, so
$G$ is a string of GST graphs as well. \endproof
%
%   here comes Figure 8
%
\centerline{\figplace{fig8}{-.2in}{0in}}

%
%      four-legged case
%
\noindent
We now turn to the four--legged case, and begin by a simple characterization
of one--particle reducible four--legged graphs. 
\Rem{\FonePI} Let $G$ be a four--legged graph and all vertices 
of $G$ have an even incidence number. If $G$ is one--particle--reducible, 
$G$ is obtained from a 1PI four--legged graph $\Gp $ by attaching strings
of two--legged diagrams to the external legs of $\Gp $.

%
%   here comes onepi4
%
\centerline{\figplace{onepi4}{-1 true in}{0in}}
\Proof Induction in the number of vertices of $G$.
Let $G$ be 1P reducible and $l$ a line such that 
cutting $l$ disconnects the graph. Upon cutting 
$l$, $G-l$ falls into two connected components. 
Their numbers of external legs must add up to six. Since by assumption, 
any subgraph of $G$ must have an even number of external legs.
one of them must be four--legged and the other one two-legged. 
Apply the inductive hypothesis to the four--legged subgraph, 
then the statement follows for $G$ itself. \endproof  
\vskip-.6in
\centerline{\figplace{fig9}{0 truecm}{0in}}
\vskip-.9in
\Def{\DBC}  
\leftit{$(i)$} 
A GSF graph is a graph $G$ with four external legs,
all joining to a single vertex $v_\one$ of $G$, 
such that upon deletion of two of the external legs, 
$G$ becomes a GST graph.
\leftit{$(ii)$} 
A dressed bubble chain (DBC) of length $r\geq 0$ 
is a four--legged graph as follows. 
There are $r+1$ GSF graphs $G_\one , \ldots G_{r+1}$,
such that for all $i \in \{ 1, \ldots , r \} $ 
$G_i$ is joined to $G_{i+1}$ by exactly two strings of GST graphs, 
and the external legs of $G_\one $ and $G_{r+1}$ 
are connected to the external legs of $G$
by strings of GST graphs 
(which may consist of only a single line).

\Rem{\dummy}  
If an external vertex $v$ of a \NOL\ four--legged diagram 
has at least two external legs, 
joining them to form a self--contraction 
gives a \NOL\ two--legged diagram which must be a GST string. 
This is used in the proof of the following Lemma.
An example for a GSF graph is shown in Figure 9 (a). 
The thick lines in this figure  
stand for strings of GST diagrams. 
An example of a DBC with $r=2$ is given in Figure 9 (b), 
again denoting strings of GST diagrams by thick lines 
and denoting GSF graphs by four--legged vertices with a box. 
An example with $r=1$ where all vertices and lines are drawn 
is shown in Figure 9 (c). A DBC of length $r=0$ is a GSF 
with strings of GST diagrams attached to the external vertex 
of the GSF.
\Lem{\FourL} {\Lesty Let $G$ be a connected graph 
whose vertices all have an even incidence number, 
and with number of external lines $E(G)=4$. 
If $G$ is \NOL, then $G$ is a DBC.
More precisely, let $V_E \in \{1,2,3,4 \}$ be 
the number of external vertices of $G$
(a vertex $v$ is called external if an external leg of $G$
joins to $v$).
If $G$ is 1PI and \NOL, then $V_E \leq 2$ with
$G$ a GSF for $V_E=1$ and a DBC of length $r \geq 1$  for $V_E=2$.}
\Proof 
For $V_E \le 3$, one of the external vertices, $v_\one $,  
must have at least $E_\one \geq 2$ external lines going in. 
Two of the external lines of $v_1$ can be joined to a self-contraction $l^*$. 
By Remark \BasGra\ $(iv)$, the resulting two--legged graph $G^*$ 
is still \NOL, so 
by Lemma \TwoL, it is a string of GST graphs.
Cutting $l^*$, we see that $G$ itself is a DBC.
This is proven by the same induction process as is used to define GST. 
See Figure 10 for an example of how a DBC 
is generated when $l^*$ is cut.
If $E_\one \geq 3$, $G$ is a DBC of length $r=0$.
If $V_E =3$, the two--legged graph $G^*$ constructed from $G$
has two external vertices. Since it is \NOL, it must be 
1P reducible by Lemma \TwoL\ and Definition \GSTDef\ $(iii)$, 
so $G$ is also 1P reducible. Thus $V_E =3$ is impossible 
if $G$ is 1PI.
\centerline{\figplace{fig10}{0 truecm}{0in}}

\noindent
If $V_E=4$, we use Remark \FonePI\ to decompose $G$ into 
the 1PI graph $\Gp$
and the strings of two--legged subdiagrams attached to $\Gp$.
By Remark \BasGra\ $(i)$, $\Gp $ must also be \NOL\ and the 
strings must consist of GST diagrams. 
If $V_E(\Gp )\leq 2$, we know by the above that 
$\Gp $, and hence $G$, is a DBC. $V_E(\Gp ) =3$ is impossible
since $\Gp$ is 1PI. Thus, to complete the proof, we only have 
to show that $V_E(\Gp ) = 4$ is impossible as well
for a four--legged \NOL\ 1PI graph $\Gp$.
So assume that $V_E = 4$, let $v$ be an external vertex of $\Gp$,
and let $S=G'-v$. Since $V_E = 4$, $v$ binds only one external 
leg of $G'$, so $v$ connects to $S$ by an odd number of lines. 
Let $C_\one , \ldots , C_p$ be the connected components of $S$. 
One of them must connect to $v$ by an odd number $n^*$ of lines. 
But if $n^*=1$, 
$G' $ is reducible, contrary to our assumption, and if 
$n^* \geq 3$, $G'$ is \OL\ by Remark \BasGra $(ii)$ and $(v)$, 
again a contradiction.

The alternatives are sketched in Figure 11. The black box $K$ consists 
of $v$, together with all connected components of $S$ that do not contain
an external vertex. In the Figure $K$ is drawn four--legged; in general
it may have a larger incidence number. (a) is the case $n^*=3$. The two loops joining $K$ and $S$ overlap. (b) and (c) are cases where $n^* =1$. The figure
can be disconnected by cutting a line leaving $K$. 
\centerline{\figplace{fig11}{0 truecm}{0in}}

 \endproof

%
%   Tree-Decomposition
%
\sect{Decomposition of the Tree of a Labelled Graph}
\noindent
We now consider labelled
graphs and show how to decompose the associated tree into subtrees
corresponding to \OL\ and \NOL\ graphs. 
It was mentioned in the motivation of the classification of graphs into
\OL\ and \NOL\ ones that the bound for the value of
\OL\ graphs contains as a factor the function $I_2$, defined in
\queq{\Itwodef}.  
As discussed in Section 1, this factor arises because the propagators of scale
$j<0$ are supported in a shell of thickness $M^j$ near the Fermi
surface, and the intersection
of such a shell with its translate by some momentum  $\ve{p}$
is transversal for all $\ve{p}$ outside a set whose volume 
shrinks with the thickness of the shell. Therefore, the arguments
$\varepsilon_k$ in $I_2$ will be $M^{j_k}$, where $j_k$ are scales
of the lines involved. 
The volume improvement factor might arise only at a relatively high scale, 
and to exploit it as much as possible, it is therefore
very important in our analysis to keep track of the scale at which this
volume improvement factor arises. We do this by decomposing the tree of the
labelled graph $G^J$ into maximal subtrees corresponding to \NOL\ 
subgraphs. 

We start with an example to illustrate the idea behind the procedure.
In Figure 12, a graph with scale assignments $0 > h' > h > j$ is shown 
from top to bottom on decreasing scales. The interaction lines appear 
only on scale zero. On the lowest scale $j$ (root scale), the graph is 
\NOL, since all lines of higher scale are collapsed into effective
vertices. On scale $h > j$, the graph is \OL, and the volume improvement 
factor arises at scale $h$ in this example. 
In general, the strategy 
will be to go from lower to higher scales (from bottom to top in Figure 12), 
resolving (i.e. expanding) the effective vertices until either scale zero 
or a scale on which the graph overlaps is reached. With a properly chosen 
spanning tree for the graph, the volume gain is then extracted.
\midinsert
\centerline{\figplace{fig12}{0 truecm}{0in}}
\endinsert

\Def{\GtDef} 
\leftit{$(i)$} Let $G^J$ be a labelled graph with tree $t$. For a 
(connected) subtree
$\tp$ of $t$ (we shall denote this as $\tp\subset t$), rooted at a fork
$\phi_{\tp}$, define the projected graph $\tG (\tp )$ as a quotient
graph of $G^J_{\phi_{\tp}}$ as follows. 
If $\fdp\notin \tp $ is a fork directly above $\tp$, i.e. there is a fork
$\fp\in \tp$ such that $\pi (\fdp )=\fp$, replace $G^J_{\fdp }$
by a vertex with the same external legs as $G^J_{\fdp }$, and with
vertex function $Val(G^J_{\fdp })$. The lines in subgraphs $G^J_f$
with $f\in \tp$ join these vertices to form the graph $\tG (\tp )$
(leaves of $t$ that are also leaves of $t'$ remain the same 
vertices they were before).
\leftit{$(ii)$}For a subset $A$ of the set of forks and leaves of $t$, define
$$\eqalign{
\si(A)&=\{\fp\in t\setminus A:\fp\ \hbox{fork}\;,\
\exists f\in A:\pi(\fp)=f\}\cr
\cA(A)&=\{\fp\in t\setminus A:\fp\ \hbox{fork}\;,\
\exists f\in A:\fp\ge f\}\;
.\cr}$$
Thus $\cA(A)$ is the set of all forks of $t\setminus A$ that are above $A$
and $\si(A)$ is the set of all forks of $t\setminus A$ that are 
immediately above $A$.
\leftit{$(iii)$}For $f\in t$, denote by $t_f$ the subtree of $t$ rooted at $f$ 
that contains all forks and leaves in $\cA\bigl(\{f\}\bigr)$. 

\goodbreak

\Rem{\dummy} \leftit{$(i)$} $\tG (\tp )$ is the graph where all subdiagrams
belonging to forks above $\tp$ are collapsed to effective vertices,
and where all subdiagrams belonging to forks in $\tp $ remain subdiagrams.
\leftit{$(ii)$} $\tG (\tp )$ is connected.
\leftit{$(iii)$} The mapping $G^J_{\phi_{\tp}}\to\tG (\tp )$ also acts 
naturally on sets $\vth $ of lines of  $G^J_{\phi_{\tp}}$.
Those lines in $\vth$ corresponding to $\tp$-forks are left unchanged. 
All others are absent in the projection. The projections of
paths $L$ etc. will be denoted as $\tilde L$, or $\tilde L(\tp )$
if necessary. Note that the projection of a path need not be a path
in the sense of Definition \PathDef\ because it may  visit a vertex 
more than once and hence fail to be injective.
\leftit{$(iv)$} $\tp$ may be trivial, that is, consist only of its root fork; 
then we write $\tp=\ph_{\tp}$ and $\tG (\tp ) = \tG (\ph )$.
\sni
To do the tree decomposition, we need some more facts about
\NOL\ graphs which we state in the following Lemma.
If $G$ is a graph and $H$ a connected subgraph with $2m$ external legs, we 
denote by $G/H$ the quotient graph obtained by replacing 
$H$ by a vertex with incidence number $2m$. 
In our convention, external legs of a connected
graph are not counted as lines of the graph, and the statement
that two subgraphs $A$ and $B$ of a given graph are disjoint 
means that they share no vertex 
(so an external vertex of $A$ may be connected to an external vertex of $B$
by a line which belongs neither to $A$ nor to $B$).
\Lem{\mehr} {\Lesty 
Let $G$ be a connected graph. 
\leftit{$(i)$} Let $G$ have only vertices with an even 
incidence number, let $T$ be a connected two--legged subgraph of $G$, 
and assume that $G/T$ is \NOL. Then 
$$
G \hbox{ is \OL } \Longleftrightarrow T \hbox{ is \OL }
.$$
\leftit{$(ii)$} Let $G_\one $ and $G_\two $ be disjoint connected 
subgraphs, and assume that $\tG_\one = G/G_\one $ and 
$\tG_\two = G/G_\two $ are \NOL. Then $G$ is \NOL. 

}

\Proof $(i)$
``$\Leftarrow$'' is obvious by Remark \BasGra\ $(iii)$. 
``$\Rightarrow$'':  
There are two independent \OL\ loops $K$ and $L$ in $G$. 
Since $G/T$ is \NOL, their traces must differ in $T$. 
If $T$ were \NOL, $T$ would be a string of GST, and by 
the structure of GST graphs and the condition that any path
may visit a given vertex at most once, both $K$ and $L$ 
would have to step over the same lines in $T$. So then 
$\th (K) = \th (L)$, which is a contradiction. 

$(ii)$ Assume $G$ to be \OL. Then there are independent loops
$K$ and $L$ such that the set of lines which are part of 
both loops is not empty. Thus there exist `splitting points',
which are vertices as follows: $v$ is a splitting point if 
$v$ is endpoint of a line $\ell_\zer $ that is part of both 
$K$ and $L$, and of lines $k$ and $\ell$ such that $k$ is a 
line of $\th(K)$ but not of $\th (L)$, and $\ell $ is a line 
of $\th(L) $, but not of $\th (K)$. In other words, a splitting 
point is a vertex at which the two paths deviate after going 
over the same line(s). Let $v$ be such a splitting point, and 
$\ell_\zer, \ell $ and $k$ be as defined above. 
Also, denote the second endpoint of $\ell_\zer $ by $w$.

If $v\in G_\one$, we will construct loops 
$K_\two $ and $L_\two $ in $\tG_\two$ as follows.
First we reparametrize $K$ and $L$ (using the shifts and inversions 
described in Remark \Paths) so that they start at $w$ and the 
first line is $\ell_\zer$, and the second is $k$ for $K$ 
and $\ell$ for $L$, etc. Since $v \in G_\one$, and since
$G_\one $ and $G_\two $ are disjoint, none of $\ell_\zer, 
\ell $ and $k$ can be in $G_\two$, so $\ell_\zer, 
\ell $ and $k$ are all in $\tG_\two$.

If $w$ is in $G_\two$, we take $K_\two $ to 
be the restriction of $K$ up to the first point when a vertex of 
$G_\two $ is hit by $K$; this is a loop in $\tG_\two$. 
$L_\two $ is defined similarly. By construction, 
$\th (K_\two )$ contains $k$ but not $\ell$, and 
$\th(L_\two )$ contains $\ell$ but not $k$, so these loops 
are independent, and both contain $\ell_\zer$. 
So $\tG_\two $ is \OL, which is a contradiction. 

If $w$ is not in $G_\two$, we take $K_\two $ to be identical
to $K$ up to the first point where $K$ hits $G_\two$; then we continue 
it to be $K$ from the last time  $K$ visits a vertex of 
$G_\two$ (if $K$ does not visit $G_\two$, $K_\two =K$). 
$L_\two $ is defined similarly. Again, these loops 
are independent, and overlap at $\ell_\zer$, 
which is a contradiction. 

If $v \not \in G_\one$, we construct loops  $K_\one$ and 
$L_\one $ in $\tG_\one$ starting again at $w$, and going over 
$\ell_\zer $ and $k$ or $\ell$, this time taking out the 
parts between first and last visits of $G_\one$, to avoid multiple
visits at the vertex of $\tG_\one$ that replaces $G_\one$.
Since $v$ is not in $G_\one$, 
the lines $\ell_\zer, \ell$ and $k$ are all in $\tG_\one$, 
so $K_\one $ and $L_\one$ are again independent \OL\ loops 
in $\tG_\one$. This contradicts the assumption that 
$\tG_\one $ is \NOL. \endproof
\Rem{\oddVert} If the vertices are allowed to have odd incidence 
numbers, $(i)$ does not hold, as can be seen from the following graph
(the subgraph $H$ is the part of $G$ inside the dashed circle).

%
%   here comes inc3
%
\centerline{\figplace{inc3}{0in}{0in}} 
%
%    tree decomposition lemma
%
\Lem{\Treedec} {\Lesty 
Let $G^J$ be a labelled graph, $t$ its tree,
and $f\in t$ a fork.
\leftit{$(i)$} Let $f_1,\ldots,f_n\in t$ be forks or leaves such that
$\pi(f_i)=f\ \forall i$,
and assume that for all $i\in\{1,\ldots,n\}$, 
$\tG \left(\matrix{ f_i \cr  
f \cr}\right)$ is \NOL. 
Then
$\tG \left(\matrix{f_1 & f_2 \ldots f_n\cr
& f & \cr}\right)$ 
is \NOL\ as well.
\leftit{$(ii)$} Let $\tG (f)$ be \NOL. Then there is a
unique maximal tree $\ta_f\subset t$, rooted at $f$, such that
$\tG (\ta_f)$ is \NOL, i.e. if $\ta $ is such that $\tG (\ta )$
is \NOL\ and such that $\phi_\ta =f$, then $\ta \subset\ta _f$.
\leftit{$(iii)$} There is  
$$
\cN \subset \{ \ph \} \cup \{ f: E(G_f^J)=2, 
\tilde G ( f ) \hbox{ 1PI and \NOL }\}
\eqn $$
such that{\parindent =1in
\item{(a)} $\ph\in\cN$ if and only if $\tG(\ph)$ is \NOL
\item{(b)} if $f,f'\in\cN$ with $f\ne f'$ then $\ta_f$ and $\ta_{f'}$
are disjoint
\item{(c)} if $f\notin\cN$ but  $E(G_f^J)=2$ and  $\tG(f)$ is 1PI and \NOL\ 
then there is an $f'\in\cN$ with $\ta_f\subset\ta_{f'}$

}
\leftit{}Here $\ta_f,\ta_{f'}$ are the maximal trees associated to 
$f,f'$ in part $(ii)$. 
}

\Proof $(i)$ follows as in Lemma \mehr $(ii)$.
\pni
$(ii)$ Let $S=\{\tp\subset t_f:\tG (\tp )$ \NOL $\}$.
Since $\tG (f)$ is \NOL, $S\ne\emptyset$. 
Build up the tree $\ta_f$ recursively as follows: 
for all forks or leaves $f_1,\ldots,f_n$ with $f=\pi(f_i)$, 
add $\matrix{f_i\cr f\cr}$
to $\tau_f$ if $\tG \pmatrix{f_i\cr f\cr}$ is \NOL\ ( note that if
$\si$ is a leaf then $\tG \pmatrix{\si\cr f\cr}$ is always \NOL\
if $\tG (f)$ is \NOL\ since
$\tG \pmatrix{\si\cr f\cr}=\tG (f)$.
The resulting tree $\tG \pmatrix{f_{i_1} &f_{i_2}\ldots f_{i_k}\cr
\noalign{\pni} &\ f\hfill\cr}$ is then \NOL\ by $(i)$. If for all forks
$f_1,\ldots,f_n$, $\tG \pmatrix{f_i\cr f\cr}$ is \OL, then
$\tau_f={v_1\ldots v_b\atop f}$ where $v_1,\ldots,v_b$ are the leaves with
$\pi(v_i)=f$, or $\tau_f= f$ if $b=0$, and the process stops.
Otherwise, repeat the procedure for every 
$\fp\in\{f_{i_1},\ldots,f_{i_k}\}$
that is a fork, add branches $\matrix{\fdp \cr f_{i_\ell}\cr}$ if
the corresponding graph is \NOL, add all branches to leaves and stop if
there are no forks with \NOL\ graphs $\tG \pmatrix{\fdp \cr f_{i_\ell}\cr}$.
Repeating this, the process ends after a finite number of steps.
It is obvious by construction that the so obtained tree is maximal in $S$
and therefore unique.
\pni
$(iii)$ Put $\phi$ into $\cN$ if $\tG (\phi )$ is \NOL, and in that 
case construct $\ta_\ph$ using $(ii)$. Let 
$M_2(G^J)=\{f\in t: f>\ph, \tG (f)$ is \NOL, 1PI and two-legged
$\}$.
We construct $\cN$ by induction on
the number of forks $N$ of $M_2(G^J)$. If $N=0$, i.e.  $M_2(G^J) = \emptyset$, then
 $\cN = \emptyset $ or $\cN = \{ \phi \}$, 
depending on whether $\tG (\ph ) $ is \OL\ or not.  
Let $N\ge 1$ and assume that the family has been
constructed for all $N^\prime\le N-1$. Let
 $\{f_1,\ldots,f_n\}$ be the set of all minimal forks of $M_2(G^J)$
(in the partial ordering of $t$).
For all $k$, construct $\tau_{f_k}$ by $(ii)$. Because of the tree structure,
we can consider each $k$ separately. Let $g=f_k$ for some $k$.
${\cA}(\tau_g)$ is a disjoint union of trees rooted at forks
$f\in\si(\tau_g)$ (or ${\cA}(\tau_g)=\emptyset$, in which case we
are done with $g$). Each of these trees has
$N^\prime\le N-1$ forks in $M_2$, so the inductive hypothesis applies.
Add to $\cN$ the forks that have been selected
by the inductive hypothesis from each of the trees.
This really gives a family of disjoint trees in the sense that 
no element of $\cN$ is directly above a fork in a tree $\ta_f$ 
of $\cN$, as is implied by the Remark following this Lemma.

Suppose now that $f\in t\setminus\cN$ but
$E(G_f^J)=2$ and  $\tG(f)$ is 1PI and \NOL. By the construction of $\cN$
the set $\{f''\in\cN : f''<f\}$ is nonempty. Let $f'$ be the maximal element
of this set. Also by construction $f\in\ta_{f'}$. To complete the proof
it suffices to show that if $\ta_f\not\subset\ta_{f'}$ then the tree
$\ta_{f'}$ is not maximal in the sense of $(ii)$. To see this, first
observe that $\tG\big((\ta_{f'})_f\big)$ is \NOL\ by Remark \BasGra(i).
So the maximality of $\ta_f$ implies $(\ta_{f'})_f\subset\ta_f$, which in turn
implies $\tG(\ta_f\cup\ta_{f'})$ is gotten by replacing the two--legged
\NOL\ subgraph $\tG\big((\ta_{f'})_f\big)$ of $\tG\big(\ta_{f'}\big)$
by the two--legged \NOL\ graph $\tG(\ta_f)$. Remark \BasGra(ii) and the 
following Remark ensure that $\tG(\ta_f\cup\ta_{f'})$ is \NOL.
So the assumption that $\ta_{f'} \neq \ta_f \cup \ta_{f'}$
contradicts the maximality of $\ta_{f'}$. Thus $\ta_f \subset \ta_{f'}$.
\endproof

\Rem{\ftOL} Let $G^J$ be a labelled graph of our model, $t(G^J)$ its tree, 
and let $\ta $ be a subtree of $t$ such that $\tG (\ta )$ is \NOL.
Let $f$ be a fork directly above $\ta$, i.e. 
$f\in \si (\ta )$, such that $E(G_f) = 2$. Then 
$$
\tG \left(\matrix{ f\cr | \cr \ta \cr}\right) \hbox{ \OL } 
\Longleftrightarrow
\tG (f) \hbox{ \OL}.
\eqn $$
\Proof Apply Lemma \mehr $(i)$.

\endproof
\Rem{\Passauf} Note that Remark \ftOL\ holds only for two--legged
subdiagrams. For example, the graph at scale $j$ in 
Figure 12 is not overlapping while that at scale $h$ is overlapping.
None--the--less, to go from the graph at scale $j$ to that at scale $h$
one replaces the six--legged vertex by the tree diagram
\hbox to 0pt{\hglue 3pt
\vbox to 0pt{\vglue 7 pt\includegraphics{sixtree.ps}\vss}
\hss}
\hskip 15pt  $^h$ \hskip 16pt
, which has no loops and hence is not overlapping. So, it is possible to 
replace one vertex in a non--overlapping graph by a non--overlapping
subgraph and produce an overlapping graph.
This plays a role for estimates of derivatives.
 
\sect{Improved Power Counting }
\noindent
We now extract the volume improvement factor in the value of 
any graph that overlaps at some scale, 
and use it to show an improved
power counting bound that holds for every such graph.
We also give a natural routing prescription for the external momentum 
suited to bounding derivatives. 
In this section, let $G \in Gr(n,m; m_\one , \ldots , m_n)$, 
$J:\ell\mapsto j_{\ell}$ be a labelling of $G$
and $Val G^J(C, \cU _{v_\one}, \ldots ,\cU _{v_n})$ be given 
by \queq{\ValU}.
A loop basis for the graph is a basis for $H^1(G,\Z)$. 
Since for an overlapping graph, the two overlapping loops 
(in the sense of Definition \PathDef)
define linearly independent cycles, we may use both of them 
as basis elements. Recall that there is a natural basis for  $H^1(G,\Z)$ 
associated to any spanning tree $T$ for $G$. 
It contains one loop for each element of $L(G)\setminus L(T)$. 
The loop associated to $\ell\in L(G)\setminus L(T)$
consists of $\ell$ and the path in $T$ joining the ends of $\ell$. 
Also recall that $T$ is consistent with $J$ if $T\cap \tG(t_f)$ is a 
spanning tree for $\tG(t_f)$ for all forks $f\in t_\ph$.

By definition, a graph $G$ is \OL\ if there exist two independent
loops in $G$ which share a line. A priori, the specific loops 
determined by a spanning tree for $G$ are not required to overlap.
But of course they do. This is proven in the following Lemma.
 
\Lem{\ArbTree} { \Lesty 
\leftit{$(i)$} If $G$ has a spanning tree $T$ without any associated \OL\ 
loops, then no spanning tree of $G$ has any associated \OL\ loops.
\leftit{$(ii)$} If $G$ is \OL\ and $T$ is an arbitrary spanning tree 
of $G$, then there are two \OL\ loops associated to lines $\ell_\one $ and
$\ell_\two \in L(G) \setminus L(T)$.

}

\Proof 
$(i)$ 
Let $T$ be a spanning tree for $G$ such that all of the loops
$L_1,\cdots,L_n$ associated to $T$ do not overlap each other.
Define  $T_{NL}=G\setminus \cup_{i=1}^n L_i.$ $(i)$ is 
a consequence of
\leftit{(a)} if $\ell\in T_{NL}$ then $G - \ell$ is not connected
\leftit{(b)} for each $1\le i\le n$  every spanning tree for $G$ must
contain all of $L_i$ save exactly one line. 
\pni 
(a) and (b) imply that 
any spanning tree for $G$ must consist of $G$ minus exactly one line from
each of $L_1,\cdots,L_n$.
\pni
{\it Proof of} (a): First note that $T_{NL}\subset T$ because,
by definition every line of $G$ that is not in $T$ generates one 
of $L_1,\cdots,L_n$. If $G - \ell$ is  connected, there is a path in 
$G - \ell$ that joins the two vertices at the ends of $\ell$.
There is always such a path that is also contained in $T$, because $T$
contains all of $G$ save one line from each of $L_1,\cdots,L_n$. If the 
path uses the missing line from $L_i$ we can always replace the missing
line by the rest of $L_i$. Hence $\ell$ union the path is a loop in $T$,
which is impossible.
\pni
{\it Proof of} (b): Delete two lines $\ell_1,\ell_2$ from $\th(L_i)$
($\th(L)$ is the subgraph corresponding to $L$; 
see Definition \PathDef\ $(iii)$). Then 
$\th(L_i) - \ell_1 - \ell_2$ consists of two connected pieces $\La_1,\La_2$.
In the event that $\ell_1$ and $\ell_2$ are nearest neighbours on $L_i$,
$\La_1$ and/or $\La_2$ is a trivial graph consisting of a single vertex.
Suppose that there is a path $P$ in $G - \ell_1 - \ell_2 $ 
connecting a vertex $v_1$ of $\La_1$ to a vertex $v_2$ of $\La_2$.
We can assume without loss of generality that this path contains no lines of $L_1$. 
As in the proof of (a), we can also arrange for the path to be contained in $T$. 
One of $\ell_1$ and $\ell_2$ must be in $T$, so we
can construct a loop in $T$ using $P$ and part of $L_1$. This is impossible,
so $P$ cannot exist. So no spanning tree for $G$ can be contained in
$G - \ell_1 - \ell_2 $.
 
$(ii)$ It suffices to construct one spanning tree $T$ for $G$ that 
has two \OL\ loops associated, because by $(i)$, any other spanning tree
for $G$ will then have the same property. 
Let $L_1$ and $L_2$ be independent overlapping loops in $G$. Let 
$\ell_\one $ be a line in $L_\one $ that is not in $L_\two$. $\ell_\one $
exists because if $\th(L_1)\subset \th(L_2)$, then
either $\th(L_1)=\th(L_2)$ or $L_2$ is self-intersecting. 
Put $\th(L_1) - \ell_1$ in $T$. 
Note that, regardless of how we complete $T$ the loop associated to
$\ell_1$ will always be $L_1$. Let $\ell_2$ be a line that is in $L_\two $ 
but not in $L_\one$. Denote by
$v_1$ and $v_2$ the vertices at the ends of $\ell_2$. Add to $T$ the unique  
connected subgraph of $\th(L_2)$ that does not contain $\ell_2$, that has
$v_1$ as one terminating vertex, that has a vertex $w_1$ of $\th(L_1)$ as its
other terminating vertex and that contains only one vertex of $\th(L_1)$.
Similarly, add to $T$ the unique  
connected subgraph of $\th(L_2)$ that does not contain $\ell_2$, that has
$v_2$ as one terminating vertex, that has a vertex $w_2$ of $\th(L_1)$ as its
other terminating vertex and that contains only one vertex of $\th(L_1)$.
Note that the two pieces of $L_2$ that have just been added to $T$
contain no lines of $L_1$ and that $w_1\ne w_2$, because $L_2$ must overlap
$L_1$ and cannot be selfintersecting. Regardless of how we complete $T$,
the loop $M_\two$ associated to $\ell_2$ will contain $\ell_2$, 
continue along $L_2$ from $v_1$ to $w_1$, 
continue along $L_1$ from $w_1$ to $w_2$ and finally
continue along $L_2$ from $w_2$ to $v_2$. Thus the loops associated
to $\ell_1$ and $\ell_2$ overlap. Complete $T$ any way you like. \endproof

\Lem{\ImpPow} {\Lesty 
Let $G$ be \OL. Let $J$ be an assignment of scales to $G$. 
\leftit{$(i)$}
Let $\ta_\ph$
be the maximal subtree of $t(G^J)$ rooted at $\ph$ for which $\tG(\ta_\ph)$ 
is \NOL.
Let $j^*=\min\{j_f : f\in\si(\ta_\ph)\}$. Then for any tree $T$ consistent with 
$J$ there is a line $\ell^*\in T$ with $j_{\ell^*}\le j^*$ 
which is contained in two independent loops associated to 
lines $\ell_\one $ and $\ell_\two \in L(G) \setminus L(T)$.
In the case that $\tG(\ph)$ is \OL\ $j^*=j_\ph$.
In the assignment of momenta to lines of $G$ given by $T$, 
$$
p_{\ell^*} = \pm p_{\ell_\one} \pm p_{\ell_\two} + Q
\EQN\pmLico $$
where $Q$ is a linear combination of loop and external momenta 
independent of $p_{\ell_\one}$ and $p_{\ell_\two}$.
\leftit{$(ii)$} 
Assume that the propagators assigned to the lines of $G$ satisfy 
$$
\abs{C_{j_\ell}(p_\zer , e(\ve{p}))} \leq z_\ell M^{-j_\ell +2}
1( \abs{ip_\zer - e(\ve{p} )} 
\in \lbrack M^{{j_\ell}-2},M^{j_\ell}\rbrack )
,\eqn $$
with factors $z_\ell > 0$.
Let $K_\zer $ be as in Lemma \Simpl\ $(ii)$, $\ep $ be as in Proposition 
\NoNest, 
$A$ be as in Lemma \Simpl\ $(ii)$, and let 
$$
K_\one = C_{vol} {{u_\zer}^2 \over A^2}
.\EQN\Konedef$$
Then
$$ 
\abs{Val G^{J}}_{\zer} \leq K_\one 
\prod\limits_{\ell \in L(G)} (4 K_\zer z_\ell ) \;
\prod\limits_{v \in V(G)}\abs{\cU_v}_{\zer} \;
 M^{\ep  j_{l^*}} \; M^{D_{\phi }j_{\phi }}\;
\prod\limits  _{f>\phi } M^{D_f(j_f - j_{\pi (f)} )}
\EQN\ImpEst$$
}
\Proof  $(i)$ Let $T$ be any tree consistent with $J$. For example,
$T$ may be built by first building spanning trees for the topmost
forks of $t(G^J)$, then extending these to spanning trees for  the next
level of forks of $t(G^J)$ and so on. Let $b$ obey 
$j_b=\min\{j_f : f\in\si(\ta_\ph)\}$. Then
$\tilde T(\ta_\ph\cup\{b\})\equiv T\cap\tG(\ta_\ph\cup\{b\})$
is a spanning tree for $\tG(\ta_\ph\cup\{b\})$, because if you
collapse a {\it connected} subgraph of a tree, you get another tree.
By the maximality of $\ta_\ph$, $\tG(\ta_\ph\cup\{b\})$ is \OL.
By Lemma \ArbTree $(ii)$, any spanning
tree for any \OL\ graph has associated at least two overlapping
loops. So there is an $\ell^*\in \tG(\ta_\ph\cup\{b\})$ that is in two
independent loops $\tilde L_1$ and $\tilde L_2$ associated to 
$\tilde T(\ta_\ph\cup\{b\})$. These loops expand to two independent loops
$L_1$ and $L_2$  associated to $T$, both of which contain $\ell^*$.
\queq{\pmLico} follows for $\ell^*$ because any line that is part of 
the loop $L_i$ has the momentum $p_{\ell_i}$ flowing through it, i.e.\
the linear combination of momenta making up $p_{\ell^*}$ contains 
summands $\pm p_{\ell_\one }$ and $\pm p_{\ell_\two}$, the $\pm $ 
depending on the relative orientation of the lines. 
\pni
$(ii)$ After fixing of the momenta on the lines of $T$, 
the expression \queq{\ValU} for $Val(G)$ becomes
$$\eqalign{
Val(G^J) (\et_\one , \ldots , \et_{2m-1}, \be_{2m}) & =
\sum _{spins\; \al} \;
\int \prod\limits  _{l \in L(G)\setminus  L(T)} \db^{d+1}p_{l} 
\; \prod\limits  _{l \in L(G)}
\bigl( C_{j^{l}} (p_{l}) \bigr)_{\al_{l} \al_{l}^{\prime}} \cr
& \prod\limits_{i=1}^{n} \; \bigl( \cU _{v_i} 
(p_\one^{(i)}, \ldots , p_{2m_i-1}^{(i)}) \bigr)
_{{\al _\one^{(i)}}, \ldots , \al _{2m_i}^{(i)}} 
\cr} \EQN\FixVal $$
where $\et_k = (q_k, \be_k )$, and $2m_i = E_{v_i} $, and
the momenta on lines and in the vertex functions
$\cU_v$ match up according to the fixing of the momenta described 
in $(i)$, and for each $t \in L(T)$, $p_t$ is a linear combination 
of the loop momenta $(p_l)_{l \in L(G)\setminus L(T)}$ and the 
external momenta $q_\one , \ldots , q_{2m-1}$.

We bound the spin sum at both ends of every line 
$l\in L(G)$ by a factor $2$ times the maximum over spins 
and take the sup norm of all $\cU _{v_i}$, to get
$$
\abs{Val G(C, \cU _{v_\one}, \ldots ,\cU _{v_{n}})}_\zer
\le 4^{\vert L(G) \vert} X\;\prod\limits  _{k=1}^{n} 
\abs{\cU _{v_{k}}}_\zer  
\eqn $$
where
$$
X = \sup\limits_{q_\one, \ldots , q_{2m-1}} 
\int \prod\limits_{l \in L(G)\setminus L(T)} \db^{d+1}p_{l} 
\prod\limits_{l \in L(G)} 
\abs{C_{j_l}((p_{l})_\zer, e(\ve{p}_l))} 
\eqn $$
By hypothesis, upon integration over the $(p_{l})_\zer$,
$$
X \le Y \; \left( \prod\limits_{l \in L(G)} z_l M^{-j_l+2} \right)
\prod\limits_{l \in L(G)\setminus L(T)} (2 M^{j_l})  
\eqn $$
where
$$
Y = \sup\limits_{q_\one, \ldots , q_{2m-1}}
\int \prod\limits_{l \in L(G)\setminus L(T)}
\Bigl(\db^{d} \ve{p}_{l} \; 
1 \bigl( \vert e(\ve{p}_{l}) \vert \le M^{j_l} \bigr) \Bigr) 
\prod\limits_{l \in T} \; 
1 \bigl( \vert e(\ve{p} _{l}) \vert \le M^{j_l} \bigr) 
\EQN\Ytreefo $$
Ordinary power counting, Lemma \Powco\ $(i)$, would be obtained by omitting
the last product over $l \in T$. Improved power counting is
obtained by keeping only one factor, that from $l=l^{*} \in T$, of this
product, to use the volume improvement estimate, Proposition \NoNest. 
Applying $(i)$, integrating over the loop momenta $p_{l_\one}$ and
$p_{l_\two}$ first and recalling \queq{\Itwodef},
$$
Y \le  I_\two (M^{j_{l_\one}} , M^{j_{l_\two}} ,M^{j_{l^*}})
\int\prod\limits_{l \in L(G)\setminus L(T)
\atop {l \not\in\{ l_\one, l_\two\}}}
\db^{d} \ve{p}_l \;  1 (\abs{e(\ve{p}_l)} \le M^{j_l})
\eqn $$
By Proposition \NoNest\ and Lemma \Simpl\ $(ii)$,
$$\eqalign{ 
Y &\leq C_{vol} 
 M^{\ep  j_{l^*}} M^{j_{l_1}} M^{j_{l_2}} 
\prod\limits_{l \in L(G)\setminus L(T)
\atop {l \not\in\{ l_\one, l_\two\}}}
{A \over u_\zer } M^{j_l}
\cr
& \leq K_\one  M^{\ep  j_{l^{*}}} 
\prod\limits_{l \in L(G) \setminus  L(T)} {A \over u_\zer} M^{j_{l}}
\cr}\eqn $$
We insert this  bound for $Y$, use 
$\vert L(G) \setminus  L(T) \vert \le \vert L(G) \vert$
and $A/u_\zer \ge 1$, and reorder the product over scales by 
the usual telescope formula $j_l = j_\ph + 
\sum\limits_{f > \ph \atop l \in G_f} (j_f - j_{\pi (f)})$. 
\endproof
\Rem{\Nbene}
Apart from a constant, the improved power counting bound is the
ordinary power counting bound times an improvement factor $M^{j^* \ep }$
where $j^*$ is the scale at which the graph overlaps.
By Lemma \Simpl\  $(i)$, the propagators $C_j$ given by \queq{\Cjxydef}
satisfy the hypothesis of $(ii)$ with $z_\ell =1$. Derivatives with 
respect to $p$ or $e$ satisfy a bound with $z_\ell = \Const M^{-j_\ell}$
by Lemma \Simpl\ $(iii)$.

\someroom\noindent
We now want to prove that for any labelled graph $G^J$ with (scale )
tree $t(G^J)$, there is a spanning tree such that the external momentum
does not enter any of the lines in $\tG(\ta_f)$, for all $f \in \cN$.
We first explain why there is anything to prove. 
Any two--legged 1PI \NOL\ graph has only one external vertex $v_\one$. 
The external momentum can trivially avoid all internal lines of such a graph.
However, even if $\tG(f)$ is \NOL, $G(t_f)$ may be \OL.  
In fact, the image of a poorly chosen spanning tree for $G$ 
under the projection onto 
$\tG$ may not even be a tree. Consider, for example, the graph drawn in 
Figure 12. If the leftmost line carrying 
scale $j$ is in the spanning tree of $G$ at scale zero (top of the figure), 
what remains of $T$ in the projection of $G$ on scale $j$ (bottom of 
the figure) is certainly not a tree graph. 
The way to avoid this problem is to start at the bottom, i.e. at root scale, 
to construct a spanning tree for $\tG (\ph )$ and then go upwards on scales, 
constructing spanning trees for all the subgraphs that appear as 
effective vertices, and combine them to a spanning tree for $G$ 
using the following simple fact. 

\Rem{\Baumeli} Let $G$ be a graph, 
$G'$ a connected subgraph of $G$ and
$\tilde G$ the quotient graph of $G$ obtained by replacing $G'$ by
a vertex. Let $T'$ be a spanning tree for $G'$ and $\tilde T$ a spanning tree for
$\tilde G$. Note in particular that $T'$ necessarily consists only of
{\it internal} lines of $G'$. Let $T=T' \cup \tilde T$.  Then $T$ is
a spanning tree of $G$.

\Lem{\MomRout} {\Lesty 
Let $G^J$ be a labelled two-legged 1PI graph and let
$(\ta_f )_{f\in \cN}$ be the family of subtrees of Lemma \Treedec\ $(iii)$.
Then there is a spanning tree of $G$ such that for all $f \in \cN$, 
the external momentum enters in no line of $\tilde G (\ta _f)$, 
and such that for all  $f \in \cN$, there is an improvement factor 
$M^{\ep j^*_f}$ with
$j^*_f\le\min\{ j_b : b $ leaf of $\ta _{f}\}$.}
\Proof  We first construct a suitable spanning tree for
$\tilde G(\ta _{f})$, for any $f\in \cN$. This is easy.
On each loop of $\tilde G (\ta_{f})$ delete a line, $l^\prime$, of 
lowest scale. Since $\tilde G(\ta _{f})$ is a $GST$
graph, this does not disconnect it. 
It leaves a tree $T_f$, which
is already the desired spanning tree for $\tilde G (\ta _{f})$.
In particular it is consistent with $\ta_f$. This means the following. 
Let $f'$ be a fork of $\ta_f$ and $t_{f'}$ the subtree of $\ta_f$ 
consisting of $f'$ and all forks of $\ta_f$ above $f'$. Then
$T_f\cap\tG(t_{f'})$ is also a spanning tree for $\tG(t_{f'})$. 
To see consistency, it suffices to check that $T_f\cap\tG(t_{f'})$
connects all pairs of vertices $v,v'$ of  $\tG(t_{f'})$,
because clearly $T_f\cap\tG(t_{f'})$ can contain no loops.
As $\tG(t_{f'})$ is connected it contains some path from $v$ to $v'$.
The only problem is that this path may use the one line $\ell'$ of some
loop $L$ that is omitted from $T_f$. But because $\ell'\in\tG(t_{f'})$
and $j_{\ell'}\le j_{\ell''}$ for all $\ell''\in L$ we necessarily
have $L\subset\tG(t_{f'})$. But then we may use $L\setminus\ell'\subset T_f$
instead of $\ell'$ in the path.

Since the loops determined by $T_f$ do not overlap and $\tG (\ta_f)$
is 1PI, all lines on the same loop 
carry precisely the loop momentum $p_{l^\prime}$. 
The external momentum enters only in the vertex function 
of the one external vertex,
but not in any internal line of $\tilde G(\ta_f)$.
Now we combine them, going upwards from the lowest forks $f\in \cN$.
Choose a leaf $b$ of $\ta_f$ such that $j^*=j_b$ is minimal. 
By the maximality of $\ta_f$,
$G_f' = \tilde G \left(\matrix{b\cr \ta_f }\right)$ is \OL.
Grow a spanning tree for $\tG (b)$. Combine it with 
the spanning tree $T_f$ of $\tG (\ta_f )$ by Remark \Baumeli\ to get
a spanning tree for $G_f'$. Lemma \ImpPow\ applies, so there is a 
volume improvement factor on scale $j^*$ or below.  
To get a spanning tree for $G$, we do the above procedure for all
$\tilde G (\tau _{f})$, $f \in \cN $, then choose an appropriate spanning
tree for the remaining subgraphs of $G$ and put them together using
Remark \Baumeli\ to obtain a spanning tree $T$ for $G$.
This is possible because, by Lemma \Treedec(iii) and Remark \ftOL, 
all the $G_f'$'s are disjoint.
\endproof
\Rem{\nixDrei} Note that the external momentum does enter internal 
lines of \NOL\ 1PI two--legged graphs if vertices with odd incidence number
are there; see, e.g.\ the graph of Remark \oddVert.
%
% Improved power counting theorem
%
\The{\IPCT} {\Thsty 
(Improved power counting) 
Let $G^J$ be a labelled graph contributing to the sum 
\queq{\GjIform} for $G_{2m,r}^I$, 
let $t$ be the tree associated to $G^J$ and $\ph $ its root. 
Let $\cN$ be as in Lemma \Treedec\ $(iii)$ and for $f \in \cN $, let
$$
j^*(f)=\min\{ j_b: b \hbox{ leaf of } \ta_f\} 
\EQN\Jstar $$
Then
$$\eqalign{
\abs{Val(G^J)}_\zer &\leq (4K_\zer)^{\abs{L(G)}} \; 
\left( K_\one M ^{\ep j}\right)^{1(\ph \not \in \cN )}\;
\prod\limits_{f \in \cN} \left( K_\one  M^{\ep\, j^*(f)} \right) \cr
& M^{jD_\ph} \; 
\prod\limits_{f \in t \atop f > \ph } M^{D_f (j_f - j_{\pi (f)} )}
\cr
& \prod_{v\; two-legged} \abs{\th_v}_\zer
\prod_{v\; four-legged} \abs{F_v}_\zer
.\cr}\eqn $$
}
\Proof  Choose the spanning tree of Lemma \MomRout, fix the momenta, 
collect the improvement 
factors given in Lemma \MomRout, going upwards from root scale, 
don't forget the one at root scale if the graph is \OL\ 
on root scale, that is, $\ph \not\in \cN$. 
This works because the higher $\tG (\ta_f )$ appear as vertex functions 
in the lower ones, so that one can indeed apply Lemma \ImpPow\ separately 
for all $\tG (\ta _f)$, $f \in \cN$, and because Remark \ftOL\ assures that 
no improvement factors are counted twice.\endproof

\noindent
If $\tG (\ph )$ is \NOL\ on root scale, then there is 
no improvement factor $M^{\ep j}$. For general \NOL\ graphs, e.g.\
four--legged ones, there is no further improvement without more
specific assumptions on the band structure $e$. However, for 
two--legged 1PI graphs, one can use a refined bound, that exploits
sign cancelations, to show that 
their root scale behaviour does contain another factor of $M^{\ep j}$
even if they are \NOL\ on root scale. This bound, which we now prove, 
is more subtle than 
the previous ones and we will have to use it with care when proving 
the statements about the derivative with respect to $e$ in Chapter 3. 
We first give the explicit formula for $Val(G)$ for \NOL\ two--legged graphs.

\Rem{\NOLVal}  
Let $G$ be a \NOL, $1PI$, two-legged graph.
By Lemma \TwoL, $G$ is a GST graph. By definition, 
these graphs have an obvious recursive
structure: let $v_\one$ be the external vertex of $G$, 
with incidence number $2m_\one $. Let $v_{i_\one},
\ldots , v_{i_{r_\one}}$ be the vertices of $G$ that have incidence
number $\geq 4$ and that are on one of the self-contraction loops
of $v_\one$. By definition of a GST graph, each
vertex $v_{i_k}$ is again an external vertex of a GST (or ST)
graph $G_{i_k}$. 
Choose a spanning tree for $G$ as in Lemma \MomRout.
Then the value of G takes the form
$$\eqalign{
\bigl( Val G^{J} & (C, \cU _{v_\one }, \ldots , \cU _{v_{n}}) \bigr)
_{\be \be '} (q) = 
\sum _{\al _\one ,\ldots , \al _{2 m_\one  -2}}
\int \prod_{i=1}^{m_\one  -1} 
\Bigl(\dbar ^{d+1} p_i \bigl( S_{i}(p_i) \bigr) 
_{\al _{m_\one -1 +i} \al _{i}} \Bigr) \cr
& (\cU _{v_\one })_{\al_\one \ldots \al_{m_\one - 1} \be
\al_{m_\one }\ldots\al _{2m_\one -2} \be '}
(p_\one , \ldots , p_{m_\one - 1} , q,  
p_\one , \ldots ,p_{m_\one -1}) 
\cr}\EQN\Explicit $$
where $S_i(p) \in M_\two  (\C)$ are strings of subdiagrams,
$$
S_i(p)= \left( \prod_{k=1}^{w_i -1} C_{j_k} (p) \cP_{k} T_{k} (p) \right)
C_{j_{w_i}}(p)
\EQN\Strgdef $$
with $w_i$ the number of lines of the string $S_i$,
$\cP _k \in \{ 1,\ell ,1-\ell \} $, 
and $T_{k} (p) $ the kernel either of 
a two-legged vertex or of the (G)ST graph
$G_{i_k}$ if it is associated to one of 
$v_{i_\one }, \ldots ,v_{i_{r_\one }}$. 
Because of \queq{\Nullzwei}, there are 
$j^{(i)}$ such that for all $k \in \nat{w_i}$, 
$j_k \in \{ j^{(i)}, j^{(i)}+1\}$.
%
%     bound for improved PC for GST diagrams
%
\Lem{\besser} {\Lesty
Let $j < 0$, $n \geq 1$, $k \in\natz{n-1}$, 
$m \in \nat{n}$. Let 
$T_\one \ldots T_{n-1} \in C^2(\R\times \cB)$ and 
$g \in C(\R \times \cB)$.
Let 
$$\eqalign{
I_j  = 
\int_\R dp_\zer \; \int_\cB d^d\ve{p} 
\;  & C_j (p_\zer , e(\ve{p} ))^m 
\;  C_{j+1} (p_\zer , e(\ve{p} ))^{n-m} 
\cr
& g (p_\zer , \ve{p} ) \prod_{w=1}^k (1-\ell ) T_w (p) 
\; \prod_{w=k+1}^{n-1} \ell T_w (p) 
\cr}\eqn $$
Then there are constants $U_\one\leq U_\two$, $U_s$ depending on $M$, $u_\zer$, 
$\abs{u}_s$, $d$ and $\de $, such that
$$
\abs{I_j} \leq U_\one M^{2n}  M^{j}
\abs{g}_\zer \prod_{w=1}^k \abs{T_w}_\one 
\prod_{w=k+1}^{n-1} ( \abs{T_w}_\zer M^{-j} )
\EQN\Ordin $$ 
and, if $g \in C^1(\R \times \cB)$,
$$\eqalign{
\abs{I_j}  &\leq U_\two M^{2n} M^{2j}
\prod_{w=k+1}^{n-1} ( \abs{T_w}_\zer M^{-j} ) 
\Bigg( 
\abs{g}_{\one,j} \prod_{w=1}^k \abs{T_w}_\one 
+ 
\abs{g}_{\zer,j} \sum_{v=1}^k \abs{T_v}_\two
\prod_{w \neq v} \abs{T_w}_\one \Bigg) 
.\cr}\EQN\Bettr $$
where
$$
|g|_{s,j}=\sum_{\al:|\al|\le s}\sup\{ |\partial^\al g(p)| : |p_\zer|\le M^{j},
|e(\ve{p})|\le M^{j}\}
.\EQN\jNormdef $$
}

\Proof
We change variables to $(\rh , \om )$, as given in Lemma \uProps,
and denote 
$\ga (p_\zer, \rh , \om ) = g(p_\zer, \ve{p} (\rh , \om )) $
and 
$\th_w (p_\zer , \rh , \om ) = T_w (p_\zer, \ve{p} (\rh , \om ))$. Then 
$$\eqalign{
\ell T_w ( p ) & = \th_w (0,0,\om ), \cr
(1-\ell ) T_w (p) & = 
\th_w (p_\zer , \rh , \om ) - \th_w (0,0,\om )
\cr}\EQN\goodcoord $$
and 
$$
I_j = \int\limits_\R dp_\zer \int\limits_\R d\rh 
\;  C_j (p_\zer , \rh )^m 
\;  C_{j+1} (p_\zer , \rh )^{n-m} 
\int\limits_S d\om 
(J\ga ) (p_\zer , \rh , \om )
\prod_{w=1}^{n-1} \cP_w \th_w (p_\zer , \rh , \om )
.\eqn $$
Here $J(p_\zer, \rh, \om ) = J (\rh, \om )$ is the Jacobian of the 
change of variables, see Lemma \uProps.
In polar coordinates $(r, \vphi)$ such that 
$\rh = r \sin \vphi $ and $p_\zer = r \cos \vphi $, 
$dp_\zer \; d\rh = r \; dr \; d\vphi$, 
$$
C_j (p_\zer , \rh ) = 
{f(M^{-2j} r^2 ) \over i r e^{i\vphi }}
.\eqn $$
Since $m \geq 1$,  
$$
f(M^{-2j} r^2 ) ^m f(M^{-2j-2} r^2 )^{n-m} \leq f(M^{-2j} r^2) \leq 
1 (r \in \lbrack M^{j-2}, M^{j} \rbrack )
.\eqn $$
Noting that $\ell T_w$ is independent of $\vphi $ and $r$ 
by \queq{\goodcoord}, and writing the difference
$$
\th_w (r \cos \vphi , r\sin\vphi , \om ) - 
\th_w (0,0,\om ) = r \De_w (r, \vphi , \om )
\eqn $$
with 
$$
\De_w(r, \vphi , \om ) = 
\int\limits_0^1 dt \; 
(\cos\vphi \; \del_\zer + \sin\vphi \; \del_\one) 
\th_w ( tr\cos\vphi , tr \sin\vphi ,\om ) 
\EQN\thatsDelta $$
we obtain
$$
\abs{I_j} \leq H_j M^{2n+1} M^j \int\limits_S d\om \; 
\prod_{w=k+1}^{n-1}
(\abs{\th_w}_\zer M^{-j} ) 
\eqn $$
where 
$$
H_j = \sup\limits_{\om \in S} 
\sup\limits_{r \in [0, M^{j}]}
\abs{\int\limits_0^{2\pi } d \vphi \; e^{-in\vphi }
\ph (r, \vphi , \om )}
\eqn $$
with 
$$
\ph (r, \vphi , \om ) = 
(J\ga ) (r \cos \vphi , r \sin \vphi , \om )
\prod_{w=1}^k \De_w (r, \vphi , \om )
.\EQN\thatsPhi $$
Bounding the $\vphi $--integral by $2 \pi \abs{\ph}_\zer$  
results in the ordinary power counting bound \queq{\Ordin}.

But we can do better than that by being more careful about the 
integral over $\vphi $. After a Taylor expansion 
of $\ph $ around $r=0$, 
$$
\ph (r, \vphi , \om ) = \ph (0, \vphi , \om ) 
+ r \int\limits_0^1 dt \; 
{\del \ph \over \del r} 
(tr , \vphi , \om )
\eqn $$
$H_j$ splits into two terms. The first term contains 
$$
\ph (0, \vphi , \om ) = 
J(0, \om ) \ga (0,0,\om ) 
\prod_{w=1}^k 
(\cos\vphi \; \del_\zer \th_w (0,0,\om ) + 
\sin\vphi \; \del_\one \th_w (0,0,\om ))
\eqn $$
which is a polynomial of degree $k \leq n-1 $ in  
$e^{i \vphi }$ and $e^{-i \vphi }$, so 
$$
\int\limits_0^{2\pi } d\vphi \; e^{-in\vphi } 
\ph (0, \vphi , \om ) = 0
.\eqn $$
In the second term, we bound 
$$
\abs{{\del \ph \over \del r} 
(tr , \vphi , \om )} \leq 
\abs{J\ga}_\zer \sum_{v=1}^k \abs{T_v}_\two
\left( \prod_{w\neq v} \abs{T_w}_\one \right) 
+ \abs{J\ga}_\one \prod_{w=1}^k \abs{T_w}_\one
.\eqn $$
The factor $r$ from the Taylor expansion gives the additional $M^j$. 
Collecting the constants $\abs{J}_\one$, and others, coming 
from the relation between $\ga $ and $g$, we obtain 
the Lemma with constants $U_s$ that depend on
$\de_\zer$, $\abs{u}_s$, $u_\zer $ and $d$. 
\endproof
\Rem{\Whythat} In the application to the value of a \NOL\ graph, 
the $g$ in Lemma \besser\ will be the vertex function 
$\cU_{v_\one}$, which may depend on other momenta.  
Replacing $\abs{g}_\zer$ and $\abs{g}_\one$ by the corresponding
norms of the restriction of $g(p)$ to $p$ obeying 
$\abs{p_\zer},\abs{e(\ve{p})}\le M^{j+1}$  retains  information about
the support of the propagator $C_j$. As the example of Remark \Passauf\ 
shows, this information is necessary for volume improvement. 
In fact, Lemma \ImpPow\ applies
to that expression, in which the string $S_\one $ is replaced by a 
propagator that satisfies the Hypothesis of Lemma \ImpPow\ $(ii)$ 
with $z_l = M^{j_l}$. 
We shall also need the expression for $I_j$ itself; it is
$$\eqalign{
I_j & = i^{-n} M^{j(n-k-1)} 
\int\limits_0^\infty dr\; r^{k-n+2} 
f(M^{-2j} r^2)^m f(M^{-2j-2}r^2)^{n-m} \cr
& \int\limits_S d \om \int\limits_0^{2\pi} d \vphi \; e^{-in\vphi } 
\int\limits_0^1 dt \; \frac{\del \phi}{\del r} (t\, r, \vphi, \om )
\prod_{w=k+1}^{n-1} M^{-j} \th_w (0,0,\om)
\cr}\EQN\noSup $$
with $\ph $ given by \queq{\thatsPhi}.

\sect{Convergence of the Renormalized Green Functions}
\noindent
The power counting bounds show that divergences in the 
scale sums of graphs contributing to the Green functions
come from unrenormalized two--legged insertions, 
as discussed at length
in the Introduction. In this section we show that the 
renormalized Green functions converge in the limit 
$I \to -\infty$ in every order in perturbation theory, 
i.e. that the scale sum for the value of any graph 
converges. The bound for this value depends on the 
order of perturbation theory $r$, and for some graphs 
it contains a factor $r!$. We show that under the stated 
assumptions, in particular because of the non--nesting condition
\AThr, these factorials in bounds
for single graphs can arise only from the lack of decay of those
forks $f$ with $E(G_f^J) = 4$ for which the graph
$\tG (f) $ is \NOL. For the \OL\ graphs such
factorials do not arise even if $G_f^J$ is four--legged 
because the improved power counting always produces enough 
decay to make the scale sum convergent instead of marginal.
We state this precisely in this section and prove 
Theorems \Renorm\ and \Nice.

We shall show finiteness of the renormalized Green functions
by deriving power counting bounds for the two-- and four--legged
effective vertices that arise in the scale flow. The two--legged 
vertices correspond to the $r$-- and $c$--forks. Although dealing 
with these effective two-- and four--legged vertices is a standard
procedure of handling trees and labelled graphs, what is not standard
here is the behaviour of the $c$--forks. Normally \quref{FT1,FT2}, 
they are constants, and therefore any derivative acting on them 
gives zero (such derivatives always arise from the Taylor expansions 
used to perform renormalization cancellations). In the nonspherical case,
however, the $c$--forks are still momentum--dependent because the shape 
of $S$ is not fixed by a symmetry. Since the scales of $c$--forks are summed
downwards, the ordinary power counting bounds are insufficient to show 
convergence of a differentiated $c$-fork, and the improved power counting 
bounds are necessary.   

For $M>1$, $n \in \N $, $h \in \Z $ and $\veps > 0$ define the function
$$
\la_n (h, \veps ) = \sum\limits_{p=1}^\infty (\abs{h} + p + 1)^n
M^{-\veps p}
.\eqn $$
Obviously, $\la$ is monotonically increasing in $\abs{h}$ and
$(\abs{h}+1)^m \la_n (h,\veps ) \leq \la_{m+n} (h, \veps )$.
This function bounds the effect of $n$ of the marginal four--forks
mentioned above on the scale sum of the fork
below this. 
The following properties allow one to collect
the accumulated effect of such factors when summing scales
down a fixed tree.
\Lem{\Factori} {\Lesty 
Let $\veps >0$ and $M_\zer (\veps ) = 2^{2/\veps}$.
Then for all $M \geq M_\zer$, $a\ge\veps$, all $m,n \in \Z$, 
and all $j \in \Z, j<0$,
 
\leftit{$(i)$} $$\la_m (j,\veps ) \la_n(j,\veps ) \leq 
\la_{m+n} (j,\veps )$$

\leftit{$(ii)$} 
$$\eqalign{
\sum_{l\leq j} (\abs{l}+1)^m M^{a l} \la_n (l, \ep/2) &\leq
(1-M^{-\veps /2})^{-1} \la_{m+n} (j, \veps/2 ) M^{a j} \cr 
&\leq 2 \la_{m+n} (j,\veps/2)  M^{a j}
\cr}\eqn $$

\leftit{$(iii)$} 
$$
\sum_{h=j+1}^0 (\abs{h} +1)^m M^{ h \veps/2 } \la_n (h, \veps/2) \leq 
2\la_{m+n} (j, \veps/2 )
\eqn $$

\leftit{$(iv)$} 
$$
\sum_{h=j+1}^0 (\abs{h} +1)^m M^{ -a (h-j) } \la_n (h, \veps/2) \leq 
\la_{m+n} (j, \veps/2 )
\eqn $$
\leftit{$(v)$} At fixed $n$,
$$
\la_n \biggl( k ,{\veps \over 2} \biggr) \leq a_n k^n + b_n
\eqn $$
with
$$\eqalign{
a_n & = {2^n \over M^\veps -1} \cr
b_n & = \sum\limits_{p\geq 1} (2p+1)^n M^{-\veps p}
\cr}\eqn $$

}

\Proof  $(i)$: see \quref{FMRS}, Lemma 2.4.
%
%   explicit proof of (i)
%
%By definition,
%$$\eqalign{
%\la_m (j, \veps ) \la_n(j,\veps ) & =
%\sum_{p,q=1}^\infty (\abs{j} + p + 1 ) ^m
%(\abs{j}+q+1)^n M^{-\veps (p+q)} \leq 
%\cr
%& \leq \sum_{p,q=1}^\infty (\abs{j} + \max\{ p, q\} + 1 ) ^{m+n}
%M^{-\veps (p+q)} 
%\cr
%& \leq 2 \sum_{\mu =1}^\infty (\abs{j} + \mu +1)^{m+n}
%\sum_{\nu=1}^\mu M^{-\veps (\mu + \nu )}
%\cr
%& \leq 2 \sum_{\nu = 1}^\infty M^{-\veps \nu } 
%\la_{m+n} (j, \veps ) 
%\cr
%& \leq 2 { M^{-\veps } \over 1 - M^{-\veps }}
%\la_{m+n} (j, \veps ) 
%.\cr}\eqn $$
%Since $M^\veps \geq 4$, $(i)$ holds. 

\noindent
$(ii)$ setting $l=j-k$, $k \geq 0$, we can rewrite the sum as
$$\eqalign{
 M^{a j} &\sum_{k=0}^\infty (\abs{j}+k+1)^m M^{-a k}
\sum_{p=0}^\infty M^{-\veps p/2} (\abs{j}+k+p+1)^n  \cr
&\leq 
 M^{a j}\sum_{q=0}^\infty M^{-\veps q/2} (\abs{j}+q+1)^n
\sum_{k=0}^q M^{-\veps k/2} (\abs{j}+k+1)^m 
\cr}\eqn $$
In the sum over $k$, we estimate each term by
$\abs{j}+k+1 \leq \abs{j}+q+1$.
Extending the sum over $k$ to $\infty$, we obtain the result. 

\noindent
$(iii)$ Since for each $h$ in the sum $\abs{h} \leq \abs{j}$,
$ (\abs{h} +1)^m \la_n(h, \veps/2 ) 
\leq  (\abs{j} +1)^m \la_n (j, \veps/2) 
\leq \la_{n+m} (j, \veps /2)$, 
$$\eqalign{ 
\sum_{h=j+1}^0 (\abs{h} +1)^m M^{ h \veps/2 } \la_n (h, \veps/2) &\leq
\la_{n+m} (j, \veps /2)
\sum_{h\leq 0} M^{ h \veps/2 }  \cr 
& \leq 2\la_{n+m} (j, \veps /2)
\cr}\eqn $$

\noindent
$(iv)$ As in the proof of $(iii)$ 
$$\eqalign{ 
\sum_{h=j+1}^0 (\abs{h} +1)^m M^{ -a(h -j) } \la_n (h, \veps/2) &\leq
\la_{n+m} (j, \veps /2)
\sum_{h\geq j+1} M^{ -a(h -j) }  \cr 
& \leq \frac{M^{-a}}{1-M^{-a}}\la_{n+m} (j, \veps /2)\cr
& \leq \frac{1}{\left(2^{2/\veps}\right)^\veps-1}\la_{n+m} (j, \veps /2)
\leq \la_{n+m} (j, \veps /2) \cr
}\eqn $$

\noindent
$(v)$ 
$$\eqalign{
\la_n \biggl( k ,{\veps \over 2} \biggr) & =
\sum\limits_{p=1}^\infty (k +p+1)^n M^{-\veps p} \cr
& = \sum\limits_{p=1}^{k -1} (k +p+1)^n M^{-\veps p} +
\sum\limits_{p\geq k} (k +p+1)^n M^{-\veps p} \cr
&\leq (2k)^n \sum\limits_{p=1}^\infty M^{-\veps p} +
\sum\limits_{p\geq 1} (2p+1)^n M^{-\veps p} \cr
& \leq a_n k^n + b_n
\cr}\eqn $$
with $a_n$ and $b_n$ as given in the statement of the Lemma.
\endproof
\Rem{\JFK} Given any labelled graph $G^J$ with tree $t$
contributing to the renormalized Green functions, 
we will now construct the quotient graph $\Gp $  
mentioned in Remark \Reso(ii) and the corresponding tree $\tp $.
We recall that $\Gp $ is to have the following properties:
$\Gp$ has only two-- and four--legged vertices, with 
vertex functions that are either interaction 
vertices or values of 1PI two-- or four--legged subgraphs. The only
nontrivial two--legged subdiagrams of $\Gp$ that correspond to forks of $\tp$ 
are strings of two--legged vertices.  Any
nontrivial four--legged subdiagram of $\Gp$ that corresponds to a fork of $\tp$ 
consists of a single four--legged vertex with strings of two--legged vertices
appended. The significance of this 
in the inductive proof of finiteness of the infrared limit
is that the scale sum over the scales of forks $f \in \tp$ 
can be easily bounded once the vertex functions of $\Gp$ are
controlled -- and the latter will be covered by an 
appropriate inductive hypothesis because they are of lower order.
  
Let $\ph $ be the root of $t$, and let
$f_\one , \ldots , f_r$ be all forks of $t$ that satisfy:
for all $k \in \nat{r}$,
the number of external legs of $G_{f_k}$ is two or four, 
and $f_k$ is minimal in the sense that there is no fork $g$ such 
that $\ph < g < f_k$ and $G_g$ has two or four external legs. 
Let $\tilde t$ be the tree rooted at $\ph$ and obtained from $t$
by trimming $t$ at $f_\one , \ldots , f_r$ (i.e. by 
collapsing $t_{f_i}$, as defined in Definition \GtDef, to a leaf)
so that $f_\one , \ldots , f_r$ are leaves of $\tilde t$, 
with vertex functions $Val (G_{f_k}^J)$. 
The result is a graph $\tG$ and a tree $\tilde t$, such that no fork of 
$\tilde t$ corresponds to  a
nontrivial two-- or four--legged subdiagram. $\tilde t$ is not yet
the tree with the stated properties because the $G_{f_k}$ 
need not be 1PI. When this is the case, we  extend the tree 
further above $f_k$ to construct $t^\prime $.  

Let $f$ be one of the forks $f_\one , \ldots , f_r$.
If $G_f$ is 1PI, $f$ is a leaf of $\tp$.
If $G_f$ is 1P reducible, then in the transition from $\tilde t$ to $\tp$, $f$ is replaced by one fork  with some leaves above it.
We now specify the procedure for getting $\tp $ in the different 
possible cases.
  
If $f$ is a $c$--fork with $E(G_f)=2$, $G_f$ must be 1PI, 
since $\ell \, Val (G ) = 0$ for any 1P reducible graph
by the support properties of the propagator $C_j$.
 
If $f$ is an $r$--fork  with $E(G_f)=2$ and $G_f$ is 1P reducible, 
let $\cC$ be the set of lines $l \in L(G_f )$ 
such that $G_f$ disconnects if $l$ is cut. 
If all lines in $\cC$ are cut, what remains of $G_f$ falls into $s$ 
connected components $\th_i$. 
By the definition of $\cC$, all the $\th_i$ are two--legged graphs. 
Moreover, they are all 1PI.
Thus $G_f$ is a string of two--legged 1PI
subdiagrams $\th_\one , \ldots , \th_s$ 
joined by the lines in $\cC$, and 
$$
Val(G_f^J) (p) = S(p) = 
\left( \prod\limits_{i=1}^{s-1} T_i(p) C_{j_i}(p) \right) T_s (p)
\eqn $$
where $T_k = \cP _{\th_k} Val (\th_k)$. Note that the external lines of
$G_f^J$ must have scales $j_{\pi(f)}$ or below, while each line of $\cC$
must have scale $j_f$ or above. By momentum conservation, 
the scale assignments and \queq{\Nullzwei}, 
all $l \in \cC$ must carry scales $j_l=j_f=j_{\pi (f)}+1$.
Since $\ell C_j =0$, 
$\ell $ applied to the value of such a string is zero, 
so effectively $1-\ell $ is replaced by $1$. 
Let $\th $ be one of $\th_\one , \ldots , \th_s$. 
Then $\th$ can be $\th = G_g$ 
where $g$ is an $r$-- or $c$--fork directly above $f$, 
i.e. $\pi (g) =f$ and $\cP_\th = 1-\ell $ or $\ell$, 
or $\th $ is a two-legged graph of root scale $j_f$, 
in which case $\cP_\th =1$. Let us call this latter case a \SSI.
We continue the construction of $\tp $ by reinstalling the fork $f$ and adding,
for every $k \in \nat{s},$ a leaf $b_k$ above $f$
which has vertex function $T_k$.
Now, $G_f$ just consists of the lines of $\cC$ and the vertices
 $b_\one,\cdots, b_s$.

If $G_f$ is four--legged and 1PI, $f$ is a leaf of $\tp $. 

Finally, if $G_f$ is four--legged and 1PR, remove the strings attached to $G_f$ 
according to Remark \FonePI, and add a leaf, above the fork $f$, 
for the 1PI ``core'' of $G_f$, as well as for each 1PI two--legged 
subdiagram $\th_i$ of the strings. The strings have the same properties as the ones discussed in the 1P--reducible $r$--fork case. 

Doing this for all of $f_\one , \ldots , f_r$, we obtain the 
tree $\tp $. By construction, $\Gp = \tG ( \tp )$ has 
the desired properties.    

Finally, we note that if $G$ is 1PI, $\Gp$ is as well, since it is 
a quotient graph of $G$.

The relation between the scale sums for $G$ and $\Gp$ is 
$$
\sum\limits_{J \in \cJ (t,j)} Val (G^J) = 
\sum\limits_{J_\one \in \cJ (\tp , j)} Val ({\Gp}^{J_\one} )
\EQN\GGprel $$
In this formula, $\cJ$ is as usual, but the 
vertices $w$ of $\Gp$ carry a scale index $j_w$,
as discussed in Remark \Reso. If $j_w=0$, 
$w$ is also a vertex of $G$, and the associated vertex function 
is $\hat v$. Otherwise, $j_w$ is the root scale of a subgraph 
of $G$ whose value is a vertex function in $\Gp$
(given by \queq{\Vwdef}) and $j_w $ is summed over. 
For fixed $j_{\pi(w)}$, the summed vertex function is 
$$
F_w = \cP_w \sum_{j_w}\sum_{J \in \cJ (t_w, j_w)}
Val(\tG (t_w))
\EQN\Fw $$
where $\cP_w \in \{ 1-\ell , \ell \} $ for two--legged vertices 
associated to forks, $\cP_w=1$ for  two--legged vertices 
corresponding to  {\SSI}s and for four--legged vertices. The 
range of summation for $j_w$ is: 
a sum $I \leq j_w \leq j_{\pi (w)}$ for a $c$--fork, 
a sum $j_{\pi (w)} +1 \leq j_w < 0$ for an $r$--fork
or a four--legged vertex, and no sum at all, but $j_w = j_{f=\pi(w)}$ 
for a \SSI. The last point is important because 
these diagrams do not have $1-\ell $ in front, but the ``correct''
factor $M^{j_f}$ is there because their scale is fixed.
For a fork $f \in t$, let 
$$
n_f = \abs{ \{ \fp \in t: \fp > f, \tG (\fp )\  
\hbox {\NOL,}\ E(G_{\fp }) =4, G_{\fp} \hbox{ 1PI}\}}
\EQN\nfdef $$
$n_f$ indeed depends only on $G$ and $t$, 
but not on the scale assignment $J \in \cJ (t,j)$. 

\The{\TwoFour} {\Thsty Let $G$ be a graph with $E(G)=2m$ external legs 
and $t$ be a tree rooted at a fork $\ph$ compatible to $G$, 
so that $(t,G)$ contributes to the renormalized effective action 
at scale $j$, $G_{j,2m,r}^I$ (see Remark \Reso). 
For $I<j < 0$ and $J \in \cJ (t,j)$, 
let $Val(G^J)$ denote the value of the labelled graph $G^J$
with root scale $j_\ph =j$.  
Let $\ep$ be the volume improvement exponent of Proposition \NoNest.
Let $\abs{\;\cdot\;}_s$ as in \queq{\Sobotka} and \queq{\sNormdef};
recall that  
for $2m$--point functions with $m >1$, the supremum is taken 
over all $2m-1$ independent external momenta entering into $G$.
The numbers of vertices and of internal lines of $G$ are denoted by
$|V(G)|$ and $|L(G)|$ respectively.

\leftit{$(i)$} Let $G$ be 1PI. 
There is a constant $Q_\zer $ such that for $ s\in \{ 0,1,2\}$
$$
\sum\limits_{J \in \cJ (t,j)} \abs{Val(G^J)}_s \leq 
 {Q_\zer}^{\abs{L(G)}} |\hat v|_s^{|V(G)|}
\la_{n_\ph} (j,\sfrac{\ep}{2}) M^{j Y_s} 
\eqn $$
where 
$$
Y_s (G) = \cases{ (1+\ep -s) & if $E(G)=2m=2$ \cr
     2-m-s    & if $E(G)\geq 4$ and  $\tG(\ph )$ is \NOL \cr 
     2-m-s+\ep & if $E(G)\geq 4$ and  $\tG(\ph )$ is \OL\  \cr }
\eqn $$
\leftit{$(ii)$}
Let 
$X=1 + W_\one + W_\two $
where $W_s$ is as in Lemma \Simpl\ $(iii)$, 
let $K_\zer $ and $K_\one$ 
be as in Lemma \ImpPow, $U_\two $ as in Lemma \besser, and 
$$
K_\two = \max \Big\{ 2(2+d \abs{\prP}_\one), {2\sqrt{2} M^2 \over u_\zer}, 
M^{2(1+\ep)}, M^4U_\two \Big\}
\eqn $$
Then 
$$
Q_\zer = { 18d K_\zer K_\one K_\two  X^2 \over 1 - M^{-1}}
\EQN\Qzerodef $$
will do.
\leftit{$(iii)$} For $s \leq 1$ and $E(G)=2m \geq 4$, the estimate
$(i)$ also holds  for one--particle reducible graphs. 
\leftit{$(iv)$} As $I \to -\infty$, $\sum\limits_{J \in \cJ (t,j)} Val (G^J)$ 
converges in $\abs{\cdot }_\one$ to a function that obeys the 
bound $(i)$, and, for $G$ two--legged and 1PI, 
$\sum\limits_{j=I}^{-1} \sum\limits_{J \in \cJ (t,j)} Val (G^J)$
converges in $\abs{\; \cdot \; }_\one $.  
 
}  

\Proof We take $(i)$ -- $(iv)$ as induction hypotheses and 
do induction over the depth of the pair $(t,G)$, 
which is defined as
$$
P = \max\{ k: \exists f_\one > f_\two >  \ldots > f_k > \ph {\rm \ with\ }
E(G_{f_i}^J) \in \{ 2,4\} {\rm \ for\ }1\le i\le k\}
.\EQN\Depthdef $$
In other words, given any leaf of the tree $t$, there are at most
$P$ two--legged or four--legged forks on the unique path between
the root $\ph$ of the tree and this leaf.
Let $\cN$ be as in Lemma \Treedec\ and recall that 
$\ph \in \cN \Longleftrightarrow \tG (\ph ) $ \NOL.
Also, call $Q(G) = Q_\zer ^{\abs{L(G)}}$.  

If $P=0$, $G$ has no two-- or four--legged subgraphs associated to forks of 
$t$, so $n_\ph =0$. Since no $G_f$ is two--legged, $\cN = \emptyset $ or 
$\cN = \{ \ph \} $, depending on whether $\tilde G (\ph ) $ 
is \OL\ or not. Also, once $(i)$ -- $(iii)$ are proven, 
$(iv)$ is trivial since $Val (G^J)$ does not depend on $I$ 
at all for $P=0$. 
We note right away that the only places where $I$ 
will enter for $P > 0$ are in the values of two--legged subdiagrams through 
the lower limit of the scale sum for c--forks.
%%%%%%%%%%
\medskip\noindent
{\it Case 1: $P=0, s=0$ with $E(G) \geq 4$ or $E(G) =2$ and $\tilde G(\phi)$ \OL.}
By Theorem \IPCT, 
$$\eqalign{
\sum\limits_{J \in \cJ (t, j)} \abs{ Val (G^J )}_\zer\ \leq\ &
(4K_\zer)^{\abs{L(G)}} K_\one M ^{\ep j\,1(\ph \not \in \cN )}
\; M^{\ep j^* (\ph ) \, 1(\ph \in \cN ) } \cr
& M^{D_\ph j}\; \sum\limits_{J \in \cJ (t, j)}
\prod\limits_{f > \ph} M^{D_f (j_f - j_{\pi (f)} ) }
\prod_{V_4(G)}|\hat v|_\zer
\cr}\eqn $$
(see \queq{\Ddeldef} for the definition of the $D_f$).
Since there are no two-- or four--legged forks (except possibly $\phi$), 
$D_f \leq -1$ holds for all $f>\ph $. 
In the sum over $J \in \cJ (t, j)$, 
$j_f$ runs from $j_{\pi (f)}$ to -1, 
since there are no $c$--forks $f>\ph$ (the corresponding 
subgraph would be two--legged). Thus every scale sum is 
bounded by 
$$
\sum\limits_{j_f > j_{\pi (f) } } 
M^{D_f (j_f - j_{\pi (f)} ) }  \leq 
\sum\limits_{k \geq 0} M^{-k} \leq 
\; {1 \over 1 - M^{-1} } 
.\eqn $$
Doing the scale sums downwards from the leaves of $t$ in the 
standard way, we get a factor $(1-M^{-1})^{-1}$ for every fork 
of $t$, except for $\ph$. Since every fork $f$ corresponds to 
a subgraph of $G$, the number of forks is bounded by $\abs{L(G)}$.
Thus  
$$
\sum\limits_{J \in \cJ (t,j )} 
\abs{Val (G^J)}_\zer  \leq 
\left({4 K_\zer K_\one \over 1-M^{-1} } 
\right)^{\abs{L(G)}}
M^{D_\ph j} M^{\ep j \, 1(\ph \not \in \cN )} |\hat v|_\zer^{|V(G)|}
\eqn $$
Recalling that $D_\ph =2-m$ if $G$ has $2m$ external legs, 
and that 
$\ph \not \in \cN  \Longleftrightarrow \tilde G(\ph ) $ \OL, 
we obtain the statement for $s=0$. 
%%%%%%%%%%%%%%%
\medskip\noindent
{\it Case 2: $P=0,  s\in\{1, 2\}$ with $E(G) \geq 4$ or $E(G) =2$ and $\tilde G(\phi)$ \OL.}
Now we apply $s \leq 2$ derivatives with respect to the 
external momenta. The derivative can act on vertices (interaction lines) 
or on fermion lines in the spanning tree of the graph.
A bound for the number of targets for each derivative is 
thus $2|V(G)|-1$.  Because $G$ is connected, $|V(G)|\le|L(G)|+1$.
If the derivatives act on interaction lines, their effect 
can be  bounded by $\abs{\hat v}_s$. By Lemma \Simpl\ $(iii)$, the 
effect of $s$ derivatives acting on fermion lines can be 
bounded by an additional factor $W_s M^{-s j_\one }$ where $j_\one $ is the lowest scale at which the derivative acts. Moreover, the value of the 
differentiated graph can be bounded using Theorem \IPCT\ since 
all support properties remain the same as before.  

If $\tilde G(\phi) $ is \OL, we use $M^{-j_\one } \leq M^{- j} $
to bound, 
$$\eqalign{
\sum\limits_{J \in \cJ (t,j)} \abs{Val(G^J)}_s &\leq 
K_\one (4K_\zer)^{\abs{L(G)}} [(2|L(G)|+1)X]^2 |\hat v|_s^{|V(G)|} \cr
&\hskip.5in  M^{(D_\ph +\ep -s )j}
\sum\limits_{J \in \cJ (t, j)} \prod\limits_{f > \ph} 
M^{D_f (j_f - j_{\pi (f)} ) }
\cr
&\leq  
Q(G)|\hat v|_s^{|V(G)|} M^{(D_\ph +\ep -s )j}
\la_\zer (j, \ep ) 
\cr}\EQN\Epsbou $$
as before. 
If $\tilde G(\phi) $ is \NOL\ and $E(G) \geq 4$, a similar bound holds 
without the $\ep$. So far irreducibility has played no role, so $(iii)$ holds for $P=0$ even with $s=2$. 
%%%%%%%%%%%%
\medskip\noindent
{\it Case 3: $P=0,\ E(G) =2$ and $\tilde G(\phi)$ \NOL,\ $s\in\{1, 2\}$.}
Now $\tilde G (\ta_\ph )$ is a non--overlapping two--legged graph. 
It is 1PI because it is a quotient graph of the 1PI graph $G$.
Let $s \geq 1$. By Remark \NOLVal, the derivative 
does not act on lines of $\tilde G( \ta_\ph )$ but only on 
lines with scale $\geq j^*(\ph )$, where $j^*(\ph ) $ is 
the lowest scale above $\ta_\ph$, as in Theorem \IPCT,
so its effect can be bounded by a factor $X^s M^{-s j^* (\ph )}$. 
If $\ta_\ph =t$, then $j^* (\ph )=0$, and the derivative 
can act only on the interaction lines or lines of scale zero.
Otherwise, by Theorem \IPCT, we have a factor $M^{\ep j^* (\ph )}$.
Since $s \geq 1 \geq \ep$ and $0 \geq j^* (\ph ) \geq j$, 
$$
M^{-(s-\ep )j^*(\ph ) } \leq M^{(\ep -s ) j}
\eqn $$ 
and we again obtain the bound \queq{\Epsbou}. 
Note that if the derivative acts only on interaction lines, 
the bound is true since $ 1 \leq M^{(\ep -s )j}$.

%%%%%%%%%%%%
\medskip\noindent
{\it Case 4: $P=0,\ E(G) =2$ and $\tilde G(\phi)$ \NOL,\ $s=0$.}
We use the representation \queq{\Explicit} for $Val (G^J)$
and Lemma \besser. Pick a string $S_\one$ that contains a line of scale $j$. 
This is possible because $\tilde G(\phi)$ is \NOL. Recall that $j$ is 
the root scale, hence the lowest possible scale for any line of the graph.
Let 
$$\eqalign{
g(p_\one ,q)_{\al_\one \al_{m_\one} \be \be'} & = 
\sum _{(\al_i)_{i \not\in\{ 1, m_\one\}}}
\int \prod_{i=2}^{m_\one  -1} 
\Bigl(\dbar ^{d+1} p_i \bigl( S_{i}(p_i) \bigr) 
_{\al _{m_\one -1 +i} \al _{i}} \Bigr) \cr
& (\cU _{v_\one })_{\al_\one \ldots \al_{m_\one - 1} \be
\al_{m_\one }\ldots\al _{2m_\one -2} \be '}
(p_\one , \ldots , p_{m_\one - 1} , q,  
p_\one , \ldots ,p_{m_\one -1}) 
\cr}\EQN\needsNumber $$
then 
$$
V_j = \left( Val (G^J)\right)_{\be\be'} (q) = 
\sum_{\al,\al'} 
\int  \dbar^{d+1} p \bigl( S_\one (p) \bigr)_{\al\al'} 
g(p,q)_{\al' \al \be \be'} 
\eqn $$
The string $S_\one $ can contain only insertions at scale $j$, 
i.e.\ vertices with generalized self--contractions of scale $j$
because $P=0$. So 
$$
S_\one (p) = 
C_j (p_\zer , e(\ve{p} ))^m 
\;  C_{j+1} (p_\zer , e(\ve{p} ))^{n-m} 
 \prod_{w=1}^{n-1} T_w (p) 
\EQN\thisToo $$
with $m\geq 1$, and where the $T_w$ are values of 1PI two--legged subdiagrams 
with root scale precisely $j$, and which are \NOL\ on scale $j$  (those are 
not excluded by $P=0$ because we did not use normal ordering). 
In the notation of Lemma \besser, 
$$\eqalign{
V_j = \int\limits_0^\infty & r\; dr\int\limits_0^{2\pi} d\vphi \; 
(ire^{i\vphi})^{-n} \; f(M^{-2j} r^2)^m f(M^{-2j-2} r^2)^{n-m} \cr
& \int\limits_S d\om \; \left( \ph (r \cos \vphi, r\sin\vphi, \om )
- \ph (0,0,\om ) \right) 
\cr}\eqn $$
with 
$$
\ph (p_\zer, \rh, \om ) = J(\rh, \om ) g(p_\zer, \ve{p} (\rh, \om ), q)
\prod_{w=1}^{n-1} T_w (p_\zer, \ve{p} (\rh, \om ))
\eqn $$
(as in Lemma \besser, $\int d\vphi \; e^{-in\vphi } \ph (0,0,\om ) = 0$
because $\ph (0,0,\om ) $ does not depend on $\vphi )$. By Taylor 
expansion, 
$$\eqalign{
V_j  = \int\limits_0^\infty & r^2\; dr\int\limits_0^{2\pi} d\vphi \; 
(ire^{i\vphi})^{-n} \; f(M^{-2j} r^2)^m f(M^{-2j-2} r^2)^{n-m} \cr
& \int\limits_S d\om \; 
\int\limits_0^1 dt \; (\cos \vphi \;\del_\zer + \sin \vphi \;\del_\one ) 
\ph (t\, r \cos \vphi, t\, r\sin\vphi, \om ) 
\cr}\EQN\SoneLOOP $$
As in the proof of Lemma \besser, the extra factor of $r$ gained by 
Taylor expansion alone would improve the scale behaviour by a factor 
$M^j$. However, there are now derivatives acting on either $J$, or $g$, 
or one of the $T_w$'s. We consider all these cases separately. 

If the derivative acts on $J$, we use Lemma \uProps\ $(iv)$ to bound 
$\abs{J}_\one$. Moreover, we can use the IH (proven as Case 1) 
for the four--legged graph $F$ whose value is $g$ to bound 
$$
\abs{g}_\zer \leq Q(F) \abs{\hat v}_\zer^{|V(F)|}
\eqn $$ 
Thus this contribution to $V_j$ is 
$$\eqalign{
& \leq \abs{J}_\one \abs{g}_\zer\prod_{w=1}^{n-1} \abs{T_w}_\zer
\int\limits_0^\infty dr \; r^{2-n} \; f(M^{-2j}r^2) \int_S d\om \cr
& \leq M^{2j} M^{2n} \abs{J}_\one Q(F)\abs{\hat v}_\zer^{|V(F)|}
\prod_{w=1}^{n-1} \left( \abs{T_w}_\zer M^{-j}\right)
\cr}\eqn $$
The root scale behaviour of the two--legged graphs is (applying the IH
to their external vertex and the power counting bounds for the propagators)
$Q_w \abs{\hat v}_\zer^{|V(T_w)|}M^j$, so the statement follows for this term. 

If the derivative acts on one of the $T_w$'s, it can act only on an interaction 
line, or on a scale where the two--legged graph overlaps. Thus, bounding its
value by the IH (proven as Case 2 or 3), 
$$
\abs{T_w}_\one \leq M^{\ep j} Q_w\abs{\hat v}_\one^{|V(T_w)|}
\eqn $$
so the contribution from this term is bounded by 
$$
\leq M^{j (1+\ep) } M^{2n} \abs{J}_\zer Q(F)\abs{\hat v}_\zer^{|V(F)|}
\prod\limits_{w=1}^{n-1}\left( Q_w\abs{\hat v}_\zer^{|V(T_w)|}\right)
\eqn $$
If the derivative acts on $g$, it affects only $\cU_{v_\one}$. There, 
it can hit any line of scale $j^*$ or higher, where $j^*$ is the scale 
at which $G$ overlaps, or an interaction line. We now wish to use the argument
of Theorem \IPCT\ to extract the volume gain. The crucial step in this 
argument, as applied to the current situation, is bounding the two overlapping
momentum loop integrals
$$\eqalign{
Y =
\sup_{q,\phi,t}\int_{\cB} \hskip-5pt d^d \ve{p}_\one\int_0^\infty\hskip-7pt dr\int_Sd\om\ 
 & f(M^{-2j}r^2)1\big(\abs{e(\ve{p}_\one)}<M^{j^*}\big) \cr
& 1\big(|e(v_\one\ve{p}_\one+v_\two\ve{p}(tr\sin\phi,\om)+q))|<M^{j^*}\big) 
.\cr}\eqn $$
Here $\ve{p}_\one$ is the spatial momentum of a loop of $\cU_{v_\one}$, the
two factors $1(\ldots)$ come from the cutoffs on two lines of that loop,
the integrals over $r$ and $\om$ come from the momentum integral \queq{\SoneLOOP}
for the string $S_\one$ and the factor $ f(M^{-2j}r^2)$ comes from the cutoff of
one of the lines of $S_\one$. This two loop integral is not quite of the form of
Proposition \NoNest, with the most serious difference being the appearance
of $\ve{p}(tr\sin\phi,\om)$ in place of $\ve{p}(r,\om)$. 

However, writing $\ve{p}_\one = \ve{p} (\rh_\one, \om_\one)$, doing the 
same Taylor expansion as at the beginning of Appendix A, and using that
in the support of the integrand, $r \leq M^j \leq M^{j^*}$, we get
$$\eqalign{
1(|e(v_\one\ve{p}_\one 
& +v_\two\ve{p}(tr\sin\phi,\om)+q))|\le M^{j^*}) \leq
\cr
\leq & 1(|e(v_\one\ve{p} (0,\om_\one )+v_\two\ve{p}(0,\om_\two )+\ve{q})|
\leq (1+ 2 \sfrac{|e|_\one}{u_\zer})M^{j^*})
\cr}\eqn $$
so that 
$$\eqalign{
Y \leq \sup_{q,\phi,t}\int\limits_{-M^{j^*}}^{M^{j^*}} & d\rh_\one
\int\limits_0^{M^j} dr \int\limits_S  d\om_\one 
\abs{J(\rh_\one, \om_\one)} \int\limits_S d\om \cr
& 1(|e(v_\one\ve{p} (0,\om_\one )+v_\two\ve{p}(0,\om_\two )+\ve{q})|
\leq (1+ 2 \sfrac{|e|_\one}{u_\zer})M^{j^*}) \cr
&\le 2 \abs{J}_\zer M^j \; M^{j^*} \; 
W \left(  (1+ 2 \sfrac{|e|_\one}{u_\zer})M^{j^*}\right)
\cr}\eqn $$
with $W$ the function defined in Appendix A. So by Lemma A.1, the integral
is bounded by $C_{vol} M^{j}M^{j^*}M^{\ep j^*}$. Substituting this
into the proof of Theorem \IPCT\ , we find that the term in which 
the derivative acts on $g$ is bounded by 
$$\eqalign{
& \leq M^{\ep j^*} M^{-j^*} W_\one K_\one
\left(\sfrac{4K_\zer}{1-M^{-1}}\right)^{|L(F)|}\abs{\hat v}_\one^{|V(F)|}
M^{2n} M^{(3-n)j} 
\prod_{w=1}^{n-1} \abs{T_w}_\zer \cr
&\leq M^{(1+\ep)j} M^{2n} W_\one  K_\one
\left(\sfrac{4K_\zer}{1-M^{-1}}\right)^{|L(F)|}\abs{\hat v}_\one^{|V(F)|}
\prod_{w=1}^{n-1} \left( Q_w\abs{\hat v}_\zer^{|V(T_w)|}\right)
\cr}\eqn $$
Collecting the constants into the $U_s $ mentioned in Lemma \besser, 
this proves the statement for Case 4.
%%%%%%%%%%%%%%%%%
\medskip
Now assume $P \geq 1 $ 
and $(i)$ -- $(iv)$ to be proven for all $P^\prime < P$. 
Construct the graph $\Gp $ and its tree $\tp $ as in Remark \JFK.
Recall that, by construction, 
$\Gp $ has only two-- and four--legged vertices. 
Furthermore any two--legged subgraph corresponding to a fork must
be a string of two--legged vertices and any four--legged subgraph 
corresponding to a fork must consist of a single four-legged vertex with
some strings of two--legged vertices appended. 
Recall the definition of the vertex functions $F_w$ and the 
scale sums involved therein from Remark \JFK. 
By construction of $\Gp $, all graphs $\tG (t_w)$, 
whose values $V_w$ appear in the definition of $F_w$, 
are 1PI and two-- or four--legged, and they are of depth 
at most $P-1$, so the inductive hypothesis applies to them.
Our procedure is 
to estimate the norms of $\abs{F_w}_s$ for $s \in \{ 0,1\}$
(and, when $F_w$ is four--legged, for $s=2$)
first, using the inductive hypothesis, 
and then to apply this to complete 
the induction step using the case $P=0$, since by construction, 
$\Gp $ has depth zero. We abbreviate 
$Q_w = Q(\tG (t_w ))|\hat v|_\two^{|V(\tG (t_w ))|}$
and call $n_w=n_f$ if $w$ comes from the fork $f \in t$.

Let $F_w$ belong to a $c$--fork. Then 
$$
\abs{F_w}_s = \bigg|\ell \sum_{j_w = I }^{j_{\pi (w)}} V_w\bigg|_s
.\eqn $$
For $s=0$, by $\abs{\ell T}_\zer \leq \abs{T}_\zer $, 
the inductive hypothesis (IH), and Lemma \Factori\ $(ii)$,
$$\eqalign{
\abs{F_w}_\zer & \leq
Q_w \sum_{j_w = I }^{j_{\pi (w)}}
\la_{n_w} (j_w , \sfrac{\ep}{2} ) 
M^{j_w Y_\zer } \cr
&\leq 
2 Q_w 
\la_{n_w} (j_{\pi (w)} , \sfrac{\ep}{2} ) M^{j_{\pi (w)}(1+\ep )}
.\cr}\EQN\cForkopt $$
If $s=1$, we use $\abs{\ell T}_\one \leq (1+ d\abs{P}_\one )
\abs{T}_\one $, where $d$ is the spatial dimension, $P$ is the projection onto $S$, 
and $\abs{P}_\one = \max\limits_i \abs{P_i}_\one $, 
then 
$$\eqalign{
\abs{F_w}_\one & \leq
(1+ d\abs{P}_\one ) \sum_{j_w=I}^{j_{\pi (w)}} \abs{V_w}_\one
\cr
& \leq 
Q_w (1+ d\abs{P}_\one )  \sum_{j_w=I}^{j_{\pi (w)}}
\la_{n_w} (j_w , \sfrac{\ep}{2} ) 
M^{j_w Y_\one } 
\cr
& \leq 2 Q_w (1+ d\abs{P}_\one ) M^{\ep j_{\pi (w)}}  
\la_{n_w} (j_{\pi (w)} , \sfrac{\ep}{2} )
\cr}\EQN\cForkopts $$
by Lemma \Factori\ $(ii)$ since $Y_\one = \ep$. 

If $F_w $ belongs to an r--fork, 
$$
\abs{F_w}_s = \bigg| (1-\ell )\sum_{j_w > j_{\pi (w)}} V_w\bigg|_s
.\eqn $$
For $s=0$, by \queq{\RTaylor}, and because 
the momentum $p$ flowing through $\tG_w$ must be in supp 
$C_{j_{\pi (w)}}$, i.e. $\abs{ip_\zer - e(\ve{p} ) } \leq 
M^{j_{\pi (w)}+2}$, 
$$\eqalign{
\abs{F_w}_\zer & \leq
\sum_{j_w > j_{\pi (w)}} M^{j_{\pi (w)}+2} \;
{\sqrt{2} \over u_\zer } \abs{V_w}_\one \cr
& \leq 
Q_w {\sqrt{2} M^2 \over u_\zer }  
M^{j_{\pi (w)}} \sum_{j_w > j_{\pi (w)}} 
\la_{n_w} (j_w , \sfrac{\ep}{2} ) 
M^{j_w \ep } \cr
&\leq
2 Q_w {\sqrt{2} M^2 \over u_\zer }  
M^{j_{\pi (w)}} 
\la_{n_w} (j_{\pi (w)} , \sfrac{\ep}{2} ) 
\cr}\EQN\rForkest $$
by Lemma \Factori\ $(iii)$.

For $s=1$, we ignore the renormalization gain, and bound  
$$
\abs{ (1-\ell ) V_w}_\one \leq  (2+d\abs{\prP}_\one )
\abs{V_w}_\one 
\eqn $$
Inserting the IH, the scale sum is as in the $s=0$ case, and 
$$
\abs{F_w}_\one \leq
 (2+d\abs{\prP}_\one ) 2 Q_w
\la_{n_w} (j_{\pi (w)} , \sfrac{\ep}{2} ) 
.\eqn $$

If $F_w $ belongs to a \SSI, $j_w = j_{\pi (w)}$, and so 
$$
\abs{F_w}_s \leq Q_w \;
\la_{n_w} (j_{\pi (w)},\sfrac{\ep}{2})
M^{j_{\pi (w)} Y_s (\tG (t_w ))}
\eqn $$
for all $s \leq 2$ follows directly from the IH. 

If $F_w $ belongs to a four--legged fork of $t$, the IH implies
$$
\abs{F_w}_s \leq Q_w
\sum_{j_w > j_{\pi (w)}} 
\la_{n_w} (j_w , \sfrac{\ep}{2} ) 
M^{j_w Y_s (\tG (t_w)) }
.\eqn $$
Bound $M^{j_w Y_s} \leq M^{-s j_{\pi (w)}} M^{Y_\zer j_w} $. 
If $\tG (w) $ is \OL, $Y_\zer =\ep$, so by Lemma \Factori\ $(iii)$, 
$$
\abs{F_w}_s \leq 2 Q_w 
\la_{n_w} (j_{\pi (w)} , \sfrac{\ep}{2} ) 
M^{-s j_{\pi (w)}}
.\eqn $$
If  $\tG (w) $ is \NOL, $Y_\zer = 0 $, and the scale sum 
grows logarithmically, i.e.\ as $\abs{j_{\pi(w)}}$, and 
$$
\abs{F_w}_s \leq Q_w \la_{n_w+1} (j_{\pi (w)} , \sfrac{\ep}{2} ) 
M^{-s j_{\pi (w)}} 
.\eqn $$

In summary we have for $s\leq 1$ the bounds 
$$
\abs{F_w}_s \leq Q_w K_\two M^{ j_{\pi (w)} (1-s) }
\la_{n_w} (j_{\pi (w)} , \sfrac{\ep}{2} )
\EQN\TwoVert $$
for the vertex functions of two--legged vertices $w$ of $\Gp$, 
and for all $s \leq 2$ the bounds 
$$
\abs{F_w}_s \leq 2Q_w M^{ -s j_{\pi (w)} }
\la_{\tilde n_w} (j_{\pi (w)} , \sfrac{\ep}{2} )
\EQN\FourVert $$
for all four--legged vertices $w$ of $\Gp$, with 
$$
\tilde n_w = \cases{ n_w & if $\tG (w)$ is \OL \cr
n_w+1 & if $\tG (w)$ is \NOL 
\cr}\eqn $$

We return to $\Gp $ and complete the inductive step. 
Choose a spanning tree $\Tp $ as in Lemma \ImpPow\ $(i)$ for $\Gp $, 
and fix the momenta, 
to obtain $Val (\Gp ^J)$ in the form \queq{\FixVal}, with 
the $\cU_w$ given by the $F_w$ in the present case.
%%%%%%%%%%%%%%%%% 
\medskip\noindent
{\it Case 5: $P>0,\ E(G) \ge 2$ and $\tilde G(\phi)$ \OL,\ $s \in \{ 0,1\}$.}
Let $q$ be an external 
momentum of $G$, and denote $\del_\be = \del/\del q_\be$. 
Then 
$$\eqalign{
{\del_\be}^s Val (\Gp ^J) & = \sum_\si \sum_{spins \; \al} 
\int
\prod_{l \in L(\Gp ) \setminus L(\Tp ) } 
\dbar^{d+1}{p_l} 
\prod_{l \in L (\Gp )}
{\del_\be}^{\si_l}C_{j_l} ((p_l)_\zer , e ( \ve{p}_l ))_{A_l}
\cr
& 
\prod_{w \in V_4 (\Gp ) } 
{\del_\be}^{\si_w}
F_w (p^{(w)}_\one , p^{(w)}_\two , p^{(w)}_3)_{A_w}
\prod_{w \in V_\two (\Gp ) } 
{\del_\be}^{\si_w}
F_w (p^{(w)})_{A_w}
\cr}\EQN\DelBe $$
where the $A$ denote the spin assignments for vertex functions
and propagators, as defined in the graph rules. 
Here $\si : L(\Gp ) \cup V(\Gp ) \to \{ 0, 1 \}$ 
keeps track of which factor gets differentiated, so exactly 
one of its components is nonzero, 
$\sum_l \si_l + \sum_w \si_w =1$.
To count the number of terms in the sum over $\si$, observe 
that $\del_\be C_{j_l} =0$ if $l \not\in \Tp $. Since 
$ \abs{L(\Tp )} = V(\Gp ) -1$, 
the sum over $\si $ is bounded by 
$2 \abs{V (\Gp )}$ times the maximum of the summand. 
$\tG (\ph )$ is a quotient graph of $\Gp$, so $\Gp $ is \OL.
Applying Lemma \ImpPow\ $(ii)$, using   
\queq{\TwoVert}, \queq{\FourVert}, and bounding 
$M^{-s j_{\pi (w)} } \leq M^{-sj}$,  
$$\eqalign{
\max\limits_{\be} \abs{\del_\be Val (\Gp^J)}  \leq 
& K_\one (4 K_\zer )^{\abs{L(\Gp )}} 2 \abs{V(\Gp )} X 
\prod_{w \in V(\Gp )} Q_w \cr
 & M ^{j (\ep + D_\ph (\Gp ) -s )} 
\; 
\prod\limits_{f \in \tp \atop f > \ph } 
M^{D_f(\Gp ) (j_f - j_{\pi (f)} ) }
\prod\limits_{w \in V_\two(\Gp )} 
( K_\two M^{j_{\pi (w)}} )
\cr
& 
\prod\limits_{w \in V_4(\Gp )} 
\big(2\la_{\tilde n_w} (j_{\pi (w)},\sfrac{\ep}{2})\big)
.\cr} \EQN\Zwis $$ 
By \queq{\Ddeldef}, since $\Gp$ 
has two-- and four--legged vertices, $V(\Gp_f) = V_\two (\Gp_f ) + 
V_4 (\Gp_f )$, so 
$$
D_f (\Gp ) = L( \Gp ) - 2 (V(\Gp) -1) = 
{1 \over 2} (4 - E(\Gp_f)) - V_\two (\Gp_f)
.\eqn $$
By Lemma \Factori\ $(i)$ and the definition of $n_\ph $,  
$$
\prod\limits_{w \in V_4(\Gp )} 
\la_{\tilde n_w} (j_{\pi (w)},\sfrac{\ep}{2})
\leq
\la_{n_\ph} (j, \sfrac{\ep}{2} ) 
.\eqn $$
Using the telescope formula $(j_\ph=j)$
$$
j_{\pi(w)} = j_\ph + 
\sum_{f \in \tp \atop \ph < f \leq \pi (w)} 
(j_f - j_{\pi (f)}) 
= j_\ph + 
\sum_{f \in \tp \atop f > \ph } 
(j_f - j_{\pi (f)}) \;
1(w \in \Gp_f^J )
\eqn $$
we get
$$
\sum_{w \in V_\two (\Gp ) } j_{\pi (w)} = 
j_\ph \abs{V_\two(\Gp )} + 
\sum_{f \in \tp \atop f > \ph } 
(j_f - j_{\pi (f)})
\sum_{w \in V_\two (\Gp ) } 1(w \in \Gp_f^J )
\eqn $$
so  
$$
\prod_{w \in V_\two (\Gp ) }
M^{j_{\pi (w)}} = 
M^{j_\ph \abs{V_\two(\Gp )} } 
\prod_{f \in \tp \atop f > \ph }
M^{(j_f - j_{\pi (f)}) \abs{V_\two(\Gp_f^J )} }
\eqn $$
and we see that all $D_f(\Gp )$ get renormalized to 
$$
D_f^{(R)}(\Gp) = 
D_f(\Gp ) + V_\two (\Gp_f ) = {1\over 2} (4 - E(\Gp_f ))
.\eqn $$
Inserting these estimates into \queq{\Zwis}, summing over 
$J \in \cJ (t,j)$ and remembering \queq{\GGprel},
$$\eqalign{
\sum_{J \in \cJ(t,j)} \abs{Val (G^J)}_s 
\leq & 
(2 d\abs{V(\Gp )} X +1 )K_\one 
(4 K_\zer ) ^{\abs{L(\Gp )}} 
{K_\two}^{\abs{V (\Gp )}}
\prod_{w \in V(\Gp )} Q_w  \cr
&M^{j (\ep -s + D_\ph^{(R)} (\Gp )) }
\la_{n_\ph} (j, \sfrac{\ep}{2}) 
 \sum_{J \in \cJ(\tp ,j)}
\prod_{f \in \tp \atop f > \ph }
M^{ D_f^{(R)} (\Gp ) (j_f - j_{\pi (f)})}
.\cr}\eqn $$
We are now back to the case of zero depth, since 
by construction, if $f$ is a  two--legged or four--legged fork
for $\Gp $, then $j_f=j_{\pi(f)}+1$ by conservation of momentum. So there
is no corresponding scale sum and
the last scale sum is now identical to that of the case with 
zero depth. It produces a factor
$(1-M^{-1})^{-1}$ for all forks of $\tp$ with $E(G_f)>4$, except for $\ph$. 
The product of the various constants is now again bounded 
by $Q(G)$ since the number of lines of the subgraphs $G_w$
and that of $\Gp$ add up to $\abs{L(G)}$. 
Finally, we note that $D_\ph^{(R)} (\Gp ) =
{1 \over 2} (4 - E(\Gp )) = 2-m$. This proves $(i)$ and $(ii)$ for $s\leq 1$ and also $(iii)$ since we have used 
the assumption that $G$ is 1PI only to bound 
$\abs{V(G)} \leq \abs{L(G)}$. For general connected graphs, 
$\abs{V(G)} \leq \abs{L(G)}+1$, which only changes the constant.
For $s=2$ we will need the 1PI assumption 
since we cannot afford to have two derivatives acting on 
$c$-forks.
%%%%%%%%%%%%%%%%%%%
\medskip\noindent
{\it Case 6: $P>0,\ E(G) \ge 2$ and $\tilde G(\phi)$ \OL,\ s=2.}
For $s=2$, we have to apply another momentum derivative to 
\queq{\DelBe}. It can act again on at most $2V(\Gp )$ targets. 
However, we have to avoid having two derivatives act on an $F_w$
coming from a c--fork because the scale sum down to $I$ 
would diverge as $I \to -\infty $ in that case. 
Whenever the second derivative acts on the same two--legged vertex 
as the first (no matter whether this vertex comes from a 
$c$-- or an $r$--fork), we remove
it by integration by parts as follows. Since $G$ is 1PI, so is 
$\Gp$, so the momentum through any two--legged vertex is a 
linear combination of momenta, at least one of which is 
a loop momentum $p$. So we can rewrite the derivative 
$$
{\del \over \del q_\ga } T(p \pm q ) = 
\pm {\del \over \del p_\ga } T(p \pm q )
\eqn $$
Integrating by parts with respect to $p$ distributes the 
derivative on at most $\abs{V}-1$ other lines and at most 
$V$ vertices. 
Using \queq{\TwoVert} and \queq{\FourVert}, the estimate 
follows as in the case $s \leq 1$, but with the constant $(2 \abs{V} X)$
replaced by $(2 \abs{V} X)^2$
because there are more terms in the sum, 
and because the derivatives can now act on two $C$'s. 
%%%%%%%%%%%%%%%%% 
\medskip\noindent
{\it Case 7: $P>0,\ E(G) \ge 4$ and $\tilde G(\phi)$ \NOL.} Just delete all the
$M^{\ep j}$'s from cases 5 and 6.
%%%%%%%%%%%%%%%%% 
\medskip\noindent
{\it Case 8: $P>0,\ E(G) =2$ and $\tilde G(\phi)$ \NOL,\ s=0.}
We proceed as in the case $P=0$. The value of $G$ again takes 
the form of \queq{\needsNumber} -- \queq{\thisToo}, but with the 
$T_w$ replaced by 1PI insertions with values $F_w$ (see \queq{\Fw})
on the strings $S_i$ of $G(t_\phi)$, 
where $F_w$ may now belong to a $c$--fork, an $r$--fork, or an SSI. 
If $F_w$ belongs to a $c$--fork, the additional $M^{\ep j}$ can be read
off \queq{\cForkopt}. Since the scale of an SSI is fixed, the additional
$M^{\ep j} $ follows directly from the IH. Thus if any $c$--fork or SSI 
is on the string, the statement follows immediately. 

There remains the case where all insertions on the string are $r$--forks, 
i.e.\ $\cP_w = 1-\ell $ in \queq{\Fw}. Then the value of $G$ is given by
\queq{\noSup}, with $k=n-1$ (only $r$--forks), and with $\phi$ and $\De_w$ 
given by \queq{\thatsPhi} and \queq{\thatsDelta}. 
The derivative $\sfrac{\del}{\del r} $ acting on $\phi $ in
\queq{\noSup} can act on $J$, or on $\ga$, or on the $\De_w$, 
similarly to Case 4 above.

If the derivative acts on $J$, $\abs{J}_\one \leq \sfrac{A_\one}{u_\zer^2}$, 
and by Theorem \TwoFour\ $(i)$ (proven as Case 5 and 6 above),
$$
\abs{\ga}_\zer \leq \abs{g}_\zer \leq Q(F) \abs{v}_\two^{\abs{V(F)}}
M^{\ep/2} \la_{n_f} (j,\sfrac{\ep}{2})
\eqn $$
with $F$ the four--legged graph whose value is $g$. The insertions 
to which the $\De_w$ belong are of depth $\leq P-1$, so by the IH
and \queq{\thatsDelta}
$\abs{\De_w}_\zer$ is bounded. Thus this contribution is bounded 
by $Q(G)   \abs{v}_\two^{\abs{V(G)}}\la_{n_\ph} (j,\sfrac{\ep}{2}) M^{2j}$.

The case of the derivative acting on $\ga (r, \vphi, \om ) = g(r \cos \vphi, 
\ve{p}(r \sin \vphi, \om ))$ is completely similar to that given in Case 4.

If the derivative acts on one of the $\De_w$, we apply the IH for $s=2$
to the two--legged graph of depth $\leq P-1$ that produces $\De_w$. 
The gain of $M^{\ep j}$ follows immediately from the IH. This completes the
induction step for Case 8.

%%%%%%%%%%%%%%%%%% 
\medskip\noindent
{\it Case 9: $P>0,\ E(G) =2$ and $\tilde G(\phi)$ \NOL,\ $s\in\{1,2\}$.}
If $s=1$, we choose the spanning tree constructed in Lemma \MomRout, 
then the derivative with respect to the external momentum 
can act only on the vertex function $\cU_{v_\one}$. Since the 
volume improvement is at the same scale, the statement follows
even without an application of Lemma \besser.
If $s = 2 $, at most one derivative
can act below $j^* (\ph )$, and that happens only if it has to be
rerouted to avoid having two derivatives act on a single $c$--fork. 
Bounding it by $M^{-j}$, and then proceeding as in the case 
$s=1$, we arrive at a similar bound.  

\medskip
Finally, $(iv)$. The effective vertices in $\Gp $ 
have vertex functions that depend on $I$, 
and that converge as $I \to -\infty$ by the IH. 
Since the scales of r-forks and four--legged diagrams 
are summed over a region that does not depend on $I$, 
and since a \SSI\ has no scale sum at all, 
$(iv)$ will be proven if we can show 
that the scale sum for a c-fork, 
which runs from $I$ to $j_{\pi (f)}$, 
also converges as $I \to -\infty$. 
Let $\cC$ be the Banach space 
$(C^1 ([-1,1]\times \cB , \C ), \abs{ \cdot }_\one )$. 
The sequence $g^I=(g_n^I)_{n \leq 0}$ given by 
$$
g_n^I  =\cases{\sum_{J \in \cJ (t_w , n )}  Val (G_w^J)   &
if $n \in \{ I, \ldots , j_{\pi (w)} \}$ \cr
0 & otherwise \cr} 
\eqn $$
is an element of the space $\ell^1 ( \Z_- , \cC )$ by $(i)$. 
By the IH applied to $G_w$, there is $(g_n)_{n \leq 0}$ such that 
$g_n^I \to g_n $ in $\abs{\cdot }_\one$ as $I \to - \infty $, 
in other words, $g^I \to g$ pointwise as a sequence. 
Let $f \in \ell^1 ( \Z_- , \cC )$ be the sequence 
$f_n = M^{\ep n} Q(G_w) \la_{n_w} (n, \ep/2 )$, then 
$$
\norm{g^I}_{ \ell^1 ( \Z_- , \cC )} 
\leq \norm{f}_ {\ell^1 ( \Z_-  )}
\eqn $$
for all $I < 0$. By dominated convergence, 
$g \in  \ell^1 ( \Z_- , \cC )$ and 
$$
\norm{g^I -g}_{\ell^1 ( \Z_- , \cC )} = 
\sum\limits_{n \leq 0} \abs{g_n^I -g_n}_\one \to 0
\eqn $$ 
as $I \to -\infty$, so 
$\sum\limits_{n \in \{ I, \ldots , j_{\pi (w)} \}} F_n $
converges in $\abs{\cdot}_\one$, 
which shows that $\ga_j^I = \sum\limits_{J \in \cJ (t,j)} Val (G^J)$
converges in $\abs{\; \cdot \; }_\one$ as $I \to -\infty$. 
Moreover, if $G$ itself is two--legged and 1PI,  
$$
\abs{\ga_j^I} \leq \vphi_j = M^{\ep j} Q(G) \la_{n_\ph } (j , \ep/2)
.\eqn $$
Now repeat the dominated convergence argument for the $\ga_j^I$
to see that $\sum\limits_{j \geq I} \ga_j^I$  also converges as 
$I \to -\infty $.  
\endproof
Theorem \TwoFour\ contains the most important information, that
of the renormalization flow of the two--and four--legged, i.e.\
relevant and marginal, effective vertices. For \OL\ four--legged 
graphs, the bounds show that the scale behaviour is not marginal, 
but irrelevant in the usual language of the renormalization group. 
The convergence as $I \to -\infty$ allows us to view the 
flow of effective actions to the uncutoff limit 
$G_{j,2m,r} = \lim\limits_{I \to -\infty}G^I_{j,2m,r}$.
\The{\AllGF} {\Thsty 
Let $\abs{ \; \cdot\; }'$ be as in \queq{\priNormdef}, 
$n_\ph $ as in Remark \JFK, and let $(t,G)$ be fixed. 
Let
$$
\norm{\;\cdot\; } = \cases{
\abs{\;\cdot\; }_\one & if $E(G)=2$ and $G$ is 1PI \cr
\abs{\;\cdot\; }_\zer & if $E(G)=2$ and $G$ is 1P--reducible, \cr 
                      & or $E(G)=4$ and $\tG (\ph )$ is \OL \cr
\abs{\;\cdot\; }' & otherwise \cr}
\eqn $$
Then $V_I (t,G) = \sum\limits_{j=I}^{-1} \sum\limits_{J \in \cJ (t,j)} 
Val (G^J) $ converges in $\norm{\cdot}$ and satisfies
$$
\norm{\lim\limits_{I \to -\infty} V_I(t,G)} \leq n_\ph ! \Const^{\abs{L(G)}}
.\eqn $$
}
\Proof
For $E(G)=2$, or $E(G)=4$ and $\tG (\ph )$ \OL, the statement follows 
from Theorem \TwoFour\ by summation over $j$, noting that for $\al > 0$, 
$$
\sum\limits_{j < 0} M^{\al j} \la_n (j, \ep/2) \leq 2\la_n(0,\ep/2)
\le \Const^n n!
\eqn $$
It remains to show the bound in $\abs{\cdot}'$.
Construct $\tG $ and $\tilde t$ as in Remark \JFK. 
The two--legged vertices of $\tG$ are either strings of vertices of $G'$ 
or $c$--forks. By Theorem \TwoFour, the scale behaviour is 
$T_w \leq \Const^{L_w} \la_{n_w} (j_{\pi (w)},\ep/2) M^{j_{\pi (w)}}$ 
for two--legged vertices and 
$F_w \leq \Const^{L_w} \la_{\tilde n_w} (j_{\pi (w)},\ep/2) $ 
with $\tilde n_w = n_w $ if $\tG (w) $ is \OL\ and 
$\tilde n_w = n_w +1$ if it is \NOL.
Inserting the second part of Lemma \Powco\ and Lemma \Factori\ $(i)$, 
it follows now as in \quref{FT1,FT2} that the scale sums 
$\sum\limits_{J \in \cJ (t,j)} Val (G^J) $ converge as $I \to \infty$, 
and that they are uniformly (in $I$) bounded by a summable function in $j$.  
The convergence as $I \to -\infty$ now follows by imitating the proof
of Theorem \TwoFour\ $(iv)$, using $L^1 (([-1,1]\times \cB)^{n-1} \times
\{ \uparrow, \downarrow \} ^n , \C  )$ instead of $\cC$. \endproof

\Rem{\dummy} The convergence statements of Theorem \Renorm\ follow 
directly from this, recalling that the graphs contributing to $\Si$ 
and $K$ are 1PI and two--legged, and that the sum over trees $t$ at 
fixed $G$ is always finite. Under the hypotheses of Theorem \Nice,
$n_\ph = 0$ for all $t$ that are compatible with $G$, so, taking into 
account the $1/n_f!$ to bound the sum over trees by $\Const ^r$, 
Theorem \Nice\ also follows. The local Borel summability bound 
requires an adapted induction scheme that combines the summation 
over trees with the bounds of Theorem \TwoFour, using Felder's Lemma. 
We will not repeat the proof here; it is similar to the one given in 
\quref{FT1}. 

\Rem{\rubitin} Theorem \AllGF\ states that the value of every 
four--legged graph that is \OL\ on root scale converges in the sup norm
to a continuous function. The only four--legged graphs that may produce 
a singularity in the four--point function are thus the \NOL\ four--legged
graphs. By Lemma \FourL, these are the ladder diagrams. 
The `dangerous divergences' mentioned in many places in the literature 
are those of the four--point function. Their `danger' is that they 
can produce factorial growth of the value of individual diagrams
when they appear as subdiagrams and thus may prevent convergence of 
the renormalized expansion in $\la$ (even though every order is now finite). 
Theorem \AllGF\ shows that for 
our class of models with a non--nested Fermi surface,  
these `dangerous divergences' can only be produced by dressed ladder diagrams,
so that it suffices to investigate them to see whether $r$ factorials 
in the value of individual diagrams of order $r$ appear.

\vfill\eject
\chap{The Derivative with Respect to the Band Structure}
\noindent
Let $D_h$ be the directional derivative with respect to $e$, 
as defined in \queq{\Dhdef}. 
It is obvious from the formula for the value of graphs and the
way $e$ appears in the propagators and in the projection that for
a fixed infrared cutoff $I > - \infty $, all Green functions
have bounded $D_h$, and moreover, their multiple derivatives with respect
to $e$\  exist as multilinear operators. However, the norms of these 
operators diverge as $I \to -\infty$. In this chapter we show that 
$D_h K^{I}_r (e)$ converges as $I \to -\infty$, and that 
$K_r = \lim\limits_{I \to -\infty} K^I_r $ is differentiable in $e$
in the sense of Fr\' echet. 
To get bounds that are suitable for removal of the cutoff $I$, 
we have to rearrange some contributions that appear divergent at first.

To motivate why there is a problem taking this derivative, 
we first explain how it affects power counting.  Abbreviating
$f( M^{-2j} ( p_\zer ^2 + e(\ve{p} )^2 ) ) =f_j (p)$, the derivative
$$
D_h C_{j}( p_\zer , e (\ve{p})) = \left(
{f_j (p) \over ( ip_\zer -e(\ve {p})) ^2}
+ {2M^{-2j} e(\ve {p}) f_j^\prime (p) \over ip_\zer - e(\ve{p}) }
\right) h (\ve{p} ) 
\EQN\Cdot $$
obeys
$$ \abs{D_h C_{j}( p_\zer , e (\ve{p})) } \leq \Const M^{-2j} 
1(\abs{ip_\zer - e(\ve{p} )} \in [M^{j-2},M^j ] )\abs{h}_\zer
\eqn $$
which is a factor $M^{-j}$ worse than the usual scaling behaviour of $C_j$
(Lemma \Simpl\ $(iii)$).
By power counting, a two-legged graph on scale $j$ behaves as $M^j$, 
so $D_h$ removes the decay and seems to make the scale sum
marginally divergent. Similarly, $D_h$ also acts on the projection $\ell $
and can upset renormalization cancellations. 

In brief, taking a derivative of $1/(ip_\zer -e(\ve{p}))$ effectively 
produces a square of the denominator, which, as discussed in the 
introduction, is not locally integrable. The problem that the infrared 
singularity gets stronger when derivatives are applied also appears, for
example, in Euclidean field theory with propagators singular at zero momentum, 
when differentiating with respect to a mass in the infrared.

However, the singularity of the propagator is on a surface in our case 
and this makes a big difference under the non--nestedness assumption 
\AThr. The improved power counting estimate
implies that for contributions from graphs that are \OL\ on root scale, 
the scale sum is actually still convergent
because of the volume improvement factor $M^{\ep j}$, so that for these
graphs, $\sum\limits_J D_h Val (G^J)$ converges. 
For contributions from \NOL\ graphs,
one has to apply an integration by parts similar to Lemma \besser\ 
to show that the scale sum still converges. So the derivative of $K$
is convergent without any further insertions or counterterms,
because of the geometry of the singularity.  
The two above observations will be used to treat general
labelled graphs using Lemma \Treedec. Note that in contrast to
derivatives with respect to the external momentum, where momentum routing
implied that lines in the \NOL\ parts of the graph are never hit by such a
derivative, derivatives with respect to $e$ will affect all lines, 
and it requires
a separate argument to remove $D_h$ from lines in the \NOL\ parts of the 
tree. When taking norms, there will be several subtleties which we discuss 
in detail below.

\sect{Integration by Parts}
We start with the simple observation that if $F \in C^1(\R ,\C )$,
then $F\circ e : \cB \to \C $ satisfies
$$
D_h F \bigl( e (\ve{p} ) \bigr) = F^\prime \bigl( e (\ve{p} ) \bigr)
h (\ve{p})
\eqn $$
and
$$
\nabla F \bigl( e (\ve{p}) \bigr) = F^\prime \bigl( e (\ve{p} )
\bigr) \nabla e (\ve{p})
\eqn $$
Thus, choosing $\de$ as in Lemma \uProps (iii), for
$\ve{p} \in \cU _{\de } (S)$,
$$
D_h (F \circ e) = \left( {h \over \cD _{u} e} \cD _{u} F \right) 
,\eqn $$
where $\cD _{u} = u \cdot \nabla $, as in Lemma \Lieder.
Since supp $C_{j} \subset \cU _{\de} (S)$ for
all $j \le -1$,
$$
D_h C_j \bigl( p_\zer , e (\ve{p} ) \bigr) =
{h(\ve{p} )\over \cD_u e (\ve{p}) } (\cD_u C_j )
\bigl( p_\zer ,e(\ve{p}) \bigr)
,\EQN\Cjdot $$
and this rewriting introduces no singularities since
$\abs{u \cdot \nabla e (\ve{p})} \ge u _\zer$.
Obviously, then, for $X \in C^1 ( [-1,1] \times \cB, \C )$,  
$$
\int d^d \ve{p}\ X (\ve{p}) D_h C_j \bigl( p_\zer , e(\ve{p})\bigr) 
= - \int d^d \ve{p}\ C_j \bigl( p_\zer ,e(\ve{p} )
\bigr) \nabla\!\cdot\! \biggl( {h u \over \cD _{u} e} X \biggr) (\ve{p})
\EQN\threesix $$
and thus
$$
\int _{\cB } D_h (C_{j} X) = - \int _{\cB } C _{j} X 
\nabla\!\cdot\! \biggl( { h u \over \cD _{u} e} \biggr)
+ \int _{\cB} C _{j} \de_h X
\EQN\IBP $$
where
$$
\de_h X = D_h X - {h \over \cD _{u} e } \cD _{u} X
.\EQN\deGl $$
By the definition of $\de_h $, $\de_h F = 0 $ for all 
$F(p)=b\big(p_\zer,e(\ve{p})\big)$ and all $h$. In particular  
$\de_h C_j(p) = 0 $. Note, however, that
for $F(p) = C_{j} \bigl(p_\zer + q_\zer , e(\ve{p} + \ve{q}) \bigr)$,
$\de_h F$ will not be zero if $q \ne 0$.
We shall only use integration by parts for \NOL\ graphs, so by Remark
\NOLVal\ such shifts by additional momenta $q$ will not occur. Moreover,
$X$ will be given by \queq{\Strgdef}, so, to get the integration by
parts-formula, we only have to give the derivative of $\ell$.

\Lem{\Pdot} {\Lesty 
Let $u$ be fixed independently of $e$, so that $D_h u =0$.
Then 
$$\eqalign{
D_h \prP(q) &= -{h\big(\prP(q)\big) \over \cD _{u} e\big(\prP(q)\big)} u\big(\prP(q)\big)
=- \ell \biggl( {h u \over \cD _{u} e} \biggr) (q)\cr
D_h(\ell T) &= \ell \de_h T
}\EQN\lTdot 
$$
On $U_{\de} (S)$, $\de_h \ell = \ell \de_h $ and thus 
$\de_h (1 - \ell ) = (1-\ell )\de_h $. }

\Proof Fix $\ve{q} \in \cU _{\de_\zer} (S)$. Changing $e$ to $e+\al h$
moves the Fermi surface. The new suface is $\tilde S = \{ \ve{p} : (e + \al h)
(\ve{p} ) = 0 \} $. $h$ is bounded, so for $\al$ small enough, $\tilde S
\subset \cU _{\de} (S)$. By assumption, changing $e$ does not change
the curve $\ga$ used to define $\prP (\ve{q})$ since $\ga$ is an integral curve
determined by $u$ and $\ve{q}$. What changes is the intersection point
of $\ga$ with $\tilde S$. Since this point moves on $\ga$, and 
$\ga $ is an integral curve of $u$,  
$$
D_h \prP (\ve{q} )= \be \;  u \bigl( \prP (\ve{q}) \bigr)
\eqn $$
with a function $\be = \be (e,h,u,\ve{q})$.
Inserting this into the equation 
$$
0= D_h (e (\prP (\ve{q} )) = h(\prP(\ve{q})) + \nabla e (\prP(\ve{q}))
\cdot D_h \prP (\ve{q} )
,\eqn $$
we obtain $\be = - \ell \bigl(
{h \over \cD_u e} \bigr)$ and thus the statement for $\prP $.

Since $(\ell T) (q) = T\bigl( 0, \prP(\ve{q} ) \bigr)$,
$$\eqalign{
D_h \ell T (q) &= (D_h T) \bigl( 0, \prP (\ve{q} ) \bigr) + \nabla T
\bigl( 0, \prP (\ve{q}) \bigr) \cdot D_h \prP (\ve{q} ) = \cr
& = (\ell D_h T + \ell \nabla T \cdot D_h \prP ) (q) 
\cr}\eqn $$
which, by the formula for $D_h P$, implies \queq{\lTdot}.
On $U_{\de} (S)$,
$\cD_u \ell = 0$ by Lemma \Lieder. 
Thus for $\ve{q} \in U_{\de _\zer} (S)$,
$$
(\de_h \ell T) (q) = (D_h \ell T) (q) = (\ell \de_h T) (q)
.\eqn $$
The statement for $1-\ell $ is obvious. \endproof

\Rem{\dummy} Allowing $u$ to vary with $e$ would have given an
additional term parallel to the surface. Since that term vanishes
linearly on $S$, it can be included without any problems, but the
resulting expressions are more complicated. For our purposes, keeping
$u$ fixed is enough. Indeed, for the bound without $\nabla$ acting
on $h$ it is necessary since the additional term contains $\nabla h$.
\someroom
To use this for strings of two--legged insertions, 
we write the string of \queq{\Strgdef}
as $S_i (p) = C _{j_{i,1}} (p) \cdot Y_i (p)$ with
$$
Y_i = \prod ^{w_i}_{k=1} (\cP _{i,k} T _{i,k} ) C_{j_{i,k+1}}
\EQN\Ydef $$
and note that for $j_{i,k} \le -1$, the momentum $\ve{p} \in \cU_
{\de_\zer} (S)$, so that Lemma \Pdot\ applies. Since
$\de_h C_{j_{i,k}} = 0$ for all $k$,
$$
\de_h Y_i = \sum ^{w_i} _{l =1} \prod^{w_i} _{k=1} ( \cP _{i,k}
T^\prime_{i,k,l}) C_{j_{i,k+1}}
\eqn $$
where
$$
T^\prime _{i,k,l} = \cases{\de_h T_{i,k} & if $ k=l $ \cr
T_{i,k} & if $ l \neq l $ \cr }
\eqn $$
We denote
$$
S^\prime _i (p) = C_{j_{i,1}}(p) \cdot \de_h Y_i (p)
.\eqn $$
\Lem{\GSTdot} {\Lesty 
Let $G$ be a \NOL, 1PI, two--legged graph as in 
Lemma \MomRout\ and $v_\one$ its external vertex. Then 
$$\eqalign{
D_h \bigl( Val \, & G^J (C, \cU_{v_\one} , \ldots , \cU_{v_n} )
\bigr)_{\be \bep} (q) =
\left( Val \, G^J (C, D_h \cU_{v_\one} , \ldots , \cU_{v_n} )
\right)_{\be \bep} (q) + \cr
&+ \sum_{l=1}^{n_\one -1}  
\sum_{(\al_i)_i}
\int\prod\limits_{i=1\atop i \neq l}^{n_\one -1} 
\left( \dbar^{d+1}p_i\, (S_i (p_i))_{\al_{n_\one +i} \al_i} \right)
\cr
& \left( (S_l^\prime (p_l))_{\al_{n_\one+l}\al_l} 
- (S_l(p_l))_{\al_{n\one+l}\al_l}
\nabla \cdot \left( {h u \over \cD_u e} \right) (p_l)
- {h \over \cD_u e} u(p_l) \cdot \nabla_{p_l}
\right) \cr
& \left( \cU_{v_\one}\right)
_{\al_\one\ldots\al_{n_\one -1}\be 
\al_{n_\one+1}\ldots \al_{2n_\one-1}\bep}
(p_\one, p_\two, \ldots , p_{n_\one -1}, q , 
p_\one, p_\two , \ldots , p_{2n_\one -1})
\cr}\EQN\GSTderiv $$
}
\Proof  For every string $S_i$ attached to $v_\one$, use 
\queq{\Ydef} and integration by parts
$$\eqalign{
&\int dp_\ell\ X(p_\ell)\,D_hS_\ell(p_\ell)
=\int dp_\ell\ X(p_\ell)\,\de_hS_\ell(p_\ell)
+\int dp_\ell\ X(p_\ell){h(p_\ell) \over (\cD_u e)(p_\ell)} u(p_\ell)\!\cdot\!
\nabla S_\ell(p_\ell)\cr
&=\int dp_\ell\ X(p_\ell)S'_\ell(p_\ell)
-\int dp_\ell\ X(p_\ell)S_\ell(p_\ell)\nabla\!\cdot\!{h\,u \over \cD_u e} 
-\int dp_\ell\ S_\ell(p_\ell){h\,u \over \cD_u e}\!\cdot\!\nabla X(p_\ell)\cr
}$$ 
to remove all derivatives from propagators. The momentum derivatives
acting on $\cU_{v_\one}$ and the additional summands all arise
from integration by parts. \endproof
\Rem{\Meaning} Less formally, one can say that after applying integration 
by parts, no derivative acts on a propagator of $G$, 
and that all derivatives act 
as $\de_h $ on functions associated to higher forks in the tree, i.e.
one of the $T_{i,k}$ in the strings gets differentiated in the 
terms in \queq{\GSTderiv}. If $T_{i,k} = T$ is again a \NOL\ graph, 
it is a string of GST graphs; since $\de_h C_j =0$, the derivative 
only affects 1PI subgraphs, which are then GST graphs, and we can 
again apply Remark \NOLVal\ and Lemma \GSTdot. For the momentum 
derivative contained in $\de $, we will apply Theorem \TwoFour\ 
directly. For the $D_h$ contained in $\de_h$, we apply 
\queq{\GSTderiv} again, to avoid having the derivative act on lines. 
This procedure can be iterated according to the recursive structure of 
the GST graph $G$, and all of $\cU_{v_{i_\one}}, \ldots , 
\cU_{v_{i_r}}$ get differentiated in this procedure. This can be used 
to make all derivatives act on higher, \OL\ parts of the tree, where 
the factors $M^{-j}$ they produce are controlled by the 
improved power counting factors $M^{\ep j}$. So, the upshot of 
\queq{\IBP} is that things can be arranged such that the derivative 
$D_h$ also does not act on lines in the \NOL\ part of $t(G^J)$ 
(as was the case for derivatives with respect to external momenta).
 However, because of 
the way integration by parts was done here, 
the price paid for this is that $\abs{\nabla h }_\zer$, not only 
$\abs{h}_\zer$, appears in the bound. The integration by parts is 
similar to the Taylor expansion in Lemma \besser, which also 
produces $\nabla h$ terms when used on a string on which a factor 
of $h$ from a $D_h C_j$ sits. 

\sect{Bounds for the Directional Derivative}
\noindent
We now show convergence of the directional derivative and a bound 
that contains only $\abs{h}_\zer $. Some parts of the proof will 
be subtle, and therefore we illustrate the two procedures for the
lowest-order contribution, 
$$
F(q) = \sum\limits_{j=I}^{-1} F_j (q) =
\sum\limits_{j=I}^{-1} \int \db^{d+1} p \; \hat v (q-p) C_j ( p_\zer ,
e(\ve{p}))  \eqn $$
We want to bound
$$
D_h F(q) = \int \db ^{d+1} p \; \hat v (q-p) \sum\limits _{j=I}^{-1} 
D_h C_j \Bigl( p_\zer ,e(\ve{p}) \Bigr)
\eqn $$
To get the bound in $\abs{h}_\one$ , we use \queq{\Cjdot}
to write
$$
D_h F_j (q) = \int \db ^{d+1} p \; \hat v (q-p) {h(\ve{p}) \over
\cD _u e (\ve{p}) } (u \cdot \nabla ) C_j \Bigl(p_\zer ,
e(\ve{p}) \Bigr)
\eqn $$
and integrate by parts, to get
$$
D_h F_j (q) = -\int \db^{d+1} p \; C_j ( p_\zer ,e(\ve{p}) )
\nabla\! \cdot\! \bigg( \hat v (q-p) {h(\ve{p}) \over
\cD_u e (\ve{p}) } u(\ve{p})\bigg)
\eqn $$
and bound by \queq{\Cjpbd}, \queq{\Kzerodef} and Lemma \uProps\   $(ii)$
$$
\abs{D_h F_j}_\zer \leq K_\zer M^j \cdot \abs{\hat v}_\one
\cdot \abs{h}_\one \abs {u}_\one  {1 \over u_\zer }
\left(1+\frac{|e|_\two}{u_\zer}\right)
.\EQN\threefifty $$
Thus, the scale sum converges absolutely if one allows a derivative 
to act on $h$, which gives
$$
\abs{D_h F}_\zer \leq \Const \abs{h}_\one
\eqn $$
The bound in $\abs{h}_\zer $ is obtained by an integration by parts in $p_\zer$,
using
$$
D_h C_j (p) = h(\ve{p}) \left( i {\del \over \del p_\zer} C_j 
(p_\zer , e( \ve{p} )) - 2M^{-2j} f' \left(M^{-2j}
\left( p_\zer^2 + e (\ve{p})^2 \right) \right)\right)
\eqn $$
Then $D_h F_j = A_j + B_j $, where
$$
\eqalign{
A_j (q) &= \int \db^{d+1} p \; \hat v (q-p) i \del_\zer C_j
\Bigl(p_\zer ,e(\ve{p}) \Bigr) h(\ve{p})\cr
        &= -i \int \db ^{d+1} p \; C_j \Bigl( p_\zer ,
e(\ve{p}) \Bigr) (-\del_\zer \hat v ) (q-p) h(\ve{p})\cr
}
\eqn $$
(note that $h$ does not get differentiated since it does not depend
on $p_\zer$) so
$$
\abs {A_j (q)} \leq K_\zer M^j \abs {\hat v}_\one
\abs {h}_\zer
\eqn $$
and the scale sum $\sum\limits_{j=I}^{-1} A_j$ converges absolutely. However,
$$
B_j (q) =  -2 M^{-2j} \int \db ^{d+1} p \; \hat v (q-p) 
f' (M^{-2j} (p_\zer^2 + e(\ve{p})^2))
\eqn $$
is $O(1)$, so
$\sum_{j=I}^{-1} \abs {B_j (q)} \geq \Const \abs {I}$, and we
have to perform the sum over $j$ before taking $\abs {\cdot}$
to get a sharper bound. 
We write
$$
\sum\limits_{j=I}^{-1} B_j (q) = -2 \int \db^{d+1} p \hat v
(q-p) h(\ve{p}) \Biggl[ {\del \over \del x} \sum\limits_{j=I}^{-1}
f (M^{-2j}x) \Biggr] _{x=p_\zer^2 + e(\ve{p})^2}
\eqn $$
By \queq{\cutfdef}
$$
\sum\limits_{j=I}^{-1} f (M^{-2j}x) = a \bigl( M^{-2I}x\bigr)-a(x)
\eqn $$
and $a' (x) \neq 0$ only if $x\in \Bigl[ M^{-4} , M^{-2} \Bigr]$
by \queq{\cutadef}. So
$$
\eqalign{
\abs{ \sum\limits_{j=I}^{-1} B_j (q) } & \leq 2 \abs{\hat v}_\zer 
\abs{h}_\zer \int \db^{d+1} p
\Biggl( a' \bigl( {p_\zer^2 + e(\ve{p})^2 } \bigr) 
+ M^{-2I} a' \biggl( {p_\zer^2 + e (\ve{p})^2 \over
M^{2I}} \biggr) \Biggr) \cr
& \leq 2 \abs{\hat v}_\zer \abs{h}_\zer \abs{a'}_\zer 
\int \db^{d+1} p\ \biggl(1 \bigl( \abs{ip_\zer - e (\ve{p})} \leq 1\bigr) + M^{-2I}1 \bigl( \abs{ip_\zer - e(\ve{p})} \leq M^I \bigr) \biggr) \cr
& \leq 4K_\zer \abs{\hat v}_\zer \abs{a'}_\zer  \abs {h}_\zer \cr
& \leq \Const \abs{h}_\zer 
\cr}\EQN\threesixty $$
Thus, the divergence of $\sum \abs{B_j}$ as $I \to - \infty $ 
is due to terms
that depend on the scale decomposition. Once the partition of unity
is resummed, all that remains is a boundary term at $j=I$ which is
uniformly bounded as $I \to -\infty$. 
In general, the contributions to $K$ where this procedure has to be
applied, are those from graphs that are \NOL\ on rootscale. There
things are more complicated because the resummation of the partition
of unity has to be done carefully, and because there are a lot more
terms from the integration by parts. Note that the integration by
parts formula \queq{\IBP} combines nicely with the $\ell$--operations;
also, the bound in $\abs{h}_\one $ avoids boundary terms 
and therefore allows us to show convergence, not
only boundedness, of $K'$ as $I \to -\infty$.

\The{\zerBou} {\Thsty Let $G$ be a 1PI two-legged graph and $t$ 
an associated tree so that $(t,G)$ contributes to \queq{\KrIform}. Then 
there is $Q_\one > 0$ such that 
$$
\sum\limits_{J \in \cJ (t,j )} \abs{D_h Val (G^J)}_\zer 
\leq {Q_\one}^{\abs{L(G)}} \la_{n_\ph } (j, \sfrac{\ep}{2} ) 
M^{\ep j} \abs{h}_\one 
,\EQN\onebou $$
and for all $h\in C^1(\cB,\R)$
$$
V_I (h,t,G) =
\sum\limits_{j\geq I} \sum\limits_{J \in \cJ (t,j )} D_h Val (G^J) 
\quad \hbox{ converges in }\abs{\; \cdot \; }_\zer \hbox{ as }I \to -\infty
\EQN\oneconv $$ 
Moreover, there is a constant $\Const $ (depending on $G$, $\hat v$ and $u$ but 
independent of $I<0$) such that for all $h \in C^1(\cB,\R)$
$$
\abs{ \sum\limits_{j=I}^{-1} 
\sum\limits_{t \sim G} \prod\limits_{f\in t} {1 \over n_f!}
\sum\limits_{J \in \cJ (t,j)}
D_h \ell Val (G^j) }_\zer \leq \Const \abs{h}_\zer
\EQN\zerbou
.$$
}

\Rem{\dummy} Note that in \queq{\zerbou}, as in the above example, 
the norm is taken after summing 
over the root scale $j$. This, as well as the additional sum $\sum_{t\sim G}$
over trees associated to $G$, 
is necessary to resum the partition of unity properly. 
In all terms where the resummation of the partition of unity is not 
necessary, the sum over trees will be replaced by a maximum over trees
times a constant, since the number of trees compatible with a 
fixed graph is always finite. The constant appearing in \queq{\zerbou}
depends on $\hat v$, $e$, $u$ and the graph $G$ in the same way as in 
Theorem \TwoFour. In particular, it is uniform on the set $\cA$ 
given in \queq{\Atwodef}. To reduce notation a little, we are 
not going to trace the factors of $\la $
through this proof, because it will be obvious in the proof 
that the factorials are again only  
produced by the non--overlapping four--legged subgraphs.
We denote a polynomial in $\abs{j}$, 
whose coefficients may increase in inequalities and combine
with other constants, by $\Pol{j}$. In that notation, Theorem \TwoFour\
reads 
$$
\abs{\sum\limits_{J \in \cJ (t,j)} Val (G^J)}_s \leq \Pol{j} M^{Y_s j}
\eqn $$
for any $t \sim G$.
We also assume $\ep < 1 $. Note also that \queq{\zerbou} 
is not simply an application
of Lemma \besser\ because the latter will cause $\nabla h$ terms in some
cases.

\Proof By \queq{\lTdot} and \queq{\deGl},
$$
D_h \ell = \ell \biggr( D_h - {h \over \cD_u e} \cD _u \biggr)
,\eqn $$
so the left side of \queq{\zerbou} consists of two terms.
Since $\abs{\ell f }_\zer \leq \abs {f}_\zer$, the contribution to the
second term from any fixed $t,j$ is
$$
\biggl|\sum\limits_{J \in \cJ (t,j)} \ell
\biggl( {h \over \cD_u e} u \cdot \nabla Val (G^J) \biggr)\biggr|_\zer 
\leq  {\abs{h}_\zer \abs{u}_\zer \over u_\zer } \sum\limits_{J \in \cJ (t,j)} \abs{Val (G^J) }_\one
,\eqn $$
which, by Theorem \TwoFour\ $(i)$, is
$$
\eqalign{
&\leq {\abs{h}_\zer \abs{u}_\zer \over u_\zer} 
Q_\zer ^{\abs{L(G)}} \la_{n_\phi}\left( j,\sfrac{\ep}{2} \right) M^{\ep j} \cr
 \cr
}
\eqn $$
As 
$$
{\abs{h}_\zer \abs{u}_\zer \over u_\zer} \sum\limits_{j \leq -1}
Q_\zer ^{\abs{L(G)}} \la_{n_\phi}\left( j,\sfrac{\ep}{2} \right) M^{\ep j}
\le \Const |h|_\zer
$$
the second term is consistent with \queq{\onebou} and \queq{\zerbou}.

It remains to bound
$$
\abs{ \sum\limits_{j=I}^{-1} 
\sum\limits_{t \sim G} \prod\limits_{f \in t} { 1 \over n_f!}
\sum\limits_{J\in \cJ (t,j)} \ell
D_h Val (G^J)}_\zer
\EQN\Muss $$
Again, we do induction over the depth $P$ of $(t,G)$, defined in
\queq{\Depthdef}. The induction hypothesis (IH) is: if $G$ is 
two--legged and 1PI and if $(t,G)$ is of depth $P$, then 
\queq{\onebou} and \queq{\oneconv} hold. Moreover

\leftit{(a)} For all two--legged 1PI graphs $G$ and 
all $i \in \{ I,\ldots ,-1\} $,
$$\abs {\sum\limits_{j=I}^{i} 
\sum\limits_{t \sim G \atop depth (t,G) \leq P} 
\prod\limits_{f \in t} { 1 \over n_f!}
\sum\limits_{J\in \cJ (t,j)} D_h
Val (G^J) }_\zer \leq \Const \abs{h}_\zer
\EQN\zerboui $$
Moreover, for any 1PI graph $G$ with $E=2$ or $4$ external legs
and for $s \in \{ 0,1\}$,
 $$
\abs{ \sum\limits_{J \in \cJ (t,j)} D_h Val (G^j)}_s 
\leq \abs{h}_\zer \Pol{j} M^{Y'_s j}
,\eqn $$
\leftit{(b)} if $\tG (\phi )$ is \OL, $Y'_s= 1-{E\over 2} + \ep -s $
\leftit{(c)} if $\tG (\phi )$ is \NOL, 
$$
Y'_s = \cases{ -1-s & if $E=4$\cr
                0   & if $E=2$ and $s=0$ \cr
              \ep-1 & if $E=2$ and $s=1$ \cr}
\eqn $$

\noindent
The tree sum in (a) will be necessary to resum the scale decomposition
in the way illustrated in the above example. An informal restatement 
of (b) and (c) is that the effect of $D_h$ on the scale behaviour is 
similar to that of a momentum derivative. Note that the 
$h$ does not get differentiated even in the case $s=1$; this hinges on 
the 1P - irreducibility of $G$. 

If $P=0$, $G$ has no nontrivial two- or fourlegged forks
(so $G'$, constructed in Remark \JFK, is equal to $G$), there are
no $\ell$-operations, and therefore the only factors that depend on
$e$ are the propagators. By \queq{\Cdot},
$$\eqalign{
\abs{ D_h C_j (p)} & \leq M^{-j+2} (1+2 \Vert f' \Vert_\infty ) M^{-j+2}
\abs{h}_\zer 1 (\abs{ ip_\zer - e(\ve{p})} \in [M^{j-2},M^j]) \cr
& \leq \Const M^{-2j} 1_j (p) \abs{h}_\zer 
.\cr}\eqn $$
We shall use the just introduced notation $1_j (p)$ in what follows. 
Also, $\Const $ will denote constants that may increase 
in inequalities and depend on $\hat v$, $u$ and the graph $G$, but 
not on $j$ or the infrared cutoff $I$.
 
%%%%%%%%%%%%%%%%% 
\medskip\noindent
{\it Case 1: $P=0$ and $\tilde G(\phi)$ \OL.}
We use Theorem \IPCT\ to get
$$
\abs{ D_h Val (G^j)}_\zer \leq \Const \abs{h}_\zer M^{j(2-m +\ep )} 
\max_{\ell \in L(G)} M^{-j_\ell} 
\prod\limits_{{f\in t \atop f>\phi }} 
M^{D_f (j_f - j_{\pi (f)})}  
\eqn $$
Since $P=0$, $\max M^{-j_\ell} \leq M^{-j}$ for all $J \in \cJ (t,j)$, 
so
$$
\sum\limits_{J \in \cJ (t,j)} \abs{D_h Val (G^J)}_\zer 
\leq \Const \abs{h}_\zer M^{(1-m+\ep )j} 
\sum\limits_{J \in \cJ (t,j)}
\prod\limits_{f > \ph } M^{D_f (j_f - j_{\pi (f)})}
\eqn $$
and the scale sum over all $J \in \cJ (t,j)$ can be performed
as in the $P=0$ case of the proof of Theorem \TwoFour. 
This proves (b) for $s=0$. If $G$ is two--legged, the factor
$M^{\ep j}$ makes the scale sum over $j$ convergent, and 
\queq{\onebou} (with $\abs{h}_\zer $ instead of $\abs{h}_\one$)
and (a) follow. Convergence as $I \to -\infty$ 
(that is \queq{\oneconv}) is now obvious
because every summand is independent of $I$ for $P=0$ and the 
series is absolutely convergent (recall also that  
the number of terms in the sum over $t \sim G $ is finite and independent 
of $I$). Similarly, an additional derivative with respect to the external
momentum gives another factor $M^{-j_\ell} \leq M^{-j}$.
Since $G$ is 1PI, the spanning tree can always be chosen not to
contain the line where $h$ is, so that $h$ does not get differentiated
(alternatively, one can use integration by parts to remove the derivative 
from $h$). Thus
$$
\sum\limits_{J \in \cJ (t,j)} \abs{ D_h Val (G^J)}_\one \leq
\Const M^{j(\ep -m)} \abs{h}_\zer
\eqn $$
which proves (b) for $s=1$. 
%%%%%%%%%%%%%%%%% 
\medskip\noindent
{\it Case 2: $P=0,\ E(G)=4$ and $\tilde G(\phi)$ \NOL.}
The same bounds as above hold, with $\ep$ replaced by zero.
%%%%%%%%%%%%%%%%% 
\medskip\noindent
{\it Case 3: $P=0,\ E(G)=2$ and $\tilde G(\phi)$ \NOL.}
If $E=2$, $\tG (\phi ) $ is an ST or GST diagram (see
Definition \GSTDef\ $(ii,iii)$) and so is $\tilde G (\ta _\phi)$, 
where $\ta_\phi$ is as in Lemma \Treedec. Thus $Val ( \tilde G (\phi) )$
takes the form given in Remark \NOLVal, with an effective vertex
$v_\one$ with $2m$ legs. We now consider the case in which all strings
consist only of a single propagator $(w_i = 0)$. Insertions in these
strings are treated as in $P>0$. 

If $2m=4$, $v_\one$ is four-legged.
Since $P=0$, $v_\one$ is a vertex of scale zero. This can happen
only in first order. Then
$$
Val (G^J)(q) = Val ( \tilde G (\phi)) (q) 
= \int \db^{d+1}p\ \{\hat v (q-p) {\ \rm or\ }\hat v(0)\}
 C_j ( p_\zer , e ( \ve{p}))
\eqn $$
\queq{\onebou} follows from \queq{\threefifty} and 
the bound (a) follows as in \queq{\threesixty}. 
\queq{\oneconv} is again obvious from \queq{\onebou}. 
The bound $(c)$ is obvious for $s=0$. For $s=1$, the derivative
with respect to $q$ acts only on $\hat v$, so $(c)$ holds as well
(actually, with a better exponent). 

\centerline{\figplace{fig13}{0in}{0in}}

If $2m \geq 6$, $G$ and $t$ take the
form shown in Figure 13, with $n \geq 1$ and $f^* \in t$ the
fork such that 
$\tilde G \left( \matrix{{f^*}\cr | \cr {\ta_\phi}\cr}\right)$ 
is \OL\ ($f^*$ exists since otherwise $G$ would be \NOL, hence a 
GST diagram, hence a ST diagram since $P=0$, but then $2m=4$
since at scale zero there are only four-legged vertices), so
$$\eqalign{
\sum\limits_{J\in \cJ (t,j)} Val (G^J) (q) = \sum\limits_{j_\one > j} &\sum\limits_{j_\two > j_\one} \ldots 
\sum\limits_{j^* = j_n > j_{n-1}} \cr
& \int \prod\limits_{k=1}^{m^*} \db^{d+1}
p_k C_{i_k} \Bigl( (p_k)_\zer , e(\ve{p}_k) \Bigr) W (q,p_\one,
\ldots , p_{m^*} )
.\cr}\EQN\Hach
 $$  
Here
$$
W= \sum\limits_{J \in \cJ (t_{f^*}, j^* )} Val \Bigl( \tilde G
(t_{f^*}) \Bigr)
,\eqn $$
$\tilde G (t_{f^*})$ is a graph with $2m^* +2$ external legs,
where $m^* \geq \max\{ m-1,n-1\}$, and
for each $k \in \{ 1, \ldots , m^* \}$, either $i_k = j$ or there is $r \in
\{1, \ldots , n-1 \}$ such that $i_k = j_r$. By
assumption, $m \geq 3$, so there are at least two lines with 
$i_k = j$. We may choose the labelling such that they are the lines for
$k=1$ and $k=2$.
Apply $D_h$ to \queq{\Hach}. If it acts on $W$, it can act only on a
propagator of scale $j_l \geq j^*$, since $P=0$, and the net effect
is, up to constants, a factor $M^{-j_l} \leq M^{-j^*}$, and a
factor $h(\ve{p}_l)$. As mentioned in Remark \Passauf, 
$\tG (f^*)$ need neither be \OL\ nor 1PI, but 
$\tG \left( \matrix{ f^* \cr | \cr \ta_\ph }\right)$
is \OL\ and so the volume integral produces a factor $M^{\ep j^*}$
by Theorem \IPCT. Choosing the spanning tree as in Lemma \ImpPow, 
so that all $p_k$ are loop lines, Theorem \IPCT\ implies
$$\eqalign{
\abs{\int\prod\limits_{k=1}^{m^*} \db^{d+1} p_k C_{i_k} (p_k) 
D_h W (q,p_\one, \ldots , p_{m^*} )} & \leq 
\Const M^{i_\one + \ldots + i_{m^*}}
M^{(\ep -1)j^*} \cr 
& M^{D_{f^*} j^*} \sum\limits_{J \in \cJ (t_{f^*}, j^* )}
\prod\limits_{f > f^* } M^{D_f (j_f - j_{\pi (f)})}
\cr}\EQN\Hachp $$
Since $P=0$, $D_f < 0$ for all $f> f^*$ , and so the scale sum over 
$J \in \cJ (t_{f^*} , j^*)$ converges by the argument of the $P=0$ 
case in the proof of Theorem \TwoFour. 
Since $\tilde G (f^*)$ has $2m^* +2$ external legs, 
$D_{f^*}= 2-(m^* +1)=1-m^*$. 
Calling $m _r = \{ k \in \{ 1, \ldots , m^* \} : i_k = j_r \}\ge 1$,
we have
$$
m^* = \sum\limits_{r=1}^n m_r + m-1
\qquad\hbox{and}\qquad
\sum\limits_{k=1}^{m^*} i_k = \sum\limits_{r=1}^{n-1} 
m_r j_r + (m-1)j
\EQN\mSum
$$
Inserting this and using again $M^{(\ep -1 ) j^*} \leq M^{(\ep -1 ) j}$,
we have 
$$\eqalign{
\Big\vert&\sum\limits_{j^* \geq \ldots \geq j_\one \geq j} 
\int\prod\limits_{k=1}^{m^*} \db^{d+1} p_k\ C_{i_k} (p_k) 
D_h W (q,p_\one, \ldots , p_{m^*} )\Big\vert \cr
& \leq \Const
\abs{h}_\zer M^{(\ep -1 )j}
\sum\limits_{j_{n-1} > \cdots > j_\one > j}
M^{(m-1)j + \sum\limits_{r=1}^{n-1} m_r j_r} 
\sum\limits_{j^* > j_{n-1}}M^{(2-m)j^*- \sum\limits_{r=1}^{n-1} m_r j^*}
\cr
& \leq \Const\abs{h}_\zer M^{j\ep}\sum_{j^*\ge j}M^{(m-2)(j-j^*)} 
\prod\limits_{r=1}^{n-1} \sum_{i \leq j^*} M^{m_r (i - j^*)} \cr
& \leq \Const \abs{h}_\zer M^{j\ep}
\cr}\EQN\Wui $$
This proves (c) for $s=0$, \queq{\onebou}, and upon summation over $j$, (a)
and \queq{\oneconv}, 
for the contribution to \queq{\Muss} where $D_h$ acts on $W$. 
For this contribution, the only remaining case is (c) for $s=1$. 
With the spanning tree we chose, the derivative acts only on 
the functions associated to those lines in $D_h W$ in which the 
external momentum enters. When acting on a propagator, it produces 
a factor $\Const M^{-j_l} \leq \Const M^{-j}$. The only dangerous 
case is when $h$ `sits' on the path through $G_{f^*}$ through 
which the external momentum is routed, and the additional derivative
acts on $h$. In that case we use integration by parts as in the proof
of Theorem \TwoFour\ to remove the derivative from $h$. This is possible
because $G$ is 1PI (note again that $G_{f^*}$ need not be 1PI, and if
it is not, the derivative will produce a factor $M^{-j}$, not just 
$M^{-j^*}$). 
Taking absolute values, we obtain the same bound as before, only 
multiplied by $\Const M^{-j}$. This proves (c) for $s=1$ for the term with
$D_h$ acting on $W$. 

If $D_h$ acts on one of the $C_{i_k}$, $k \in \{ 1, \ldots ,m^* \}$,
we can assume without loss of generality 
that $k=1$ since $i_\one =i_\two =j$ and
$i_k \geq j$ for all $k$, and the scale behaviour degrades 
worst when the derivative occurs on the lowest scale. We have to bound
$$
\biggl|\int \db^{d+1}p_\one\ D_h C_j (p_\one) 
\sum\limits_{(j_k)}
\int \prod\limits_{r=2}^{m^*} \db^{d+1} p_r\ C_{i_r}(p_r) 
\left( \sfrac{\del}{\del q_\al }\right)^s
W(q, p_\one , \ldots, p_{m^*} )\biggr|_\zer 
\EQN\Musz $$
(where $\sum\limits_{(j_k)}$ is short for 
$\sum\limits_{j_n > \cdots > j_\one > j}$) for 
$s=0$ and $s=1$. 
To see \queq{\onebou}, we apply the integration by parts formula
\queq{\threesix} to the integral over
$p_\one $. When the derivative acts on $W$, we get the bound 
\queq{\Wui}. The term containing $\nabla h$ has the same scale behaviour 
as if there had been no derivative at all. \queq{\oneconv} follows 
by the dominated convergence argument of the proof of Theorem 
\TwoFour\ $(iv)$.
To see (a) and (c), we use Lemma \besser\ in the $p_\two$-integration
to bound this by 
$$
\leq \int \db^{d+1} p_\one \abs{D_h C_j}  \sum\limits_{(j_k)}
\int \prod\limits_{r=3}^{m^*} \db^{d+1} p_r \abs{C_{i_r}(p_r)} 
\int \db^{d+1} p_\two \; 1_j (p_\two ) 
\abs{\left(\sfrac{\del}{\del q_\al }\right)^s W}_{\one,j}
\eqn $$
Estimating
$$
\abs{D_h C_j ( p_\zer ,e(\ve{p}))} \leq 
\Const M^{-2j} \abs{h}_\zer 1_j (p)
,\eqn $$
the $C_{i_k}$ for $k \geq 3$ by Lemma \Simpl\ $(iii)$, and rearranging the product, 
\queq{\Musz } is bounded by 
$$
\leq \Const \abs{h}_\zer \sum\limits_{(j_k)}
\int \left( \prod\limits_{r=1}^{m^*} 
\db^{d+1} p_r M^{-i_r} 1_{i_r} (p_r) \right)
\abs{\left(\sfrac{\del}{\del q_\al }\right)^s W}_{\one,j}
.\EQN\reArr $$
The derivative on $W$ acts on a line in $G_{f^*}$ and 
produces a factor $\Const M^{-j^*}$.  
By Theorem \IPCT\ (and since none of the lines $k=1, \ldots , m^*$ is
in the spanning tree), the result when $s=0$ is  
$$
\leq \Const \abs{h}_\zer \sum\limits_{(j_k)} 
M^{i_\one + \ldots +i_{m^*}} M^{(\ep -1)j^*}
M^{D_{f^*} j^*} \sum\limits_{J \in \cJ (t_{f^*}, j^* )}
\prod\limits_{f > f^* } M^{D_f (j_f - j_{\pi (f)})}
.\eqn $$
The sum over $J$ is estimated as above, and also the rearrangement 
of the terms is similar to the previous case, so the bound is yet another
$$
\leq \Const \abs{h}_\zer M^{\ep j}
\EQN\gOOd $$
which proves (a), and (c) for $s=0$. For $s=1$, we need not apply 
Lemma \besser: the derivative w.r.t.\ the external momentum can act only 
on propagators associated to lines of $W$, and effectively produces a factor
$M^{-j^*}$ (it cannot act on $h$, since $h$ is in a string in the present case).
The $D_h$ acting on the string causes a factor $M^{-j}$ as compared 
to the ordinary scaling behaviour, which we take outside.
The bound now follows by the argument between 
\queq{\Hach} and \queq{\Wui}.

Some of the bookkeeping of this $P=0$ case could
have been avoided by normal ordering, but the normal ordering
prescription depends on $e$, and thus $D_h$ would have produced
similar terms there. 
Also, normal ordering does not remove the GST graphs, so
$P > 0$ has to be dealt with anyway. 

Up to now, the sum over trees 
was not really used, since the only term where a resummation 
of the partition of unity was necessary was the lowest order term. 
Let $P>0$, and assume that the IH holds for depth $P' \leq P-1$. Now there are also 
two--legged subdiagrams on which $D_h$ can act. It can 
also act on the projections $\ell $ or $1-\ell$ in front of them 
since projection on $S$ depends on $e$. For every $t \sim G$,
we  construct the graph $G'$ as in Remark \JFK. We rearrange the sum 
over trees $t$ as follows. Every $t \sim G$ that gives rise to the same $G'$ 
can be split into the tree $t'$ rooted at $\phi $ and the subtrees $t_w$ of 
$t$ associated to every vertex $w$ of $G'$. Thus, at fixed $G'$,
the sum over all $t \sim G$ splits into one over all $t'\sim G'$ 
and given $t' \sim G'$, there is a sum over trees $t_w$ rooted at $w$
for every vertex $w$ of $G'$. Blocking the sum in that way, 
we have for every $G'$ and $t'$ the vertex functions  
$$
\Ph_w = \sum\limits_{t_w \sim G_w} \prod\limits_{f\in t_w} {1 \over n_f!} F_w
\eqn $$  
with $F_w$ given by \queq{\Fw}. For two--legged vertices $w$, 
$\Ph_w$ has the same structure as the left side of \queq{\zerboui}, but 
the depth of $(t_w, G_w)$ is smaller than $P$ for all $t_w$ contributing
to the sum for $\Ph_w$. This is the basis of the induction scheme. 

%%%%%%%%%%%%%%%%% 
\medskip\noindent
{\it Case 4: $P>0,\ E(G)=2,4$ and $\tilde G(\phi)$ \OL.} 
Then $(G')^\sim (\phi )$ is 
\OL\ as well. For the derivative of a propagator we again use
$$
\abs{ D_h C_{j_l} (p)} \leq \Const 1_{j_l} (p) 
M^{-2j_l}\abs{h}_\zer 
.\EQN\DhCjp $$

If the derivative acts on an $r$--fork, we write
$\Ph_w = (1-\ell ) T$ and use
$$
D_h (1-\ell ) T = (1-\ell ) D_h T + \ell \left( 
{h \over \cD_u e} \cD_u T \right) 
\EQN\rTwoTerms $$
to isolate the term where the renormalization cancellation gets lost. 
In the second term, we associate the factor 
$$
\ze = \ell \left( {h \over \cD_u e} \cD_u T \right) 
\eqn $$
to one of the lines going into $G_w$. 
By Theorem \TwoFour, applied to $T$, 
$$
\abs{\ze}_\zer \leq {\abs{h}_\zer \over u_\zer} \abs{u}_\zer \abs{T}_\one
\leq \Const \abs{h}_\zer
\EQN\rFind $$
%
% This includes the sum over the root scale of $T$. There is no $M^{\ep j}$
% because of the $j=0$ term.
%
By the IH (b) or (c), 
$$\eqalign{
\abs{1_{j_{\pi(w)}} (1-\ell)D_h T}_\zer & \leq \Const M^{j_{\pi(w)}} \abs{D_h T}_\one 
\leq \abs{h}_\zer M^{j_{\pi(w)}}
\sum\limits_{j'\geq j_{\pi (w)}} \Pol{j'} M^{(\ep -1)j'}\cr
& \leq \abs{h}_\zer \Pol{j_{\pi (w)}} M^{\ep j_{\pi (w)}}
\cr}\EQN\rFine $$
Adding \queq{\rFind} and \queq{\rFine}, 
$$
\abs{D_h \Ph_w}_\zer \leq \Const \abs{h}_\zer
,\EQN\dhFwzer $$
so, compared to the usual behaviour \queq{\rForkest}, 
we have lost a factor $M^{j_{\pi (w)}}$, which is the same as saying that
there is an extra factor $M^{-j_\ell}$ for one of the two external legs 
of the graph $G_w$, just as if the $D_h$ had acted on that leg.

Similarly, if the derivative acts on a $c$-fork, $\Ph_w = \ell T$,
$$
D_h \ell T = \ell D_h T - \ell \Bigl( {h \over \cD_u e} \cD_u T \Bigr)
\eqn $$
where
$$
T= \sum\limits_{I \leq j' \leq j_{\pi (w)}} 
\sum\limits_{t_w \sim G_w}\prod\limits_{f\in t_w} {1 \over n_f!} 
\sum\limits_{J \in \cJ (t_w,j')} 
Val (G_w ^J)
,\eqn $$
application of the IH (a) to the first term and Theorem \TwoFour\ to
the second term yields $ \abs{D_h \ell T}_\zer \leq \Const 
\abs{h}_\zer $. Compared to \queq{\TwoVert}, the derivative has again cost 
us a factor $M^{j_{\pi (w)}}$.
Again, we associate a factor $z_\ell = M^{-j_\ell}$
to one of the external legs of $G_w$. 

For a SSI, there is the same factor.
For a four-legged vertex, there is a factor 
$\Pol{j_{\pi (w)}} M^{-j_{\pi (w)}}$ by the IH (b) and (c). 

To summarize, the effect of $D_h$ on the scale behaviour, as 
compared to the power counting behaviour \queq{\TwoVert} and 
\queq{\FourVert}, is accounted for by an additional factor 
$$
z_\ell = M^{-j_\ell} 
\EQN\Dheff $$
on a line of $G'$.  
By construction of $G'$, $M^{-j_\ell} \leq M^{-j}$, so, by Theorem \IPCT,  
(similar to \queq{\Zwis})
$$
\abs{Val(G'^J )}_\zer \leq \Const M^{j\bigl( D_\phi (G') + \ep - 
1 \bigr) } \prod\limits_{f \in t' \atop f>\phi} M^{D_{f}(G')
(j_{f} - j_{\pi (f)})} \prod\limits_{w\in V_2 (G')}
M^{j_{\pi (w)}}
\eqn $$
The proof that the scale sum over $J\in \cJ (t,j)$ converges 
is given following \queq{\Zwis}, so (b) holds for $s=0$. 
For $E=2$, the sum over $j$ converges because
of the remaining $M^{\ep j}$, which proves \queq{\onebou} and (a) and
(by the usual dominated convergence argument) also \queq{\oneconv}.
 
For $s=1$, we apply an extra derivative with respect to the
external momentum $q$ before taking $\abs{\cdot}_\zer$.
All we have to show is that its effect can be bounded by a factor 
$\Const M^{-j}$. For its action on a propagator this follows
from Lemma \Simpl\ $(iii)$, and for its action on a vertex function 
$\Ph_w$ that is not affected by $D_h$ it follows from Theorem \TwoFour.
For its action on $D_h$ of a vertex function coming from an $r$--fork or a
four-legged subdiagram of $G$, it follows from the IH (b) and (c), 
since the scales of these vertices are summed above $j_{\pi(w)}$.
However, we have to avoid two derivatives on any $c$-fork, and also
prevent the derivative from acting on $h$. The only case when two derivatives
can act on a $c$-fork is when $D_h$ acts on the $c$-fork and
${\del \over \del q_i}$ acts on the same $c$-fork. The latter
derivative can be removed by an integration by parts because
$G$ is 1PI (and then cannot act on $h$), so the bound for
$s=1$ follows as in the proof of Theorem \TwoFour.

%%%%%%%%%%%%%%%%% 
\medskip\noindent
{\it Case 5: $P>0,\ E(G)=4$ and $\tilde G(\phi)$ \NOL.}
The bound (c) is proven as above, omitting the parts used to get $\ep>0$. 
%%%%%%%%%%%%%%%%% 
\medskip\noindent
{\it Case 6: $P>0,\ E(G)=2$ and $\tilde G(\phi)$ \NOL.}
Since $G'$ is a quotient graph of
$G$ that contains all lines $\ell \in L(G)$ with $j_\ell =j$,
$\widetilde{G'} (\phi) = \tilde G (\phi)$ is a GST graph, and so is
$\widetilde{G'} (\tau'_\phi)$, where $\tau '_\phi$ is the maximal \NOL\
subtree of $t'$ rooted at $\phi$ (Lemma \Treedec\ $(ii)$). 
The value of $\widetilde{G'}(\tau '_\phi)$ is an integral of the form
given in Remark \NOLVal\ (from which we now take the notation).
By construction of $\tau '_\phi$, the vertex function $\cU _{v_1}$
belongs to a subgraph $H$ of $G'$, of scale $j^*$ such that $G'$ overlaps on 
scale $j^*$. $D_h$ can act on $\cU_{v_1}$
or on one of the strings $S_i$. In the former case, 
the bound follows by \queq{\Hachp} -- \queq{\Wui}, because 
\queq{\Dheff} applies due to the volume gain at the scale $j^*$ where
the derivative acts, and because strings of two--legged subdiagrams
satisfy $\abs{S_i(p)} \leq \Pol{j} M^{-j} 1_j (p)$ by \queq{\TwoVert}.

Let $D_h$ act on one of the strings. We call $j^{(i)}$, defined in 
Remark \NOLVal, the scale of the string $S_i$. We do the case $s=1$ 
first. We choose the spanning tree as in Lemma \MomRout, then 
the additional derivative with respect to the external momentum $q$
acts only on $\cU_{v_\one}$, i.e.\ at a scale where there is an 
improvement factor $M^{\ep j^*}$, and it cannot act on $h$.
The effect of $D_h$ is again accounted for by a factor $M^{-j^{(i)}}
\leq M^{-j}$ on one of the lines. Taking the $M^{-j}$ in front, 
(c), for $s=1$, follows by \queq{\Hachp} -- \queq{\Wui}.

For the final case $s=0$, we consider two situations, 
sketched in Figure 14, separately. 

\leftit{(A)} There is a string of scale $j$ (i.e. root scale) on which
$D_h$ does not act. (This is the case if there are at least two
strings of scale $j$ or if $D_h$ acts on a string $S'$ of higher
scale). 
\leftit{(B)} There is only one string on root scale, and $D_h$ acts on it.

\centerline{\figplace{fig14}{0in}{0in}}

\noindent
(A)  
By \queq{\Dheff} (and since all insertions in a string are 
two--legged), there is effectively an $M^{-j'}$, and a factor $h(p')$,
where $p'$ is the loop momentum of $S'$. We apply the Taylor expansion 
procedure of Lemma \besser\ to the string of scale $j$ on which $D_h$ 
did not act (see Figure 14(A)). 
Although this produces derivatives on other factors in 
the expression for $Val (G)$, there are no $\nabla h$ terms because 
$p'$ is an independent loop momentum. 
The Taylor expansion generates two kinds of terms, one from 
acting on the $r$--forks, bounded by 
$$
\Const M^{2j} \abs{T}_\two \abs{\cU_{v_\one}}_\zer \leq
\Pol{j} M^{2j} M^{(\ep -1)j} \abs{\cU_{v_\one}}_\zer
\eqn $$
by Theorem \TwoFour (this includes the sum over the scale of $T$ as well as the 
loop integral) and thus convergent, and one where $\cU_{v_\one}$ gets differentiated, 
$$
1_j(p) \abs{{\del \over \del p} \cU_{v_\one} (q,p,\ldots )}
\eqn $$
The scale balance is identical to the $P=0$ case, so 
\queq{\reArr} -- \queq{\gOOd} hold, with $\Const $ 
replaced by $\Pol{j}$. This proves (c) and (a). 

\noindent
(B) In this case, the vertex $v_\one$ is a four--legged vertex 
of $G'$ with a vertex function $\cY = \cU_{v_\one}$, to which the IH applies.
We thus have to bound $\sum\limits_{j\geq I} X_j (q)$, where
$$
X_j (q)= \int d^{d+1}p \cY (q,p) D_h S(p)
.\EQN\Hydro
$$
In some of the following cases, we need to resolve 
the four--legged graph to which 
the $\cY$ is associated (unless it is a vertex of scale zero, 
which behaves as a vertex with improvement factor $M^j$
whenever a derivative acts on it).
The vertex function $\cY$ is given by the scale sum
$$
\cY (q,p) = \sum\limits_{i\geq j+1} \cY_i (q,p)
\eqn
$$
with
$\abs{\cY _i}_s \leq \Pol{i} M^{-si}$ (see Figure 14). Note
that by Theorem \IPCT\ and by construction of $\tau'_\phi$, there is
a volume gain $M^{\ep j^{*}}$ in the entire integral for $X_j$.  $S$
is a string of length $n$,
$$
S(p) = \sum\limits^{j+1}_{j_1 ,\ldots , j_{n-1} = j \atop
\min \{ j_1 ,\ldots , j_{n-1} \} =j }
\left( \prod\limits^{n-1}_{k=1} C_{j_k} ( p_\zer , e(\ve{p})) 
\cP _k T_k (p) \right) C_{j_n} \bigl( p_\zer , e(\ve{p}) \bigr)
.\EQN\StreeSum $$
The $T_k$ are scale sums of 1PI two-legged insertions and thus
dependent on $j$ and $\cP_k \in \{ 1,\ell ,1-\ell \} $. The $T_k$ obey
the bounds of Theorem \TwoFour\ and, as graphs of depth $P'\leq P-1$,
also the IH. The undifferentiated string obeys $\abs{S(p)} \leq 
1_j(p) \Pol{j} M^{-j}$ by \queq{\TwoVert}.

The form of the scale sum in \queq{\StreeSum} is due to the 
sum over all trees $t' \sim G'$, which contains a sum over all these 
assignments. This is the point where the tree sum is necessary, 
as will be seen when the partition of unity is resummed.

If $D_h$ acts on an insertion from an $r$-fork, 
we get the two terms of \queq{\rTwoTerms}. The first one is 
bounded using \queq{\rFine}, so  
in this term the $D_h$ changes the root scale behaviour
from $M^j$ to $M^{\ep j}$. For the
second term, call $U = {h \over \cD_u e} \cD_u T $.
By \queq{\rFind}, $\abs{U}_\zer\le\Const\abs{h}_\zer$, including
the sum over the scale of $T$. This amounts to a loss of $M^{-j}$.
Apply Lemma \besser\ to the string $S$ to extract another $M^{j-j^*}$ and
use the volume gain of $M^{j^*\ep}$. 
Note that the Taylor
expansion does not produce any derivatives of $h$ since (in the
notation of Lemma \besser ) $\ell U (p) = \tilde U (0, 0, \om )$ does
not depend on $r$ and $\varphi$. The loss of the renormalization
cancellation would make the scale sum over $j$ marginal, but the extra
$M^{\ep j}$ makes it convergent. This proves (c) and (a) for this
contribution. 

The case of $D_h$ acting on a $c$-fork is similar since
$ D_h \ell T = \ell (D_h T) - \ell \biggl( {h \over \cD_u e}
\cD_u T \biggr)$.
The second term is bounded by $\abs{h}_\zer \Pol{j} M^{\ep j}$ 
by \queq{\cForkopts}. The IH applies to the first term, but we have
to make up for the loss of the usual factor $M^{(1-s)j}$ of \queq{\TwoVert}. 
Therefore, after applying $D_h$, we use Lemma \besser\ in the same way 
as for the $\ell U$--term of the just treated $r$--fork case. 
The Taylor expansion is such that the term $\ell D_h T$ is treated
like a constant. Thus, after Taylor expanding and collecting the 
gain $M^{\ep j}$, $\abs{\ell D_hT}_\zer $ appears in the bound. 
The IH applies and implies (c) and (a) for this contribution.
For SSI, write $1=1-\ell +\ell$ and treat the two terms as above.

Finally, $D_h$ can act on a propagator of scale $j$ (or $j+1$). 
Now we may assume that there are no $c$--forks or same scale insertions or
$r$--forks of scale below, for example, $j+7$ on the 
string $S$, since otherwise (c) and (a) follow immediately from 
the improved power counting behaviour 
\queq{\cForkopt}, which suffices by itself to control the 
$M^{-j}$ from the action of $D_h$. 
The strategy is now similar to that of the lowest--order example.  
To get \queq{\onebou}, we apply \queq{\threesix}. There are three terms:
the term containing $\nabla \cdot ({h u \over \cD_u e})$ has the same scale 
behaviour as if the derivative had not acted (but contains a $\nabla h$). 
The second term is when $u \cdot \nabla = \cD_u$ acts on an $r$--fork
$(1-\ell ) T$. Since $\cD_u \ell =0$, the result is $(1-\ell ) \cD_u T + \ell \cD_u T$.
By Theorem \TwoFour, the first summand has a net 
$M^j\sum_{j'>j}M^{j'(1+\ep-2)}=M^{\ep j}$, and the second 
is treated by Lemma \besser, as above. The third term is when $\cD_u $ acts on 
$\cY$. In that case, we resolve the corresponding subgraph and proceed as 
in the case $\cU_{v_\one}$. So in all cases, the scale behaviour deteriorates from 
$M^j$ only to $M^{\ep j}$, which proves \queq{\onebou}, and, by the same argument
as in the proof of Theorem \TwoFour, also \queq{\oneconv}.
For the proof of (a) and (c) we must again consider two terms, because
$D_h C_j (p) = h(\ve{p}) \left(i\del_\zer C_j (p) - 2M^{-2j}f'_j (p)\right)$.
In the string $S$, actually a product $C_j (p)^m C_{j+1} (p)^{n-m}$
appears, where $m\geq 1$ depends on the scale assignments, and
the relevant formula is
$$
D_h \biggl( \prod\limits_{l=1}^n C_{j_\ell} (p) \biggr)  =
h(\ve{p}) i\del_\zer \prod\limits_{\ell =1}^n C_{j_\ell} (p)
- {2 h(\ve{p})  \over (ip_\zer - e\bigl( \ve{p})\bigr)^{n-1}}
\biggl[ {\del \over \del x} \biggl( \prod\limits_{\ell =1}^n
f(M^{-2 {j_\ell}} x) \biggr) \biggr]_{x=p_\zer^2 + e(\ve{p})^2} 
 \EQN\Hydra $$
In the contribution of the first term to \queq{\Hydra}, we integrate
by parts in $p_\zer$. There is no boundary term. $h$ depends only on 
$\ve{p}$, so $\del_\zer$ can act only on
the vertex function $\cY$ or on an $r$--fork. 
In the former case, we combine the known behaviour
$\abs{\cY}_1 \leq \Pol{j} M^{-j^*}$, Theorem \IPCT\ to get the volume
factor $M^{\ep j^*}$, and Theorem \TwoFour\ for the $T_k$ and the 
standard bounds for $C_j$,
$$
\abs{\int d^{d+1} p\ S(p) {\del \over \del p_\zer } \cY (q,p) h(\ve{p})}
 \leq \Pol{j} \abs{h}_\zer \cdot M^{j(\ep -1)} M^j \leq
\abs{h}_\zer\Pol{j} M^{j\ep }
\eqn $$
to see that the scale sum converges (more precisely, we apply the procedure
of \queq{\Hach} -- \queq{\Wui}, which by now should be routine).

If $\del_\zer $ acts on an $r$--fork, we get two terms from 
$\del_\zer (1-\ell )T=\del_\zer T = (1-\ell)\del_\zer T +\ell\del_\zer T$.
Recalling that $T$ is given by a scale sum $T=\sum_{i>j}T_i$, recalling
\queq{\jNormdef}, and using Lemma \Lieder, we have
$$
\abs{(1-\ell ) \del_\zer T}_{\zer,j} \leq 
\sum_{i>j}\abs{(1-\ell ) \del_\zer T_i}_{\zer,j} 
\leq \frac{\sqrt{2}}{u_\zer} M^j \sum_{i>j} \abs{T_i}_\two
.\eqn $$
By Theorem \TwoFour\ $(i)$, this is 
$$\eqalign{
& \leq \frac{\sqrt{2}}{u_\zer} M^j M^{\ep/2} Q(T) 
\abs{v}_\two^{\abs{V(T)}} \sum_{i>j} \la_{n_T} (i,\sfrac{\ep}{2})
M^{i(\ep -1)} \leq \cr
&  \leq \frac{\sqrt{2}}{u_\zer} M^j M^{\ep/2} Q(T) 
\abs{v}_\two^{\abs{V(T)}} M^{j(\ep -1)} \la_{n_T} (j,\sfrac{\ep}{2})
\frac{1}{1-M^{\ep -1}} \leq \cr
& \leq \Const M^{j\ep}
\cr}\eqn $$
The second term is not as easy because there is no reason for 
$\ell \del_\zer T$ to be small. Here we have to use another integration 
by parts, and in some terms an additional resummation, as follows.
We have to bound
$$
X = \int dp \; h(\ve{p}) \cY (q,p) (\ell\del_\zer T^{(1)})(p)
\prod\limits_{l=1}^n C_{j_l} (p_\zer,e(\ve{p}))
\prod\limits_{r=2}^{n-1} (1-\ell)  T^{(r)} (p)
\eqn $$
for $n\geq 2$ (otherwise, such a term does not occur). Superficially, the 
scale sum of this looks divergent, but we can use that $\ell \del_\zer T$ 
is independent of $p_\zer $ to do another integration by parts.
We rewrite $X$ as
$$
X = \int dp \; \frac{h(\ve{p}) \cY (q,p)}{(ip_\zer - e(\ve{p}))^{n}}
(\ell\del_\zer T^{(1)} )(p)
\prod\limits_{l=1}^n f_{j_l} (p)
\prod\limits_{r=2}^{n-1} (1-\ell)  T^{(r)}(p)
\eqn $$
and use that for $n \geq 2$
$$
\frac{1}{(ip_\zer - e(\ve{p}))^{n}}= 
\frac{i}{n-1} \frac{\del }{\del p_\zer} 
\frac{1}{(ip_\zer - e(\ve{p}))^{n-1}}
\eqn $$
to get
$$
X= \frac{-i}{n-1}\int \frac{dp}{(ip_\zer - e(\ve{p}))^{n-1}}
\frac{\del }{\del p_\zer} \left( h(\ve{p})\cY (q,p)(\ell\del_\zer T^{(1)})(p)
\prod\limits_{l=1}^n f_{j_l} (p)
\prod\limits_{r=2}^{n-1} (1-\ell) T^{(r)} \right)
\eqn $$
$h$ and $\ell\del_\zer T^{(1)}$ do not depend on $p_\zer$, so 
$$
X=  \frac{-i}{n-1}\int dp \;
\frac{h(\ve{p}) (\ell\del_\zer T^{(1)} )(p)}{(ip_\zer - e(\ve{p}))^{n-1}}
\frac{\del }{\del p_\zer} \left(\cY (q,p)
\prod\limits_{l=1}^n f_{j_l} (p_l)
\prod\limits_{r=2}^{n-1} (1-\ell)  T^{(r)}(p)\right)
\eqn $$
If the derivative were not there, the integral would obey 
the ordinary power counting bound. We analyze the effect of $\del_\zer$. 
If the derivative acts on $\cY$, we proceed as above to see that 
there remains a factor $M^{j\ep}$. We postpone the treatment of the 
term containing $\del_\zer \prod f_{j+l} (p)$ to the next (and final) case. 
The derivative acting on $\prod (1-\ell ) T^{(r)}$ produces a sum of terms
similar to the one we started with, but with number of $r$--forks on the 
string decreased by one. Thus we may proceed iteratively to remove all these
terms, so that it remains to estimate (having renumbered the $T$'s)
$$
\int dp \;\frac{\cY (q,p)h(\ve{p}) 
\prod\limits_{r=1}^k(\ell\del_\zer T^{(r)} )(p)
}{(ip_\zer - e(\ve{p}))^{n-k}}
\prod\limits_{r=k+1}^{n-1} (1-\ell)  T^{(r)} (p)
\frac{\del }{\del p_\zer} \prod\limits_{l=1}^n f_{j_l} (p_l)
\eqn $$
for $k \leq n-1$. This will be done by resummation, and the procedure is 
similar for all $k$, and similar to the procedure to deal with the second 
term in \queq{\Hydra}, which we discuss now.

We have to bound the integral
$$
\eqalign{
Y(q) =  -2 \sum\limits_{0>j\geq I} \int& d^{d+1} p\ \cY (q,p)
{h(\ve{p}) \over (ip_\zer -e(\ve{p}))^{n-1}} 
\sum\limits_{j_1 ,\ldots ,j_n =j \atop  \min\{ j_1 ,\ldots ,j_n \} =j}^{j+1} 
\left( \prod\limits_{k=1}^{n-1} (1-\ell ) T_k (p) \right)\ \times\cr
& \sum\limits_{l =1}^n M^{-2j_l} f'( M^{-2{j_l}} ( p_\zer ^2 + e(\ve{p})^2 ) )
\prod\limits_{r\neq l} f( M^{-2j_r} ( p_\zer ^2 + e(\ve{p})^2 ) ) 
\cr}\eqn $$
by resumming the partition of unity on line number $l$.
To this end, we first rearrange the sum over the $j_k$ 
by using that for all $k\in \{1,\ldots ,n\}$
$$\eqalign{
\bigcup\limits_{j=I}^{-1}  \Bigl\{ (j_1 ,\ldots ,j_n) :
j_k \in \{j,j+1\} \cap \{I,\ldots,-1\}, \min \{ j_1,&\ldots,j_n\}=j\Bigr\}\cr
 &=  Z_{k}(I) \cup \bigcup\limits_{j=I+1}^{-1}M_k (j) \cr
}\eqn $$
where
$$\eqalign{
M_k (j) &= \Bigl\{(j_1 ,\ldots ,j_n):j_k =j , j_r \in \{I,\ldots,-1\}
{\rm\ and\ for\ all\ } r, s\in \{1,\ldots,n\}, \abs{j_r - j_s} \leq 1 \Bigr\}\cr
Z_k (j) &= \Bigl\{(j_1 ,\ldots ,j_n):j_k =j {\rm \ and\ } j_r \in \{j,j+1\}
{\rm\ for\ } r\ne k \Bigr\}\cr
}\eqn$$
To see this, we note that the left side is $Z=Z_1 \cup Z_2$ with,
for each fixed $k$,
$$
\eqalign{
Z_1 & = \bigcup\limits_{j=I}^{-1} \Bigl\{ (j_1 ,\ldots,j_n ) :
j_k = j , j_r \in \{ j,j+1\} \hbox{ for } r \neq k \Bigr\} \cr
& = Z_k (I)\cup\bigcup\limits_{j=I+1}^{-1} Z_k (j)   
\cr}\eqn $$
and
$$ \eqalign{
Z_2 & = \bigcup\limits_{j=I}^{-2} \Bigl\{ (j_1 ,\ldots ,j_n) :
j_k = j+1 , j_r \in \{j,j+1\} \hbox{ for }r \neq k , \min \{j_1 ,\ldots ,
j_n \} =j \Bigr\} \cr
& = \bigcup\limits_{i=I+1}^{-1} \Bigl\{ (j_1 ,\ldots ,j_n ):
j_k =i , j_r \in \{ i-1,i \} \hbox{ for }r \neq k, \min \{ j_1 ,\ldots ,
j_n \} = i-1 \Bigr\} \cr
& = \bigcup\limits_{j=I+1}^{-1} V_k (j) \cr
}\eqn $$
Here
$$
 V_k (j) = \Bigl\{ (j_1 ,\ldots ,j_n ) : j_k =j,
j_r \in \{ j-1, j\} \hbox{ for } r \neq k 
 \hbox{\ and\ } \min \{j_1 ,\ldots ,j_n \} =j-1 \Bigr\}.
$$
We have
$$
Z = Z_k(I) \cup \bigcup\limits_{j=I+1}^{-1} \Bigl( Z_k (j)
\cup V_k (j) \Bigr)
\eqn
$$
Finally,
$$\eqalign{
Z_k (j) \cup V_k (j) = \Bigl\{ (j_1 ,\ldots ,j_n):& j_k = j {\rm\ and\ either\ }
j_r \in \{ j,j+1\} \forall r \neq k\cr 
&\hbox{or\ }j_r\in\{j-1,j\}\forall r\ne k {\rm\ with\ }\min \{j_1 ,\ldots ,j_n \} =j-1 \Bigr\}\cr 
}\eqn
$$
The condition $\min \{ j_1 , \ldots , j_n \} = j-1$ implies that
the sequence $(j ,\ldots ,j)$ appears only once (i.e. $Z_k (j)
\cap V_k (j) = \emptyset$). The
result is then obviously equivalent to $\abs{j_r - j_s} \leq 1 $
for all $r,s$.

We apply this to $Y$ as follows. Note that all $T_k$ and $\cY$
depend on $j$ (unless $\cY$ is just a vertex of scale zero). For
all $r \in \{ 1,\ldots ,n-1\}$, $T_r$ actually depends on the
scale $j_r$ of a line going into $T_r$. Fix all $j_r$ for
$r\neq k$, then the scales of all $T_s$ are fixed. If $\cY$
depends on $j$, it is of the form
$$
\cY (q,p) = \sum\limits_{i> j} \cY_i (q,p)
\eqn
$$
We interchange the sums over $i$ and $j$
$$\eqalign{
Y(q)  = -2 &\sum\limits_{0>i\geq I} \sum\limits_{i> j\geq I}
\int d^{d+1} p \cY_i (q,p) {h(\ve{p}) \over ( ip_\zer -
e(\ve{p}) )^{n-1} }
 \sum\limits_{j_n ,\ldots ,j_n =j \atop \min \{ j_1 ,\ldots ,
j_n \} =j}^{j+1} 
\left(\prod\limits_{s=1}^{n-1} (1-\ell) T_s (p) \right)\cr
&\sum\limits_{l =1}^n
M^{-2{j_l}} f' ( M^{-2{j_l}} ( p_\zer^2 + e(\ve{p})^2))
\prod\limits_{r \neq l} f ( M^{-2{j_r}} ( p_\zer^2
+e(\ve{p})^2 ) ) \cr
\cr}\eqn $$
We have
$$\eqalign{
\bigcup\limits_{j=I}^i & \Bigl\{ (j_1 ,\ldots ,j_n) : j_k \in
\{ j,j+1\} \cap \{I,\ldots ,-1\}, \min \{j_1 ,\ldots ,j_n\}
=j \Bigr\} = \cr
&= V_{k} (i+1) \cup Z_{k} (I) \cup \bigcup\limits_{j=I+1}^{i}
M_k (j)
\cr}\eqn $$
as before.
Choose $k=l +1$ or $k=l -1$, then
$$\eqalign{
\sum\limits_{(j_1 ,\ldots ,j_n)\in M_k (j)} &\prod\limits_{s=1}^{n-1}
(1-\ell) T_s (p) \prod\limits_{r\neq l} f ( M^{-2j_r} (
p_\zer^2 + e(\ve{p})^2 )) M^{-2j_l} f'(
M^{-2{j_l}} ( p_\zer^2 + e (\ve{p})^2 ) ) = \cr
& = \sum\limits_{(j_1 ,\ldots ,j_n) \in M_k (j)
\atop j_l \; not \; summed}
\left( \prod\limits_{s=1}^{n-1} (1-\ell ) T_s (p) \right)
\sum\limits_{j_l =j-1}^{j+1} \biggl[ {\del \over \del x}
f (M^{-2 {j_l}} x) \biggr]_{x=p_\zer^2 + e(\ve{p})^2} \cr
& \prod\limits_{r\not\in \{ k,l\}} f (
M^{-2{j_r}} ( p_\zer^2 + e(\ve{p})^2 ))
f ( M^{-2j} ( p_\zer^2 + e(\ve{p})^2 ) )
\cr}\eqn $$

By momentum conservation and the support properties of the $f$,
\queq{\Nullzwei}, we can now extend the sum over $j_l$ to the
entire interval $\{I, \ldots ,i\}$. Using
$$  
\sum\limits_{j=I}^i f(M^{-2j}x) = a(M^{-2I}x) - a(M^{-2(i+1)}x)
\eqn
$$
(see \queq{\cutadef} and following)
and calling $E^2 = p_\zer^2 + e(\ve{p})^2$,
$$\eqalign{ 
f(M^{-2j} E^2) & \sum\limits_{j_l =I}^i M^{-2{j_l}} 
f'(M^{-2{j_l}} E^2) = \cr
& = f(M^{-2j} E^2) \Bigl( M^{-2I} a' (M^{-2I} E^2 ) - 
M^{-2(i+1)} a' (M^{-2(i+1)} E^2 ) \Bigr)
\cr}\eqn $$
Since $a'(x) \neq 0$ only for $x \in (M^{-4},M^{-2})$
and $f(x) \neq 0$ only for $x \in (M^{-4},1)$,
$$
f(M^{-2k} x) a' (M^{-2\ell}x) = 0{\rm\  for\ all\ }x 
{\rm \ unless\ } k \in \{\ell ,\ell -1 \}.
$$
Since $j \in \{ I, \ldots ,i\}$, 
the only nonzero terms in the sum over $j$ are $j=I$ and $j=i$.
Thus we get four boundary terms in the estimate for $Y$, two at 
both the lower and upper end of the summation region; the two from
$B_{1,k}(i)$ and $B_{2,k}(I)$ are similar to the following two, 
which we discuss in detail.

If $j=I$, we estimate
$$\eqalign{
&\bigg\vert 2 \sum\limits_{0>i\geq I} M^{-2I} \int d^{d+1} p \cY_i (q,p)
{h(\ve{p}) \over ( ip_\zer - e(\ve{p}) )^{n-1}}
\sum\limits_{(j_1 ,\ldots ,j_n) \in M_k (I) \atop
j_l \; not \; summed} \prod\limits_{s=1}^{n-1} (1-\ell) T_s (p) \cr &
\prod\limits_{r \not\in \{ k,l \} } f\biggl( M^{-2{j_r}}
\Bigl( p_\zer^2 + e (\ve{p}) ^2 \Bigr) \biggr) \bigg\vert \leq \cr
&\leq \Const \abs{ h}_\zer \sum\limits_{i=I}^{-1} \int d^{d+1} p
\abs{\cY_i (q,p)} {1 \over \abs{ip_\zer - e(\ve{p})}^{n-1}}
\prod\limits_{s=1}^{n-1} \abs{(1-\ell ) T_s (p)} \cr & 1 \biggl(
\abs{ ip_\zer - e(\ve{p})} \in [M^{I-2} , M^I] \biggr)
\cr}\eqn $$
We now use Taylor expansion for all the $r$-forks as in the
proof of Theorem \TwoFour, to bound this by
$$\eqalign{
\leq \Const \abs{h}_\zer &\left(\prod\limits_{s=1}^{n-1} M^I \abs{T_s}_1 \right)
M^{-2I} M^{-(n-1)(I-2)} \sum\limits_{i=I}^{-1} \int
d^{d+1} p \abs{\cY_i (q,p)} \cr & 1 \biggl(
\abs{ ip_\zer - e(\ve{p})} \in [M^{I-2} , M^I] \biggr) \cr
\leq \Const \abs{h}_\zer & R \prod\limits_{s=1}^{n-1} \abs{T_s}_1 
\cr}\eqn $$
where
$$
R = M^{-2I} \sup\limits_{q} \sum\limits_{i=I}^{-1} \int
d^{d+1} p \abs{\cY_i (q,p)} 1 \Bigl( \abs{ip_\zer -
e(\ve{p})} \in [M^{I-2} , M^I] \Bigr)
\eqn
$$
By maximality of $\tau'_\phi$, we know that there is a volume
improvement on scale $i$, so 
$$
R \leq M^{-2I} \Const \sum\limits_{i=I}^{-1} M^{\ep i} M^{2I}
\leq \Const
.\eqn $$
It remains to be shown that the product over $\abs{T_s}_1$ stays
bounded, i.e. that there are no factors $I$. This can be
seen by estimating the last sum in \queq{\rForkest} differently:
by Lemma \Factori\ $(v)$,
$$\eqalign{
\sum\limits_{j_w>j_{\pi (w)}} \la_n \left(j_w,\sfrac{\ep}{2}\right) M^{j_w\ep} &
= \sum\limits_{l=1}^{\abs{j_{\pi (w)}}-1} \la_n \left( -l ,
\sfrac{\ep}{2} \right) M^{-\ep l} \cr
& \leq \sum\limits_{l=1}^\infty (a_n l^n + b_n ) M^{-\ep l} \cr
& \leq \Const (n, \ep)
\cr}\eqn $$
independently of $j_{\pi (w)}$, so, inserting this into \queq{\rForkest}, we get
$\abs{T_s}_1 \leq \Const $ for all $s$. Thus this term is bounded.

At the upper summation end $j=i$, we have to estimate
$$
\eqalign{
& \bigg\vert \sum\limits_{0>i\geq I} \int d^{d+1} p\ \cY_i (q,p)
{h(\ve{p}) \over \Bigl( ip_\zer - e(\ve{p}) \Bigr)^{n-1}}
\cr 
 & \qquad M^{-2i}\sum\limits_{(j_1 ,\ldots ,j_n) \in M_k (i) \atop j_l not \;
summed} \prod\limits_{s=1}^{n-1} (1-\ell) T_s (p) \prod\limits
_{r \not\in \{k,l \} } f\biggl( M^{-2j_r} \Bigl( p_\zer^2
+ e(\ve{p})^2 \Bigr) \biggr) \bigg\vert  \cr
& \leq \Const \abs{h}_\zer \sum\limits_{i=I}^{-1} M^{-2i}
\int d^{d+1} p\ \abs{\cY_i (q,p)} 1 \Bigl( \abs{ ip_\zer -
e(\ve{p})} \in [M^{i-2} ,M^i ] \Bigr) \cr 
&\qquad M^{-(n-1)(i-2)}
\prod\limits_{s=1}^{n-1} (1-\ell) T_s (p) \cr
& \leq \Const \abs{h}_\zer \sum\limits_{i=I}^{-1} M^{-2i}
\prod\limits_{s=1}^{n-1} \abs{T_s}_1 \int d^{d+1} p
\abs{\cY_i (q,p)} 1 ( \abs{ip_\zer - e(\ve{p})} \in
[M^{i-2} ,M^i])
\cr}\eqn $$

By \queq{\Hachp} -- \queq{\Wui} without the $D_h$, the last integral is bounded
by $\Const M^{(2+\ep)i}$, so this is 
$$\eqalign{
& \leq \abs{h}_\zer \sum\limits_{i=I}^{-1} M^{\ep i} \prod\limits
_{s=1}^{n-1} \abs{T_s}_1 \cr
& \leq \Const \abs{h}_\zer
\cr}\eqn $$
\hfil \endproof
\Rem{\Cauchy} Actually, the following stronger convergence statement holds.
Let $(t,G)$ be fixed, $E(G)=2$, $G$ 1PI. Let $\abs{h}_1<\infty$ and 
for $I' > I$ 
$$
V_I^{I'} (h,t,G) =
\sum\limits_{j=I}^{I'-1} \sum\limits_{J \in \cJ (t,j )} D_h Val (G^J) 
\eqn $$
Then, as $I \to -\infty $, $V_I^{I'} \to V^{I'}$ with  
$\abs{V^{I'}}_\zer \leq \abs{h}_\one W^{I'}$ and
$W^{I'}$ vanishes as $I'\to -\infty$. In particular, for all $k \geq 0$ 
$$
\abs{V_I (h,t,G) - V_{I-k}(h,t,G)}_\zer \leq  \abs{h}_\one \tilde W_I
\EQN\Meiomei $$
with $\tilde W_I \to 0 $ as $I \to -\infty $.

\sect{Convergence of the Derivative}
\noindent
In this section, we prove Theorems \Kabl, \Inject, and 
\HFbrav.   
Given all the detailed information about the 
two--legged and four-legged vertices that we have 
gathered in the last two chapters, these proofs will be
applications of elementary convergence theorems
for absolutely convergent series.
For convenience of the reader, we summarize the results 
derived so far. Recall the explicit expression \queq{\KrIform}
$$
K_r^I (\ve{p})= - \sum\limits_{G}
\sum\limits_{j=I}^{-1} \sum\limits_{t \sim G}
\prod\limits_{f \in t} {1 \over n_f !}
\sum\limits_{J \in \cJ (t,j)} Val (G^J) (0, \prP (\ve{p}))
. $$
For all $r\geq 1$,
$K_r^I$ converges as $I \to \infty $ in $\abs{\; \cdot \; }_\one $ 
to a function $K_r \in C^1 (\cB, \R )$ (see Section 2.7).
Let $\cA $ be as in Definition \Atwodef, and $\cL$ as defined thereafter.
Let $e$ in $\cA$. Since for $I > -\infty$, $K_r^I$ is differentiable 
in the sense of Fr\' echet \quref{D}, the derivative 
$\left( K_r^I \right)' \in \cL $ can be evaluated as the directional 
derivative
$$
\left( K_r^I \right)' (h) = D_h K_r^I (e)
.\EQN\direderi $$
Fix $r \geq 1$, let $I = -n$ and denote $K_r^{-n} = \ka_n$. Then by 
\queq{\onebou} (summed from $-n-m$ to $-n$), $(\ka_n)_{n \geq 1} $ satisfies 
$$
\lim_{n\rightarrow\infty}\sup_{m>0}\sup\limits_{\abs{h}_\one = 1}
{1 \over \abs{h}_\one}
\abs{\ka_n' (h) - \ka_{n+m}' (h)}_\zer 
=0
\eqn $$
for all $m\geq 1$, thus it converges
in operator norm $\norm{\;\cdot\;}_\cL$ to an operator $\ka_e' \in \cL$. 
By \queq{\direderi}, 
$$
\ka_e' (h) = \lim\limits_{n \to \infty} D_h \ka_n
\eqn $$
so by \queq{\zerbou} 
$$
\abs{\ka_e' (h)}_\zer \leq \Const \abs{h}_\zer
.\eqn $$
So far this was all at fixed $e \in \cA$. 
The constants and bounds
depend on $u_\zer$, $\ep$, $\abs{e}_\two $, $\abs{u}_\two$
and the size $\de$ of the neighbourhood of $S$. 
By definition of $\cA$, \queq{\Atwodef}, every $e \in \cA$ has a 
neighbourhood $\cU$  on which these constants are uniform, and thus 
all the above bounds hold uniformly in $e$, and also the convergence is 
uniform. It remains to be shown 
that $\ka' $ is the derivative of $\ka$ and that the map $e \to \ka'_e$
is continuous on $\cA$. We first show by an $\veps/3$--argument 
that $e \mapsto \ka'_e$ is continuous, as follows. Write 
$$
\ka'_{e_\one} - \ka'_{e_\two} = 
\ka'_{e_\one} - \ka'_n ({e_\one}) +
\ka'_n({e_\one}) - \ka'_n({e_\two})+
\ka'_n({e_\two}) - \ka'_{e_\two}
\eqn $$
Let $\veps > 0$, then there is $n \geq 1$ such that 
$\norm{\ka'_{e_i} - \ka'_n (e_i)} < \veps/3$ for all $e_i\in\cU$. 
Fix $I=-n$ with that property. At fixed $I$, 
$$
\norm{(K_r^I)' (e_\one ) - (K_r^I) ' (e_\two )} < C_I \norm{e_\one -e_\two}
\eqn $$
$C_I$ grows with $I$, but we need only one fixed value of $I$. 
So for $\norm{e_\one -e_\two} < \veps / (3C_I)$, 
$\norm{\ka_n' (e_\one ) - \ka_n' (e_\two )} < \veps $.
It is now easy to see that $\ka' $ is the derivative of $K_r= \ka$ 
since   
$$
\ka_n (e_\two ) - \ka_n (e_\one ) = 
\ka_n' (e_\one ) (e_\two - e_\one ) + 
\int\limits_0^1 ds\ ( \ka_n'((1-s)e_\one + s e_\two ) - \ka_n'(e_\one ) )
(e_\two - e_\one )
\eqn $$
so, taking the limit $n \to \infty $, and calling $h=e_\two -e_\one $, 
$$
\ka (e_\one+h ) - \ka (e_\one ) = \ka'_{e_\one } (h) + 
\int\limits_0^1 ds\ ( \ka'_{e_\one + s h} - \ka'_{e_\one} ) (h)
.\eqn $$
The second term is $o(h)$ by continuity of $\ka'$, so $\ka'$ is 
indeed the derivative of $\ka $. This proves Theorem \Kabl. 

\def\KlaR{{K_\la^{(R)}}}
\sni
{\it Proof of Theorem \Inject:} Let $e_\one $ and $e_\two $ be as stated 
in Theorem \Inject, $\KlaR (e) = \sum\limits_{s=1}^R \la^s K_s (e)$, 
and $e_\one + \KlaR (e_\one ) = e_\two + \KlaR (e_\two)$. Then
$$
e_\two -e_\one + 
\int\limits_0^1 ds \; {\del \over \del s} \KlaR ((1-s)e_\one + s e_\two )
=0
,\eqn $$
that is, 
$$
( \bbbone + {\bf L} ) (e_\two -e_\one ) = 0
\eqn $$
with 
$$
{\bf L}(h) = \int\limits_0^1 ds \; \sum\limits_{s=1}^R \la^s K_s' 
(e_\one + s (e_\two - e_\one ))(h)
\eqn $$
Since $e_\one + s (e_\two - e_\one ) \in \cA $ for all $s \in [0,1]$,
$$
\norm{K_s' (e_\one + s (e_\two - e_\one ))}_\cL \leq C_s,
\eqn $$
so $\norm{{\bf L}}_\cL <1$ and $\bbbone+ {\bf L} $ is injective. Thus 
$e_\two - e_\one =0$. \endproof

\sni
Theorem \HFbrav\ follows from the observation made earlier that
any derivative with respect to the external momentum of a Hartree-Fock 
type graph will only act on $\hat v$. Since $\hat v \in C^k$, so is
$D_h H_r^I$, and the statement follows from a standard application of 
the contraction mapping theorem. 

\vfill\eject
\appendix{A}{Volume Estimates}

\noindent
In this Appendix, we prove Proposition \NoNest\ and uniformity of 
$C_{vol}$ and $\ep$ on the set $\cA_\two$ defined in \queq{\Atwodef}.
By definition \queq{\Itwodef}, the integral $I_2$ is symmetric in 
all three arguments, so we may assume that $\veps_3 \geq 
\max \{ \veps_\one, \veps_\two \}$. By choice of $C_{vol}$ we may also 
assume that $\veps_3$ is so small that $\abs{e (\ve{p} )} \leq 2 \veps_3 $
implies $\ve{p} \in U_\de (S)$, with $\de $ given in Lemma \uProps, and 
that $\veps_3 \leq G_\zer \abs{e}_\two /2$. If $S$ has more than one 
connected component, we may also assume that $\veps_3 $ is so small 
that the neighbourhood $\abs{e (\ve{p} )} \leq 2 \veps_3 $ falls 
into the same number of connected components as $S$.
Let the coordinates $(\rh, \om )$ be as in Lemma \uProps\ $(iv)$ and denote
$\ve{p}$ as a function of these coordinates by $\ve{p} (\rh, \om )$, 
then 
$$\eqalign{
I_\two (\veps_\one, \veps_\two, \veps_3) &= 
\sup\limits_{\ve{q} \in \cB} 
\max\limits_{v_\one, v_\two \in \{ -1,1\}}
\int\limits_{-\veps_\one}^{\veps_\one} d \rh_\one
\int\limits_S d\om_\one\ J(\rh_\one, \om_\one ) \cr
&\qquad\int\limits_{-\veps_\two}^{\veps_\two} d \rh_\two
\int\limits_S d\om_\two\ J(\rh_\two, \om_\two )
1(\abs{e(v_\one \ve{p} (\rh_\one, \om_\one ) + 
v_\two \ve{p} (\rh_\two, \om_\two ) + \ve{q} )} \leq \veps_3 ) \cr
& \leq 4\left( {A_\zer \over u_\zer }\right)^2 \veps_\one \veps_\two
\sup\limits_{\ve{q} \in \cB} 
\max\limits_{v_\one, v_\two \in \{ -1,1\}}
\sup\limits_{\abs{\rh_\one}, \abs{\rh_\two} \leq \veps_3}
\int\limits_S d\om_\one\cr
&\qquad\int\limits_S d\om_\two\
1(\abs{e(v_\one \ve{p} (\rh_\one, \om_\one ) + 
v_\two \ve{p} (\rh_\two, \om_\two ) + \ve{q} )} \leq \veps_3 )
.\cr}\eqn $$
By Lemma \uProps\ and the mean value theorem
$$
\abs{e(v_\one \ve{p} (\rh_\one, \om_\one ) + 
v_\two \ve{p} (\rh_\two, \om_\two ) + \ve{q} ) -
e(v_\one \ve{p} (0 , \om_\one ) + 
v_\two \ve{p} (0, \om_\two ) + \ve{q} )} \leq 2 \abs{e}_1 \sfrac{\veps_3}{u_\zer} 
\eqn $$
for all $\rh_\one, \rh_\two $ with $\abs{\rh_i} \leq \veps_3$.
Thus 
$$
\abs{e(v_\one \ve{p} (\rh_\one, \om_\one ) + 
v_\two \ve{p} (\rh_\two, \om_\two ) + \ve{q} )} \leq \veps_3
\eqn $$
implies 
$$
\abs{e(v_\one \ve{p} (0 , \om_\one ) + 
v_\two \ve{p} (0, \om_\two ) + \ve{q} )} 
\leq\left(1+2\sfrac{\abs{e}_1}{u_\zer}\right)  \veps_3
,\eqn $$
hence
$$
I_\two (\veps_\one, \veps_\two, \veps_3 ) \leq 
4\left( A_\zer \over u_\zer \right)^2 \veps_\one \veps_\two 
W\left(\left(1+2\sfrac{\abs{e}_\one}{u_\zer}\right)\veps_3\right)
\eqn $$
with 
$$
W(\veps ) = 
\sup\limits_{\ve{q} \in \cB} 
\max\limits_{v_\one, v_\two \in \{ -1,1\}}
\int\limits_S d \om _\one 
\int\limits_S d \om _\two \; 
1(\abs{e(v_\one \ve{p} (0 , \om_\one ) + 
v_\two \ve{p} (0, \om_\two ) + \ve{q} )} 
\leq \veps )
.\eqn $$
Thus the function $W(\veps )$ contains the improvement over ordinary
power counting. The following Lemma implies the bound \queq{\ImpVol} 
for $I_\two $ with 
$$
C_{vol}= Z_3 \left(1+2\sfrac{\abs{e}_\one}{u_\zer}\right)^\ep
,\EQN\CvolForm $$ 
where $Z_3$ is a constant that depends only on 
$Z_\one, Z_\two, \rh,\ka,g_\zer $ and $\abs{e}_\two $.

\Lem{\OnlyImp} {\Lesty $W(\veps ) \leq Z_3 \veps^\ep $ }

\Proof 
Let $\be \in (0,1)$ and $\cT \subset S \times S$ be the set where 
the intersection is transversal, 
$$
\cT = \{ (\om_\one, \om_\two ) \in S \times S :\;
\sqrt{1-\left( n(\om_\one ) \cdot n(\om_\two )\right) ^2} \geq \veps^{1-\be}\}
,\eqn $$
and $\cE$ its complement, $\cE = S \times S \setminus \cT$,
and split $W(\veps ) = T(\veps ) + E(\veps )$ 
into the contributions from these two sets.
The idea is that if $(\om_\one, \om_\two ) \in \cT $, then the 
tangent spaces $T_{\om_\one } \; S$ and $T_{\om_\two} \; S$ span $\R^d$, 
$T_{\om_\one } \; S+T_{\om_\two} \; S=\R^d$, and therefore a combination 
of $\om_\one $ and $\om_\two $ will be transversal to $S$, and 
that $\cE$ has small measure because of \AThr. $\be $ will be chosen 
at the end.
 
Fix any $(\bar\om_\one,\bar \om_\two)\in \cT $, and for $i\in \{ 1,2\}$ let 
$T_i = T_{\bar\om_i }S$ as a subspace of $\R^d$ (in other words, 
$T_i = \{ x \in \R^d: n(\bar\om_i )\cdot x = 0\} $). 
By transversality, $T_\one \cap T_\two$ is a proper subspace of 
both $T_\one $ and $T_\two $. 
Choose an ONB $a_\one, \ldots ,a_{d-1}$
for $T_\one$ such that $a_\one $ is orthogonal to $T_\one \cap T_\two$,
and an ONB $b_\one, \ldots , b_{d-1} $ for $T_\two$ such that 
$b_\one $ is orthogonal to $T_\one \cap T_\two $, 
then $|a_\one \cdot b_\one| =| n(\bar\om_\one ) \cdot n(\bar\om_\two )|$.
Let 
$$
a_d = { b_\one - (b_\one \cdot a_\one ) a_\one  
\over \sqrt{ 1 - (n(\om_\one ) \cdot n(\om_\two))^2}}
.\eqn $$

For $(\al'_1,\al_2,\cdots,\al_{d-1})$ in a neighbourhood of the origin let
$\om_1(\al'_1,\al_2,\cdots,\al_{d-1})\in S$ be the image of
$\al'_1a_1+\sum_{i=2}^{d-1}\al_i a_i\in T_{\bar\om_1}S$ under the 
exponential  map. Similarly let $\om_1(\be_1\cdots,\be_{d-1})\in S$ be the 
image of $\sum_{i=1}^{d-1}\be_i b_i\in T_{\bar\om_2}S$. The Jacobian
$\sfrac{\partial(\om_1,\om_2)}{\partial(\al'_1,\al_2,\cdots,\be_{d-1})}$
is bounded by a constant in a neighbourhood of the origin. Make the further
change of variables substituting
$$\eqalign{
\al'_1&=\al_1-\frac{a_1\cdot b_1}
{\sqrt{1-(n(\om_1)\cdot n(\om_2)\big)^2}}\al_d\cr
\be_1&=\frac{v_1 v_2}
{\sqrt{1-(n(\om_1)\cdot n(\om_2)\big)^2}}\al_d\cr
}$$
for $(\al'_1,\be_1)$. The Jacobian is
$$\eqalign{
\left|\sfrac{\partial(\al'_1,\al_2,\cdots\al_{d-1},\be_1,\cdots,\be_{d-1})}
{\partial(\al_1,\al_2,\cdots,\al_d,\be_2,\cdots,\be_{d-1})}\right|
&=\left|\det
\pmatrix{1&-\frac{a_1\cdot b_1}{\sqrt{1-(n(\om_1)\cdot n(\om_2)\big)^2}}\cr
         0&\frac{v_1v_2}
{\sqrt{1-(n(\om_1)\cdot n(\om_2)\big)^2}}\cr}\right|\cr
&=\frac{v_1 v_2}{\sqrt{1-(n(\om_1)\cdot n(\om_2)\big)^2}}
\le\veps^{\be-1}
}$$

Define
$$
\rho_3=e(v_1\ve{p}(0,\om_1)+v_2\ve{p}(0,\om_2)+\ve{q})
$$
viewed as a function of $\al_1,\cdots,\al_d,\be_2,\cdots,\be_{d-1}$.
Note that for $1\le  i\le d-1$
$$
\frac{\partial \rho_3}{\partial \al_i}\bigg|_0
=v_1\nabla e(v_1\ve{p}(0,\bar\om_1)
+v_2\ve{p}(0,\bar\om_2)+\ve{q})\cdot a_i
$$ 
and
$$\eqalign{
\frac{\partial \rho_3}{\partial \al_d}\bigg|_0
&=v_1\nabla e(v_1\ve{p}(0,\bar\om_1)
+v_2\ve{p}(0,\bar\om_2)+\ve{q})\cdot 
\frac{-a_1\cdot b_1}{\sqrt{1-(n(\om_1)\cdot n(\om_2)\big)^2}}a_1\cr
&\hskip.5in+v_2\nabla e(v_1\ve{p}(0,\bar\om_1)
+v_2\ve{p}(0,\bar\om_2)+\ve{q})\cdot 
\frac{v_1v_2}{\sqrt{1-(n(\om_1)\cdot n(\om_2)\big)^2}}b_1\cr
&=v_1\nabla e(v_1\ve{p}(0,\bar\om_1)+v_2\ve{p}(0,\om_2)+\ve{q})\cdot a_d\cr
}$$
 
Because $a_1,\cdots, a_d$ is an orthonormal basis, there is a  $j$
such that
$$
\left|\frac{\partial \rho_3}{\partial\al_j}\right|\bigg|_0
\ge\frac{1}{\sqrt{d}}
|\nabla e(v_1\ve{p}(0,\bar\om_1)+v_2\ve{p}(0,\om_2)+\ve{q})|
\ge\frac{1}{\sqrt{d}}g_\zer
$$
and $\left|\frac{\partial\rho_3}
{\partial\al_j}\right|\ge\frac{g_0}{2\sqrt{d}}$
in a neighbourhood of the origin. Make a final change of variables
replacing $\al_j$ by $\rho_3$. The Jacobian for the composite change
of variables from $(\om_1,\om_2)$ to
$(\al'_1,\al_2,\cdots,\al_{d-1},\be_1,\cdots,\be_{d-1})$
to $(\al_1,\cdots,\al_{d},\be_2,\cdots,\be_{d-1})$ to
$\big((\al_i)_{1\le i\le d\atop i\ne j},
(\be_i)_{2\le i\le d-1},\rho_3)$ is bounded by $\Const\veps^{\be-1}$.
Covering $S\times S$ by a finite number of such coordinate patches we have
$$
T(\veps ) \leq \Const \veps^{\be -1} 
\int\limits_{-\Const\veps}^{\Const\veps} d\rho_3
\leq \Const \veps^\be
.\eqn $$

The contribution from the set of exceptional momenta $\cE $ is bounded 
using \AThr: fix $\om_\one \in S$, let $\cD (\om_\one )$ be as in 
\queq{\cDomdef}, and let 
$$
\cE_{\om_\one} =\Big \{ \om \in S: 
\sqrt{1-\left( n(\om_\one ) \cdot n(\om_\two )\right) ^2} <
\veps^{1-\be}\Big\}
\eqn $$
Then by \AThr\ $(ii)$, $\cE_{\om_\one} \subset U_r (\cD (\om_\one ))$ with 
$r = (\veps^{1-\be} /Z_\one )^{1/\rh }$ (this $\rh$ is now that from \AThr, 
not the `radial' coordinate), so by \AThr\ $(i)$, 
$$
E(\veps ) \leq  \int\limits_S d \om_\one 
\int\limits_{U_r(\cE_{\om_\one})} d \om_\two 
\leq \left( Z_\zer (Z_\one)^{-1/\rh} \right)\; \veps^{\ka (1-\be )/\rh }
.\eqn $$
The optimal bound is when $\ka (1-\be )/\rh =\be $, that is,  
$\be = \ka / (\ka + \rh )$. \endproof

\Lem{\UnifVol} {\Lesty Let $\cA = \cA_\two (\si,\cN,g_\zer,g_\two,g_3)$ be
as in \queq{\Atwodef}. Then $\cA $ is open, and $\rh, \ka, Z_\one, Z_\two$, 
and thus $C_{vol}$ can be chosen uniformly on $\cA$, i.e.\ \queq{\ImpVol}
holds with the same $\ep $ and $C_{vol}$ for all $e \in \cA$. } 

\Proof It is obvious by definition of $\cA$ that it is an open set. 
Let $\om \in S=S(e)$. Since $\abs{n(\om )}=1$, $dn(\om)h$ is orthogonal to
$n(\om)\in \R^d$ for all $h$ in the tangent space at $\om$. $\;\cD (\om)$,
defined in \queq{\cDomdef}, is the zero set of $\phi_\om: S \to \Om^{(2)}(S)$, $\om' \mapsto n(\om') \wedge n(\om )$. $\; d\ph_\om (\om') =
dn(\om') \wedge n(\om )$, the mentioned orthogonality and rank $dn \geq \si$
imply that $\cD(\om)$ is a $C^{k-1}$--submanifold of $S$ of codimension 
$\geq \si$. It is now clear that \AThr\ holds, with $\ka=\si$, $\rh=1$, 
and with $Z_\zer$ and $Z_\one$ depending on the smallest eigenvalue 
(in absolute value), hence bounded by a function of $g_3$. Uniformity of 
$C_{vol}$ on $\cA_\two (\si,\cN,g_\zer,g_\two,g_3)$
follows from \queq{\CvolForm}, that of $\ep$ from Proposition 
\NoNest. \endproof
\goodbreak

\def\bw#1{\phi_{\raise-.75ex\hbox{$#1$}}}  %Bloch wave
                 %dispersion relation
\def\an#1{a_{\raise-.75ex\hbox{$#1$}}}     %annihilation operator
\def\ct#1{a^\dagger_{\raise-.75ex\hbox{$#1$}}}  %creation operator

\def\<{\left<}
\def\>{\right>}
\def\Lone{\raise-.5ex\hbox{$\scriptstyle L^1$}}
\def\Linfty{\raise-.5ex\hbox{$\scriptstyle L^\infty$}}

\appendix{B}{The One--Fermion Problem}
\noindent
Let $d\ge 2$, $\Ga$ be a lattice of maximal rank in $\R^d$, and
$$
\Ga^\# =\{b\in\R^d\ |\ \langle b,\ga\rangle \in 2\pi \Z \hbox{ for all }
\ga\in \Ga\}
\eqn $$
its dual. Let $q(\ve{x})$ be a smooth potential that is periodic with respect to 
$\Ga$. Then the Schr\"odinger operator $-\sfrac{1}{2m}\De+q(\ve{x})$ commutes with 
the unitary lattice translation operators
$$
(T^\ga\phi)(\ve{x})=\phi(\ve{x}+\ga)\ \ \ \ \ \ \ga\in\Ga
\eqn $$
so that the spectrum of $-\sfrac{1}{2m}\De+q(\ve{x})$ is the union over 
$\ka\in\R^d/\Ga^\#$ of the spectra of the boundary value problems
$$\eqalign{
\left(-\sfrac{1}{2m}\De+q\right)\phi&=\la\phi\cr
\phi(\ve{x}+\ga)&=e^{i<\ka,\ga>}\phi(x)\ \ \ \ \ \ \forall\,\ga\in\Ga.\cr
}\eqn $$
Label the eigenvalues of this problem, in increasing order, $\ep_\nu(\ka),
\ \nu\in\N $. Denote the corresponding eigenfunctions $\phi_{\ka,\nu}(\ve{x})$ 
and normalize them by the condition that
$$
\int_{\R^d/\Ga}d\ve{x}\,|\phi_{\ka,\nu}(\ve{x})|^2=V_\Ga:\hskip-.5pt
={\rm Vol}(\R^d/\Ga)\ .
\eqn $$
This normalization is chosen so that when $q=0$, 
$\{\phi_{\ka,\nu}\ |\ \nu\in\N\}=\{e^{i<\ve{k},\ve{x}>}\ |\ \ve{k}\in
\ka+\Ga^\#\}$ and $\ep_\nu(\ka)=\sfrac{1}{2m}\ve{k}^2$.
The boundary value problem
$$\eqalign{
\left(-{1\over 2m}\De+q\right)\phi&=\la\phi\cr
\phi(\ve{x}+\ga)&=e^{i<\ka,\ga>}\phi(x)\ \ \ \ \ \ \forall\,\ga\in\Ga\cr
}\eqn $$
is unitarily equivalent to
$$\eqalign{
\left(\sfrac{1}{2m}(-i\nabla+\ka)^2+q\right)u&=\la u\cr
u(\ve{x}+\ga)&=u(x)\ \ \ \ \ \ \forall\,\ga\in\Ga.\cr
}\eqn $$
As $(-i\nabla+\ka)^2$ is an analytic relatively bounded perturbation of $-\De$, 
the eigenvalues $\ep_\nu(\ka)$ and eigenfunctions $\phi_{\ka,\nu}(x)$ depend
analytically on $\ka$ at every $\ka$ for which $\ep_\nu(\ka)$ is a simple
eigenvalue. The Fermi surface
$$
{\cal S}_\mu=\{\ka \ |\ \exists \nu{\rm \ such\ that\ }\ep_\nu(\ka)=\mu\}
\eqn $$
for chemical potential $\mu$ is smooth at every $\ka$ for which
$$\eqalign{
\ep_\nu(\ka)=\mu\Rightarrow&(a)\ \ep_\nu(\ka){\rm\ is\ a\ simple\ eigenvalue}\cr
&(b)\ \nabla\ep_\nu(\ka)\ne 0
}\eqn $$
Since  $\nabla\ep_\nu(\ka)= 0$ is a system of $d$ equations in $d$ unknowns
$\ka$, condition (b) is generically violated only at isolated points in 
$\R^d/\Ga^\#$. In this paper, we exclude it by assumption \ATwo.
We also restrict to one band since for bands separated by a gap, 
the band index plays no interesting role.

The free two point Schwinger function
$$\eqalign{
C(\xi,\xi')&=-\<\Phi_0,T[\psi(\xi)\bar\psi(\xi')]\Phi_0\>\cr
&\hskip-22pt=-{1\over (V_\Ga L^d)^2}\sum_{\ve{k},\ve{k}'}\bw{\ve{k}}(\xi)\bar\bw{\ve{k}'}(\xi')
e^{-(\ep_\nu(\ka)-\mu)\tau}e^{(\ep_{\nu'}(\ka')-\mu)\tau'}
\<\Phi_0,T[\an{\ve{k},\si}\ct{\ve{k}',\si'}]\Phi_0\>\cr
&\hskip-22pt=-{1\over V_\Ga L^d}\sum_{\ve{k},\ve{k}'}\bw{\ve{k}}(\xi)\bar\bw{\ve{k}'}(\xi')
e^{-(\ep_\nu(\ka)-\mu)\tau}e^{(\ep_{\nu'}(\ka')-\mu)\tau'}
\de_{\ve{k},\ve{k}'}\de_{\si,\si'}\cases{-\chi(\ep_\nu(\ve{k})<\mu)&$\tau\le\tau'$\cr
                          \chi(\ep_\nu(\ve{k})>\mu)&$\tau> \tau'$\cr}\cr
&\hskip-22pt=\de_{\si,\si'}{1\over V_\Ga L^d}\sum_{\ve{k}}\bw{\ve{k}}(\xi)\bar\bw{\ve{k}}(\xi')
e^{-(\ep_\nu(\ka)-\mu)(\tau-\tau')}
\cases{\chi(\ep_\nu(\ve{k})<\mu)&$\tau\le\tau'$\cr
      -\chi(\ep_\nu(\ve{k})>\mu)&$\tau> \tau'$\cr}\cr
&\hskip-22pt=\de_{\si,\si'}{1\over V_\Ga L^d}\sum_{\ve{k}}\int {dk_0\over 2\pi}\,
\bw{\ve{k}}(\xi)\bar\bw{\ve{k}}(\xi')e^{-ik_0(\tau-\tau')}
{1\over ik_0-e_\nu(\ka)}\cr
}\eqn $$
where
$$
e_\nu(\ka)=\ep_\nu(\ka)-\mu
\eqn $$
and for $\tau=\tau'$ the limit $\tau-\tau'\nearrow 0$ is implied. In the
infinite volume limit
$$
C(\xi,\xi')=\de_{\si,\si'}\sum_{\nu}\int_{\R\times\R^d/\Ga^\#}
{dk_0\over 2\pi}{d\ka\over(2\pi)^d}\,
\bw{\ve{k}}(\xi)\bar\bw{\ve{k}}(\xi')e^{-ik_0(\tau-\tau')}
{1\over ik_0-e_\nu(\ka)}
\eqn $$
Since we consider only one band, we drop $\nu $ and set $\ka = \ve{k}$.

\Refskip
\centerline{\bfe References}
\someroom
{\parskip=0.5cm
\Ref{AGD}{A.A.\ Abrikosov, L.P.\ Gorkov, I.E.\ Dzyaloshinski, {\sl Methods of 
Quantum Field Theory in Statistical Mechanics}, 
Dover 1975}
\Ref{BR}{O.Bratteli, D.W.Robinson, {\sl Operator Algebras and 
Quantum Statistical Mechanics II}, Springer Texts and Monographs
in Physics}
\Ref{D}{J.\ Dieudonn\' e, {\sl Foundations of Modern Analysis}, vol. 1} 
\Ref{FKLT}{J.\ Feldman, H.\ Kn\" orrer, D.\ Lehmann, E.\ Trubowitz, 
{\sl Fermi Liquids in Two Space Dimensions,} in {\sl Constructive Physics},
V. Rivasseau (ed.), Springer Lecture Notes in Physics, 1995}
\Ref{FMRS}{J.Feldman, J.Magnen, V.Rivasseau, R. S\' en\' eor, 
{\sl Bounds on Renormalized Feynman Graphs, Commun.Math.Phys.
\bf 100} (1985) 23}
\Ref{FMRT}{J.Feldman, J.Magnen, V.Rivasseau, E.Trubowitz, {\sl
An Infinite Volume Expansion for Many Fermion Green's Functions,
Helvetica Physica Acta \bf 65} (1992) 679}
\Ref{FT1}{J.Feldman and E.Trubowitz, {\sl Perturbation Theory for
Many--Fermion Systems, Helvetica Physica Acta \bf 63} (1990) 156}
\Ref{FT2}{J.Feldman and E.Trubowitz, {\sl The Flow of an Electron--Phonon
System to the Superconducting State, Helvetica Physica Acta \bf 64}
(1991) 213}
\Ref{GJ}{J.Glimm and A.Jaffe, {\sl Quantum Physics}, Springer 1987}
\Ref{L} {E.H.Lieb, {\sl The Hubbard Model: Some Rigorous Results
and Open Problems}, in the proceedings of the conference
``Advances in Dynamical Systems and Quantum Physics'',
Capri, May 1993 (World Scientific)}
\par}

\end